\documentclass{article}
\usepackage{stmaryrd}
\usepackage{latexsym}
\usepackage{amsmath}
\usepackage{amssymb}
\usepackage{amsfonts}
\usepackage{ifthen}
\usepackage{mathrsfs}               
\usepackage{bbm}                    
\usepackage{bbold}                  
\usepackage{textcomp}               
\usepackage{amstext}
\usepackage{enumerate}
\usepackage{amstext}
\newcommand{\CC}{\mathbb{C}}

\newcommand{\NN}{\mathbb{N}}

\newcommand{\RR}{\mathbb{R}}

\newcommand{\ZZ}{\mathbb{Z}}
\newcommand{\supp}{\mathrm{supp}}
\newcommand{\const}{\mathrm{const}}

\newcommand{\Ran}{\mathrm{Ran}}

\newcommand{\loc}{\mathrm{loc}}
\newcommand{\el}{\mathrm{el}}
\newcommand{\nr}{\mathrm{nr}}
\newcommand{\np}{\mathrm{np}}
\newcommand{\Pf}{\mathrm{PF}}
\newcommand{\id}{\mathbbm{1}}
\newcommand{\klg}{\leqslant} 
\newcommand{\grg}{\geqslant}          
\newcommand{\ve}{\varepsilon}
\newcommand{\vp}{\varphi}

\newcommand{\vr}{\varrho}
\newcommand{\vt}{\vartheta}
\newcommand{\vs}{\varsigma}
\newcommand{\wt}[1]{\widetilde{#1}}
    
\newcommand{\SPn}[2]{\langle \,#1\,|\,#2\, \rangle} 
\newcommand{\SPb}[2]{\big\langle \,#1\,\big|\,#2\, \big\rangle}
\newcommand{\RSP}[2]{( \,#1\,|\,#2\, )} 
\newcommand{\RSPb}[2]{\big( \,#1\,\big|\,#2\, \big)} 
\newcommand{\SPB}[2]{\Big\langle \,#1\,\Big|\,#2\, \Big\rangle}
\newcommand{\rrb}{\rrbracket}
\newcommand{\llb}{\llbracket}
\newcommand{\ol}[1]{\overline{#1}} 
\newcommand{\ul}[1]{\overline{#1}} 
\newcommand{\wh}[1]{\widehat{#1}}  
\newcommand{\mr}[1]{\mathring{#1}} 
\newcommand{\bigO}{\mathcal{O}}    
\newcommand{\fourier}{\mathcal{F}} 
\newcommand{\V}[1]{\mathbf{#1}}
\newcommand{\valpha}{\boldsymbol{\alpha}}
\newcommand{\vxi}{\boldsymbol{\xi}}

\newcommand{\vsigma}{\boldsymbol{\sigma}}

\newcommand{\veps}{\boldsymbol{\varepsilon}}

\newcommand{\vnu}{\boldsymbol{\nu}}

\newcommand{\LO}{\mathscr{L}}      

\newcommand{\HP}{\mathscr{K}}
\newcommand{\HR}{\mathscr{H}}
\newcommand{\HRp}{\mathscr{H}^+_{\mathbf{A}}}
\newcommand{\Fock}{\mathscr{F}_{\mathrm{b}}}
\newcommand{\core}{\mathscr{D}}
\newcommand{\dom}{\mathcal{D}}
\newcommand{\form}{\mathcal{Q}}
\newcommand{\ball}[2]{\mathcal{B}_{#1}(#2)}
\newcommand{\spec}{\mathrm{\sigma}}

\newcommand{\Hfme}{H_{\mathrm{f},m,\varepsilon}} 
\newcommand{\Hfmg}{H_{\mathrm{f},m}^>} 
\newcommand{\PAm}{P^-_\mathbf{A}}               
\newcommand{\PA}{P^+_{\mathbf{A}}}

\newcommand{\PApm}{P^{\pm}_{\mathbf{A}}}
\newcommand{\PAepm}{P^{\pm}_{\mathbf{A}_{m,\ve}}}
\newcommand{\PAemp}{P^{\mp}_{\mathbf{A}_{m,\ve}}}
\newcommand{\PAmgpm}{P^{\pm}_{\mathbf{A}_{m}^>}}
\newcommand{\PO}{P^+_{\mathbf{0}}}
\newcommand{\POpm}{P^{\pm}_{\mathbf{0}}}
\newcommand{\POm}{P^-_{\mathbf{0}}}
\newcommand{\SO}{S_{\mathbf{0}}}
\newcommand{\SA}{S_{\mathbf{A}}}

\newcommand{\DA}{D_{\mathbf{A}}}                
 
\newcommand{\DAme}{D_{\mathbf{A}_{m,\varepsilon}}}
\newcommand{\DO}{D_{\mathbf{0}}}

\newcommand{\R}[2]{R_{#1}(#2)}                  
\newcommand{\RA}[1]{R_{\mathbf{A}}(#1)}

\newcommand{\Hf}{H_{\mathrm{f}}} 
\newcommand{\Hft}{\widetilde{H}_{\mathrm{f}}}         
\newcommand{\HT}{\check{H}_{\mathrm{f}}}      
\newcommand{\NPO}[1]{H_{#1}^{\mathrm{np}}}      
\newcommand{\PF}[1]{H_{#1}^{\mathrm{PF}}}       
\newcommand{\Ynp}[1]{Y^{\mathrm{np}}_{#1}}

\newcommand{\whPF}[1]{\widehat{H}_{#1}^{\mathrm{PF}}}
\newcommand{\YPF}[1]{Y^{\mathrm{PF}}_{#1}}
\newcommand{\tgV}{\tfrac{\gamma}{|\V{x}|}}
\newcommand{\BR}[1]{B^{\mathrm{el}}_{#1}}
\newcommand{\Hs}[1]{H_{#1}^{\sharp}}
\newcommand{\pe}{\mathbf{p}}
\newcommand{\pf}{\mathbf{p}_{\mathrm{f}}}
\newcommand{\pp}{\mathbf{P}}
\newcommand{\bts}{\mathbf{t}_{{\star}}}
\newcommand{\ad}{a^\dagger}                     
\newcommand{\gc}{\gamma_{\mathrm{c}}} 
\newcommand{\gcnp}{\gamma_{\mathrm{c}}^{\mathrm{np}}} %
\newcommand{\gcPF}{\gamma_{\mathrm{c}}^{\mathrm{PF}}} 
\newcommand{\UV}{\Lambda}
\newcommand{\Thnp}{\Sigma^{\mathrm{np}}}
\newcommand{\ThPF}{\Sigma^{\mathrm{PF}}}
\newcommand{\cA}{\mathcal{A}}\newcommand{\cN}{\mathcal{N}}
\newcommand{\cB}{\mathcal{B}} 
\newcommand{\cC}{\mathcal{C}}

\newcommand{\cT}{\mathcal{T}}

\newcommand{\cK}{\mathcal{K}}
\newcommand{\cL}{\mathcal{L}}         
\newcommand{\cM}{\mathcal{M}}       
 
\newcommand{\sC}{\mathscr{C}}
\newcommand{\sD}{\mathscr{D}} 

\newcommand{\sR}{\mathscr{R}}
\newcommand{\sS}{\mathscr{S}}

\newcommand{\sK}{\mathscr{K}}
\newcommand{\sX}{\mathscr{X}}

\renewcommand{\Im}{\mathrm{Im}\,}
\renewcommand{\Re}{\mathrm{Re}\,}
\newcommand{\proof}{{\sc Proof: }}
\newcommand{\qed}{\hfill$\Box$}
\makeindex

\makeatletter
\newcommand\chaptercontents{
{\global
\@topnum\z@ 
\@afterindentfalse 
\if@twocolumn
\@restonecoltrue
\onecolumn 
\else 
\@restonecolfalse 
\fi 
\vspace*{10pt}
\noindent 
{\small\bf Contents}\par 
\vskip1em 
\nobreak} 
{\small
\@starttoc{toc}%
}\if@restonecol
\twocolumn
\fi}
\renewcommand*\l@section[2]{%
\ifnum \c@tocdepth >\z@
\addpenalty\@secpenalty 
\setlength\@tempdima{2.5em}%
\begingroup 
\parindent \z@ 
\rightskip
\@pnumwidth \parfillskip -\@pnumwidth 
\leavevmode 
\advance\leftskip\@tempdima 
\hskip -\leftskip 
#1\nobreak\leaderfill\nobreak
\hb@xt@\@pnumwidth{\hss #2}\par 
\endgroup 
\fi}
\renewcommand*\l@section{\@dottedtocline{1}{0.1em}{1.3em}}       
\renewcommand*\l@subsection{\@dottedtocline{2}{1.5em}{2em}}      
\renewcommand*\l@subsubsection{\@dottedtocline{3}{3.5em}{2.6em}} 

\makeatother

\newtheorem{theorem}{Theorem}[section]
\newtheorem{lemma}[theorem]{Lemma}
\newtheorem{proposition}[theorem]{Proposition}
\newtheorem{corollary}[theorem]{Corollary}
\newtheorem{hypothesis}[theorem]{Hypothesis}
\newtheorem{remark}[theorem]{Remark}
\numberwithin{equation}{section}
\begin{document}

\begin{titlepage}
\begin{center}{\Large{\bf Hydrogen-like atoms in relativistic QED}}
\end{center}
\markboth{M.~K\"onenberg, O.~Matte, and E.~Stockmeyer}{Hydrogen-like atoms in relativistic QED}

\begin{center}{\sc Martin K\"onenberg}
\end{center}
\begin{center}
Fakult\"at f\"ur Mathematik und Informatik,
FernUniversit\"at Hagen,\\
L\"utzowstra{\ss}e 125,
D-58084 Hagen, Germany.\\
{\it Present address:}\\
Fakult\"at f\"ur Physik,
Universit\"at Wien,\\
Boltzmanngasse 5,
1090 Vienna, Austria.\\
{\tt martin.koenenberg@univie.ac.at}
\end{center}
\begin{center}{\sc Oliver Matte}\end{center}
\begin{center}
Institut f\"ur Mathematik,
TU Clausthal,\\
Erzstra{\ss}e 1,
D-38678 Clausthal-Zellerfeld, Germany.\\
{\it Present address:}\\
Institut for Matematik,
{\AA}rhus Universitet,\\
Ny Munkegade 118,
DK-8000 {\AA}rhus, Denmark.\\
{\tt matte@math.lmu.de}
\end{center}
\begin{center}{\sc Edgardo Stockmeyer}\end{center}
\begin{center}
Mathematisches Institut,
Ludwig-Maximilians-Universit\"at,\\
Theresienstra{\ss}e 39,
D-80333 M\"unchen, Germany.\\
{\tt stock@math.lmu.de}
\end{center}
\begin{abstract}
  In this review we consider two different models of a hydrogenic atom
  in a quantized electromagnetic field that treat the electron
  relativistically.  The first one is a no-pair model in the free
  picture, the second one is given by the semi-relativistic
  Pauli-Fierz Hamiltonian. For both models we discuss the semi-boundedness
  of the Hamiltonian, the strict positivity of the ionization energy, and
  the exponential localization in position space of spectral subspaces
  corresponding to energies below the ionization threshold. 
  Moreover, we prove the existence of degenerate ground state
  eigenvalues at the bottom of the spectrum of the Hamiltonian in
  both models.
  All these results hold true, for arbitrary values of the
  fine-structure constant, $e^2$, and the ultra-violet cut-off,
  and for a general class of electrostatic potentials including the Coulomb
  potential with nuclear charges less than (sometimes including) the critical charges
  without radiation field, namely $e^{-2}2/\pi$ for the
  semi-relativistic Pauli-Fierz operator and $e^{-2}2/(2/\pi+\pi/2)$
  for the no-pair operator.  
  Apart from a detailed discussion of diamagnetic inequalities in 
  QED (which are applied to study the semi-boundedness)
  all results stem from earlier articles written by the authors.
  While a few proofs are merely sketched, 
  we streamline earlier proofs or present alternative arguments
  at many places.
\end{abstract}

\end{titlepage}
\vspace*{12pt}   

\chaptercontents  


\section{Introduction}

\noindent
In the late 90's and the past decade 
the existence of ground states of atoms and molecules
interacting with the quantized photon field
has been intensively studied by mathematicians
in the framework of non-relativistic
quantum electrodynamics (QED).
The corresponding Hamiltonian is the 
{\em non-relativistic Pauli-Fierz operator}
which, in the case of a hydrogen-like atom, 
is given as
\begin{equation}\label{def-Hnr}
H_\gamma^\nr\,:=\,\big(\vsigma\cdot(-i\nabla_\V{x}+\V{A})\big)^2
-\frac{\gamma}{|\V{x}|}+\Hf\,.
\end{equation}
Here $\vsigma=(\sigma_1,\sigma_2,\sigma_3)$ is a vector containing 
the Pauli spin matrices, $\V{A}$ is the quantized vector potential
in the Coulomb gauge, and $\Hf$ is the energy of the photon field.
The symbol $\V{A}$ includes a prefactor entering into the analysis
as a parameter, namely the square-root of the fine
structure constant which equals the elementary charge, $e>0$,
in the units chosen in this paper.
The Coulomb coupling constant, $\gamma>0$, is the product
of $e$ and the nuclear charge. We shall, however, always
consider it as an independent parameter since the interrelationship
between $e$ and $\gamma$ does not play any role in our work.
$\V{A}$ additionally depends on some ultra-violet cut-off
parameter, $\Lambda>0$. By now it is well-known that
$H_\gamma^\nr$ has a self-adjoint realization in the
Hilbert space $L^2(\RR^3_\V{x},\CC^2)\otimes\Fock[\sK]$
whose spectrum is bounded below.
Here $\Fock[\sK]$ denotes the bosonic Fock 
space modeled over the Hilbert space for a single photon,
$\sK=L^2(\RR^3\times\ZZ_2)$.
Proving the existence of ground states for 
$H_\gamma^\nr$ means to show that the infimum of the
spectrum of $H_\gamma^\nr$ is an eigenvalue corresponding to some
normalizable ground state eigenvector in
$L^2(\RR^3_\V{x},\CC^2)\otimes\Fock[\sK]$.
Because of the spin degrees of freedom this ground state
eigenvalue will be degenerate.
Mathematically, the study of the eigenvalue at the bottom
of the spectrum of $H_\gamma^\nr$ is very subtle because
the spectrum of $H_\gamma^\nr$ is continuous up to its
minimum and the eigenvalue is, thus, not an isolated one. 
In particular, many standard methods of spectral theory
do not apply and several new mathematical techniques had to be invented
in order to overcome this problem.

The first proofs of the existence of ground states 
for $H_\gamma^\nr$ and its molecular analogs have been given in
\cite{BFS1998b,BFS1999}, for small values of the involved
parameters $e$ and $\UV$. A few years later
the existence of ground states
for a molecular non-relativistic 
Pauli-Fierz Hamiltonian has been established,
for arbitrary values of $e$ and $\UV$, in \cite{GLL2001} by means
of a certain
binding condition which has been verified
later on in \cite{LiebLoss2003}.
Moreover, infra-red finite algorithms and spectral theoretic
renormalization group methods have been applied to various models of
non-relativistic QED to study their ground state energies and
projections
\cite{BCFS2003,BFP2006,BFS1998b,BFS1998a,BFS1999,BachKoenenberg2006,FGS2008}.
These sophisticated methods yield very precise results
as they rely on constructive algorithms rather than on
compactness arguments as in \cite{BFS1998b,BFS1999,GLL2001}.
They work, however, only in a regime where $e$ and $\Lambda$
are sufficiently small.

A question which arises naturally in this context is whether
these results still hold true when the electrons are 
described by a relativistic operator.
In this review, which summarizes results from
\cite{KM2011,KMS2009a,KMS2009b,Matte2009,MatteStockmeyer2009a,Stockmeyer2009},
 we give a positive answer to this question. We study  
two different models that seem to be natural candidates for a
mathematical analysis:
The first one is given by the following no-pair operator,
\begin{equation}\label{def-np-intro}
\NPO{\gamma}\,:=\,\PA\,\big(\DA-\tgV+\Hf\big)\,\PA\,,
\end{equation}
or more generally,
\begin{equation}\label{def-np-intro-V}
\NPO{V}\,:=\,\PA\,\big(\DA+V+\Hf\big)\,\PA\,,
\end{equation}
for some electrostatic potential $V$.
Here $\DA$ is the free Dirac operator minimally coupled to $\V{A}$
and $\PA$ denotes the spectral projection
onto the positive spectral subspace of $\DA$,
$$
\PA\,:=\,\id_{(0,\infty)}(\DA)\,.
$$
The no-pair operator is considered as an operator acting in
the projected Hilbert space, $\HR_{\V{A}}^+:=\PA\,\HR$.
It is thus acting on a space where the
electron and photon degrees of freedom are always linked together.
The analog of $\NPO{\gamma}$
for molecules has been introduced in \cite{LiebLoss2002}
as a mathematical model to study the stability of matter
in relativistic QED. Under certain restrictions 
on $e$, $\UV$, and the nuclear charges it has been shown 
in \cite{LiebLoss2002} that the quadratic form of a 
molecular no-pair operator is bounded from below 
by some constant which is proportional to the number
of involved electrons and nuclei and uniform in the
nuclear positions.
Moreover, 
the (positive) binding energy has been
estimated from above in \cite{LiebLoss2002b}.
In fact, there are numerous mathematical contributions on no-pair
models where magnetic fields
are not taken into account or treated classically; see, e.g.,
\cite{MatteStockmeyer2008a} for a list of references.
For instance, it is shown in \cite{EPS1996} that a no-pair operator
with Coulomb potential but
without quantized photon field -- which is then often called the
Brown-Ravenhall operator --
has a critical coupling constant,
$\gcnp:=2/(2/\pi+\pi/2)$, such that the corresponding quadratic
form becomes unbounded below when $\gamma$ exceeds this value.
Moreover, various molecular no-pair models 
(without quantized fields, however)
are widely used in
quantum chemistry and in the theoretical and numerical study
of highly ionized heavy atoms; see, e.g.,
\cite{CCJ2002,Jo1998,ReWo}.
In this context several different choices of the projections
determining the model
find their applications. For instance, one can include
the Coulomb or a Hartree-Fock potential in the projection.
These two choices are covered by the results 
of \cite{MatteStockmeyer2008a} where two of the present authors provide a
spectral analysis of a class of molecular no-pair Hamiltonians
with classical magnetic fields.
The choice of the projection $\PA$ which does not
contain any potential terms is referred to
as the free picture.
We remark that it is essential to include the vector potential
in the projection determining the no-pair model.
For, if $\PA$ is replaced by $\PO$,
then the analog of \eqref{def-np-intro} describing
$N$ interacting electrons becomes unstable as soon as $N\grg2$
\cite{GriesemerTix1999,LiebLoss2002,LSS1997}. Moreover, the operator in 
\eqref{def-np-intro} is formally gauge invariant
and this would not hold true anymore with
$\PO$ in place of $\PA$. Gauge invariance plays, however,
an important role in the proof of the existence of ground
states as it permits to derive 
bounds on the number of soft  photons.
In fact, employing a mild infra-red regularization
it is possible to prove the existence of ground states
for the operator in \eqref{def-np-intro} with
$\PA$ replaced by $\PO$ 
\cite{Koenenberg2004,Matte2000}.
It seems, however, unlikely that the 
infra-red regularization can be dropped 
in this case \cite{Koenenberg2004}.

The second operator treated in this review,
the semi-relativistic Pauli-Fierz operator, is given as
\begin{equation}\label{def-PF2}
\sqrt{(\vsigma\cdot(-i\nabla+\V{A}))^2+\id}-\tgV+\Hf\,,
\end{equation}
where $\vsigma$ is a vector containing the Pauli spin matrices.
Since $\sqrt{-\Delta}$ and $1/|\V{x}|$ both scale as one over the
length there will again be some critical upper bound on
all values of $\gamma>0$ for which \eqref{def-PF2} defines
a semi-bounded quadratic form. As we shall see this
upper bound is at least as big as (in fact equal to \cite{KMS2009a}) 
the critical constant in Kato's inequality, $\gcPF:=2/\pi$.
Again we shall study the semi-relativistic Pauli-Fierz operator
also for a more general class of electrostatic potentials $V$.
The latter (straightforward) generalization is relevant
in a forthcoming work of the first two authors devoted to
the enhanced binding effect \cite{KM2012b}.

Also the semi-relativistic Pauli-Fierz operator has been investigated
earlier in a few mathematical articles.
For instance it appears in the mathematical study of Rayleigh
scattering \cite{FGS2001} where the finite propagation speed
of relativistic particles turns out to be an advantageous feature
in comparison to models of non-relativistic QED.
(The electron spin has, however, been neglected in \cite{FGS2001}.)
For $\gamma=0$,
the fiber decomposition of \eqref{def-PF2} with respect to different values
of the total momentum has been studied in 
\cite{MiyaoSpohn2009}. 
Moreover, there is a remark in \cite{MiyaoSpohn2009} relevant for us
saying that
every (speculative) eigenvalue of the operator in \eqref{def-PF2}
is at least doubly degenerate since the Hamiltonian commutes
with some anti-linear involution. 
The existence of the renormalized electron
mass in the semi-relativistic Pauli-Fierz model,
i.e. twice continuous differentiability of the mass shell in balls
about zero, is proved in \cite{KM2012a}, for small values of $e$.
The last author has shown \cite{Stockmeyer2009} that,
when the speed of light, $c$, is re-introduced as a parameter
and $\gamma\in[0,\gcPF)$,
then the operator in \eqref{def-PF2} converges in
norm resolvent sense to the non-relativistic 
Pauli-Fierz operator in \eqref{def-Hnr}, as $c$ tends to infinity.
Finally, there is a contribution \cite{HiroshimaSasaki2010}
on the existence of binding in the semi-relativistic Pauli-Fierz model;
see Remark~\ref{rem-HiroshimaSasaki}. 

We should also mention that the existence of ground
states in a relativistic model describing
both the photons and the electrons and positrons
by quantized fields has been studied mathematically in
\cite{BDG2004}. To this end 
infra-red and ultra-violet cut-offs for the momenta
of all involved particles are imposed in the interaction term
of the Hamiltonian considered in \cite{BDG2004}.

In the remaining part of this introduction
we explain the organization of this review  article
and summarize briefly our main results.
In Section~\ref{sec-models} we recall the definitions 
of some operators appearing in QED and 
introduce the no-pair and semi-relativistic Pauli-Fierz 
Hamiltonians more precisely.
Although the general strategy of our whole project
relies on the methods developed in \cite{BFS1998b,BFS1999,GLL2001}
the spectral analysis of the operators treated in this article
poses a variety of new and non-trivial mathematical obstacles
which is mainly caused by their non-locality. 
In fact,
both operators do not act as partial differential operators
on the electronic degrees of freedom anymore as it is the case
in non-relativistic QED.
In this respect the no-pair operator is harder to
analyze than the semi-relativistic Pauli-Fierz operator
since also the electrostatic potential and the radiation field
energy become non-local due to the presence of the
spectral projections $\PA$.
In the last subsection of Section~\ref{sec-models}
we explain a few mathematical tools
used to overcome some of the problems posed
by the non-locality thus preparing the reader for
the proofs in the succeeding sections.   

In Section~\ref{sec-sb} we provide some basic relative bounds and
study the semi-boundedness of the Hamiltonian in both models under
consideration. We start with a discussion of various diamagnetic
inequalities for quantized vector potentials. They are employed to prove
that the quadratic form of the semi-relativistic Pauli-Fierz operator
is semi-bounded below on some natural dense domain, for a suitable class
of potentials including the Coulomb potential with coupling constants
$\gamma\in[0,\gcPF]$.  For the no-pair operator we obtain similar
results with Coulomb coupling constants $\gamma\in[0,\gcnp)$.  As a consequence, both
operators can be realized as self-adjoint operators in a physically
distinguished way by means of a Friedrichs extension. We point out
that the results on the semi-boundedness, as well as all further
results described below hold true, for arbitrary values of $e$ and
$\UV$. 

Section~\ref{sec-binding} is devoted to the study of
binding. For both models treated here we show that
the infimum of the spectrum of the Hamiltonian with 
appropriate non-vanishing potential is strictly less than its ionization threshold
which, by definition, is equal to the infimum of the spectrum
of the Hamiltonian without electrostatic potential. To this end we employ  trial
functions which 
are tensor products of electronic and photonic wave functions and work
with unitarily equivalent 
Hamiltonians in order to separate the electronic and photonic degrees
of freedom. 
The unitary transformation used here represents the free Hamiltonian
($V=0$) 
as a direct integral of fiber operators with respect to different
values of 
the total momentum. 

Typically, proofs of the existence of ground states in QED
require some information on the localization of
spectral subspaces corresponding to energies below
the ionization threshold
(or at least of certain approximate
ground state eigenfunctions). 
Here localization is understood with respect
to the electron coordinates.
We establish this prerequisite in Section~\ref{sec-exp}
by adapting some ideas from \cite{BFS1998b,Griesemer2004}.
In this section we present streamlined versions
of some of our earlier arguments from \cite{MatteStockmeyer2009a}.
Moreover, we implement later improvements \cite{KM2011}
on parts of the results of \cite{MatteStockmeyer2009a}
by providing optimized exponential decay rates 
in the case of the semi-relativistic Pauli-Fierz operator
that reduce to the typical relativistic decay rates known
for the electronic Dirac and square-root operators, when the
radiation field is turned off.
The class of potentials allowed for in Section~\ref{sec-exp}
covers Coulomb potentials with
$\gamma\in[0,\gcPF]$ in the case of the semi-relativistic
Pauli-Fierz model  and with $\gamma\in[0,\gcnp)$ in the no-pair model.
It is, however, possible to prove the exponential localization 
for the no-pair operator with Coulomb potential also in the critical case
$\gamma=\gcnp$ by another modification of the arguments \cite{KM2011}.

The main results of our joint project are the proofs
of the existence of ground states for the semi-relativistic
Pauli-Fierz and no-pair operators. As already stressed
above our proofs work, for arbitrary values of $e$ and $\UV$
and for a class of potentials including the Coulomb potential
with $\gamma\in(0,\gcPF)$ and $\gamma\in(0,\gcnp)$, respectively.
Starting from these results
it is actually possible to prove the existence of ground states
also in the critical cases $\gamma=\gcPF$ and $\gamma=\gcnp$,
respectively, by means of an additional approximation argument.
We refrain from explaining any details of the latter in the present 
article and refer the interested reader to \cite{KM2011} instead.

The proofs of the existence of ground states given here are divided
into two steps:

First, one introduces a photon mass, $m>0$, and shows that
the resulting Hamiltonians possess normalized ground state
eigenfunctions, $\phi_m$ \cite{BFS1998b,BFS1999,GLL2001}.
In this first step, which is presented in Section~\ref{sec-ex-m},
we employ a discretization of the
photon momenta as in \cite{BFS1999}. Roughly speaking,
by discretizing the photon momenta one may replace the Fock
space $\Fock[\HP]$ by a Fock space modeled over some
$\ell^2$ space. As a consequence the spectrum of the
radiation field energy becomes discrete and one can in
fact argue that the total Hamiltonian has discrete eigenvalues at the
bottom of its spectrum when all small photon momenta are
discarded. At this point 
we add a new observation based on the localization estimates
to the arguments of \cite{BFS1999}
which allows to carry through the proof, for all
values of $e$ and $\UV$. (In \cite{BFS1998b,BFS1999} 
these parameters were assumed to be sufficiently small.) 
Another technical tool turns out to be very 
helpful in order to compare discretized and non-discretized Hamiltonians
(or those with and without photon mass),
namely, certain higher
order estimates allowing to control higher powers of the
radiation field energy by corresponding
powers of the resolvent of the total Hamiltonian.
For the semi-relativistic Pauli-Fierz
operator such estimates have been established
in \cite{FGS2001}. In \cite{Matte2009} one of the present authors
re-proves the higher order estimates from \cite{FGS2001}
by means of a different and more model-independent
method which also permits to derive higher order
estimates for a (molecular) no-pair operator for the first time.
We discuss these higher order estimates in Subsection~\ref{ssec-hoe}
but refrain from repeating their proofs.
We remark that many of the arguments presented in Section~\ref{sec-ex-m}
are alternatives to those used in \cite{KMS2009a,KMS2009b}.

The second step in the proof of the existence of ground states
comprises of a compactness argument showing
that every sequence $\{\phi_{m_j}\}$ with $m_j\searrow0$ 
contains a strongly convergent subsequence. In fact, one readily verifies that
the limit of such a subsequence is a 
ground state eigenfunction of the original Hamiltonian 
with massless photons. 
This step is performed in Section~\ref{ExistenceGroundStates},
in parts by means of arguments alternative to those in
\cite{KMS2009a,KMS2009b}.
The compactness argument requires, however, a number of 
non-trivial ingredients.
First, we need two infra-red bounds, namely
a bound on the number of photons with low energy
in the eigenfunctions $\phi_m$ (soft photon bound)
\cite{BFS1999,GLL2001}
and a certain bound on the weak derivatives
of $\phi_m$ with respect to the photon momenta
(photon derivative bound) \cite{GLL2001}.
To derive the infra-red bounds one can either adapt a procedure proposed
in \cite{GLL2001} (this is carried through in earlier preprint
versions of \cite{KMS2009a,KMS2009b} available on the arXiv)
or establish a formula for
$a(k)\,\phi_m$ by means of a virial type argument
and infer the bounds from that representation. 
We outline the proof of the latter formula
and of the soft photon bound for the
semi-relativistic Pauli-Fierz operator in Section~\ref{InfraredBounds}.
The photon derivative bound and 
the infra-red bounds for the no-pair operator are derived
by very similar procedures and we refer
the interested reader to our original articles \cite{KMS2009a,KMS2009b}
for the rather dull technical details.
The arguments presented in Section~\ref{InfraredBounds}
are also intended to emphasize the role of the gauge invariance of
the models treated here. In fact, one first applies
a unitary operator-valued gauge transformation
(Pauli-Fierz transformation) and the infra-red bounds are then
derived in the new gauge. Without the gauge transformation
one would encounter infra-red divergent integrals.

As soon as the infra-red estimates are established,
the soft photon bound and the exponential localization estimates
show that the eigenvectors $\phi_m$ are localized uniformly in $m$ with
respect to the electron and photon coordinates and that
their components in all but finitely many Fock space sectors 
are negligible. Moreover, the photon derivative bound
implies that their weak derivatives with respect to the
photon momenta are uniformly bounded in a suitable $L^p$-space
and since their energies are uniformly bounded we also
know that the vectors have uniformly bounded half-derivatives  with respect
to the electron coordinates in $L^2$. 
It is an idea of \cite{GLL2001} to exploit such
information by applying compact embedding theorems
for Sobolev-type spaces to single out subsequences
that converge strongly in $L^2$.
In the semi-relativistic setting considered 
in Section~\ref{ExistenceGroundStates}
some classical embedding theorems by Nikol$'$ski{\u\i}
turn out to be useful substitutes for the
Rellich-Kondrachov theorem employed in \cite{GLL2001}. 
At the end of Section~\ref{ExistenceGroundStates} we also show that 
the ground state energies of both Hamiltonians are degenerate eigenvalues.

At last, in
Section~\ref{sec-commutators}, we present the proofs of
some technical results we have referred to in earlier
sections so that most parts of this review become
essentially self-contained.


\section{Definition of the models}\label{sec-models}

\noindent
In order to introduce the models treated in this article more precisely
we first fix our notation and recall some standard facts.

\subsection{Operators in Fock-space}

\noindent
The state space of the 
quantized photon field is the bosonic Fock space,
$$
\Fock[\HP]\,:=\,
\bigoplus_{n=0}^\infty\Fock^{(n)}[\HP]
\,\ni\psi\,=\,(\psi^{(0)},\psi^{(1)},\psi^{(2)},\ldots\;)
\,.
$$
It is 
modeled over the one photon Hilbert space
$$
\HP\,:=\,L^2(\RR^3\times\ZZ_2,dk)\,,\quad
\int dk\,:=\,\sum_{\lambda\in\ZZ_2}\int_{\RR^3}d^3\V{k}\,.
$$
$k=(\V{k},\lambda)$
denotes a tuple consisting of a photon wave vector,
$\V{k}\in\RR^3$, and a polarization label, $\lambda\in\ZZ_2$.
Moreover, $\Fock^{(0)}[\HP]:=\CC$ and
$\Fock^{(n)}[\HP]$
is the subspace of all complex-valued, square integrable functions
on $(\RR^3\times\ZZ_2)^n$ that remain invariant under permutations
of the $n\in\NN$ wave vector/polarization tuples. 
The subspace
$$
\sC_0\,:=\,\CC\oplus\bigoplus_{n\in\NN}
C_0((\RR^3\times\ZZ_2)^n)\cap\Fock^{(n)}[\HP]\,
\quad\textrm{(Algebraic direct sum.)}
$$
is dense in $\Fock[\HP]$.
The energy of the photon field, $\Hf:=d\Gamma(\omega)$, is given as the
second quantization of the dispersion relation
$\omega(k):=|\V{k}|$, $k=(\V{k},\lambda)\in\RR^3\times\ZZ_2$.
We recall that the second quantization of some real-valued
Borel measurable function $\varpi$ is given by
$(d\Gamma(\varpi)\,\psi)^{(0)}=0$ and
$$
(d\Gamma(\varpi)\,\psi)^{(n)}(k_1,\dots,k_n)=
\sum_{j=1}^n\varpi(k_j)\,\psi^{(n)}(k_1,\dots,k_n)\,,\quad n\in\NN\,,
$$
for all $\psi=(\psi^{(n)})_{n=0}^\infty\in\Fock[\HP]$ 
such that $([d\Gamma(\varpi)\,\psi]^{(n)})_{n=0}^\infty\in\Fock[\HP]$.
We further recall that the creation
and the annihilation operators
of a photon state $f\in\HP$ are given, for $n\in\NN$,
by
\begin{align*}
&(\ad(f)\,\psi)^{(n)}(k_1,\ldots,k_n)\,=\,
n^{-1/2}\sum_{j=1}^nf(k_j)\,\psi^{(n-1)}(\ldots,k_{j-1},k_{j+1},\ldots)\,,
\\
&(a(f)\,\psi)^{(n-1)}(k_1,\ldots,k_{n-1})\,=\,
n^{1/2}\int\ol{f}(k)\,\psi^{(n)}(k,k_1,\ldots,k_{n-1})\,dk\,,
\end{align*}
and $(\ad(f)\,\psi)^{(0)}=0$, $a(f)\,\Omega=0$, where
$\Omega:=(1,0,0,\ldots\;)\in\Fock[\HP]$ is the vacuum vector.
We define $\ad(f)$ and $a(f)$ on their 
maximal domains. For $f,g\in\HP$,
the following canonical commutation relations hold true
on $\sC_0$,
$$
[a(f)\,,\,a(g)]\,=\,[\ad(f)\,,\,\ad(g)]\,=\,0\,,\qquad
[a(f)\,,\,\ad(g)]\,=\,\SPn{f}{g}\,\id\,.
$$
For a three-vector of functions
$\V{f}=(f^{(1)},f^{(2)},f^{(3)})\in\HP^3$, the symbol
$a^\sharp(\V{f})$ denotes the triple of operators
$
a^\sharp(\V{f}):=\big(a^\sharp(f^{(1)}),a^\sharp(f^{(2)}),
a^\sharp(f^{(3)})\big)
$,
where $a^\sharp$ is always either $a$ or $\ad$.

\subsection{Interaction term}

\noindent
Next, we describe the interaction between four-spinors
and the photon field.
The full Hilbert space underlying our models is
$$
\HR\,:=\,L^2(\RR^3_{\V{x}},\CC^4)\otimes\Fock[\HP]\,.
$$
It contains the dense subspace
$$
\core\,:=\,C_0^\infty(\RR^3_{\V{x}},\CC^4)\otimes\sC_0\,.
\quad\textrm{(Algebraic tensor product.)}
$$
We introduce the self-adjoint Dirac matrices
$\alpha_1,\alpha_2,\alpha_3$, and $\beta$ 
that act on the four spinor components of an element from $\HR$,
that is, on the second tensor factor in
$\HR\cong L^2(\RR^3_{\V{x}})\otimes\CC^4\otimes\Fock[\HP]$.
They are given by
$$
\alpha_j\,:=\,
\begin{pmatrix}
0&\sigma_j\\\sigma_j&0
\end{pmatrix}\,,\quad j\in\{1,2,3\}\,,\qquad
\beta\,:=\,\alpha_0\,:=\,
\begin{pmatrix}
\id&0\\0&-\id
\end{pmatrix}\,,
$$
where $\sigma_1,\sigma_2,\sigma_3$ denote the standard Pauli matrices,
and fulfill the
Clifford algebra relations
\begin{equation}\label{Clifford}
\alpha_i\,\alpha_j\,+\,\alpha_j\,\alpha_i\,=\,2\,\delta_{ij}\,\id\,,
\qquad i,j\in\{0,1,2,3\}\,.
\end{equation}
The interaction between the electron/positron
and photon degrees of freedom in the Coulomb gauge
is given as 
$\valpha\cdot\V{A}:=\alpha_1\,A^{(1)}+\alpha_2\,A^{(2)}+\alpha_3\,A^{(3)}$, where
$$
\V{A}:=(A^{(1)},A^{(2)},A^{(3)}):=\ad(\V{G})+a(\V{G})\,,\quad 
a^{\sharp}(\V{G}):= \int_{\RR^3}^{\oplus}
\id_{\CC^4}\otimes   a^{\sharp}(\V{G}_{\V{x}})\,d^3\V{x}\,.
$$
The physical choice of the coupling function
${\bf G}_\V{x}=(G^{(1)}_\V{x},G^{(2)}_\V{x},G^{(3)}_\V{x} )$ is 
\begin{equation}\label{Gphys}
\V{G}_\V{x}(k)\,
:=\,-e\,\frac{\id_{\{|\V{k}|\klg\UV\}}}{2\pi\sqrt{|\V{k}|}}
\,e^{-i\V{k}\cdot\V{x}}\,\veps(k),
\end{equation}
for $\V{x}\in\RR^3$ and almost every $k=(\V{k},\lambda)\in\RR^3\times\ZZ_2$.
The parameter
$\UV>0$ is an ultraviolet cut-off
and $e\in \RR$. (In nature $e^2\approx1/137$ is Sommerfeld's
fine-structure constant which equals the square of
the elementary charge in our units\footnote{
Energies are measured in units of $mc^2$, $m$ denoting the
rest mass of an electron and $c$ the speed of light.
Length, i.e. $\V{x}$, are measured in units of $\hslash/(mc)$,
which is the Compton wave length divided by $2\pi$.
$\hslash$ is Planck's constant divided by $2\pi$.
The photon wave vectors
$\V{k}$ are measured in units of $2\pi$ times
the inverse Compton wavelength,
$mc/\hslash$.}.)
The values of $e$ and $\UV$ can be chosen arbitrarily in the whole
article. Writing
\begin{equation}\label{def-kbot}
\V{k}_\bot\,:=\,(k^{(2)}\,,\,-k^{(1)}\,,\,0)\,,\qquad
\V{k}=(k^{(1)},k^{(2)},k^{(3)})\in\RR^3\,,
\end{equation}
the polarization vectors are given as 
\begin{equation}\label{pol-vec}
\veps(\V{k},0)\,=\,
\frac{\V{k}_\bot}{|\V{k}_\bot|}
\,,\qquad
\veps(\V{k},1)\,=\,
\frac{\V{k}}{|\V{k}|}\,\wedge\,\veps(\V{k},0)\,,
\end{equation}
for $\V{k}\in\RR^3\setminus\{0\}$ with $\V{k}_\bot\not=0$.
It is sufficient to determine $\veps$ almost everywhere. 
Many of our results and estimates do not depend on 
the special choice of $\V{G}_\V{x}$. If we consider
a larger class of coupling functions at a certain point
in this review we shall explain the required properties
of $\V{G}_\V{x}$ at the beginning of the corresponding (sub)section.


\subsection{The semi-relativistic
Pauli-Fierz and no-pair Hamiltonians}

\noindent
In order to define the no-pair and semi-relativistic
Pauli-Fierz operators we recall that the
free Dirac operator minimally coupled to $\V{A}$ is given as
\begin{equation}\label{def-DA}
\DA\,:=\,\valpha\cdot(-i\nabla+\V{A})+\beta
\,:=
\sum_{j=1}^3\alpha_j\,\big(-i\partial_{x_j}+\ad(G_\V{x}^{(j)})
+a(G_\V{x}^{(j)})\big)+\beta
\,.
\end{equation}
A straightforward application of Nelson's commutator theorem 
shows that
$\DA$ is essentially self-adjoint on $\core$;
see, e.g., \cite{LiebLoss2002,MiyaoSpohn2009}.
We denote its closure starting from $\core$
again by the same symbol. 
As a consequence of \eqref{Clifford} we further have
\begin{equation}\label{dirk1}
\DA^2\,=\,\cT_\V{A}\oplus\cT_\V{A}\,,\qquad \cT_\V{A}:=
(\vsigma\cdot(-i\nabla+\V{A}))^2+\id\,,
\end{equation}
on $\core$.
In particular, $\DA^2\grg1$ on $\core$, 
and since $\DA$ is essentially self-adjoint on $\core$
we see that $\|(\DA-z)\,\psi\|\grg(1-|z|)\,\|\psi\|$,
$\psi\in\dom(\DA)$, $z\in\CC$,
whence
$$
\spec(\DA)\,\subset\,(-\infty,-1]\cup[1,\infty)\,.
$$
Contrary to the usual convention used also in the introduction
we define the
{\it semi-relativistic Pauli-Fierz operator} as an operator
acting in $\HR$ {a} priori by
\begin{align}\label{def-PF}
\PF{\gamma}\,\vp\,&:=\,\big(|\DA|-\tgV+\Hf\big)\,\vp\,,\qquad
\vp\in\core\,,
\\\label{def-PF-V}
\PF{V}\,\vp\,&:=\,\big(|\DA|+V+\Hf\big)\,\vp\,,\qquad\;\,
\vp\in\core\,.
\end{align}
We shall impose appropriate conditions on the general potential
$V\in L^2_\loc(\RR^3,\RR)$ later on.
In fact, the operator defined in \eqref{def-PF} is a two-fold
copy of the one given in \eqref{def-PF2} since
$|\DA|=\cT_\V{A}^{1/2}\oplus\cT_\V{A}^{1/2}$ by \eqref{dirk1}.
We prefer to consider the operator defined by \eqref{def-PF} 
to have a unified notation throughout the following
sections.
The
{\it no-pair operator} acts in a projected Hilbert space,
$\HRp$, given by
\begin{equation}\label{def-PA}
\HRp\,:=\,\PA\,\HR\,,\qquad \PA\,:=\,\id_{[0,\infty)}(\DA)\,,
\qquad \PAm\,:=\,\id-\PA\,.
\end{equation}
{A} priori we define it on the dense domain $\PA\,\sD\subset\HRp$, 
\begin{align}\label{def-np}
\NPO{\gamma}\,\vp^+:=\,\PA\,\big(\DA-\tgV+\Hf\big)\,\vp^+,\qquad
\vp^+\in\PA\,\core\,,
\\\label{def-np-V}
\NPO{V}\,\vp^+:=\,\PA\,\big(\DA+V+\Hf\big)\,\vp^+,\qquad
\;\,\vp^+\in\PA\,\core\,,
\end{align}
where $V\in L^2_\loc(\RR^3,\RR)$ satisfies $\dom(\DO)\subset\dom(V)$.
In the above definitions we have to take care that the right hand
sides are actually well-defined
as it is, for instance, not obvious that $\tgV\,\PA\,\core\subset\HR$.
It follows, however, from the proof
of Lemma~3.4(ii) in \cite{MatteStockmeyer2009a} 
that $\PA$ maps $\sD$ into $\dom(\DO)\cap\dom(\Hf^\nu)$, for
every $\nu>0$, so that the definitions \eqref{def-np} and
\eqref{def-np-V}
make sense.

As soon as we have shown in Section~\ref{sec-sb} that
the quadratic forms of $\PF{V}$ and $\NPO{V}$
are semi-bounded below on the dense domains
$\sD$ and $\PA\,\sD$, respectively, we may extend them to
self-adjoint operators by means of a Friedrichs extension.
As already mentioned in the introduction there will, however,
be critical values for $\gamma$ above which the quadratic
forms are unbounded below in the case of the Coulomb potential.
We prove in Section~\ref{sec-sb} that, for $\PF{\gamma}$,
this critical value is not less than
$$
\gcPF\,:=\,2/\pi\,,
$$ 
which is the critical constant in Kato's inequality,
$(2/\pi)|\V{x}|^{-1}\klg\sqrt{-\Delta}$.
In the case of the  no-pair operator we prove the
semi-boundedness of the
quadratic form of $\NPO{\gamma}$, for all $\gamma\in[0,\gcnp)$, where
\begin{equation}\label{def-gammac}
\gcnp\,:=\,2/(2/\pi+\pi/2)\,.
\end{equation} 
The instability of both models above the respective
critical values for $\gamma$ is shown in \cite{KMS2009a,KMS2009b}
by means of suitable test functions that drive the energy to minus
infinity.
For the definition of $\NPO{\gamma}$ in the case
$\gamma=\gcnp$ see \cite{KMS2009b}.

It has been shown in \cite{EPS1996} that the quadratic form
associated with the (electronic) Brown-Ravenhall operator,
\begin{equation}\label{def-Bel}
\BR{\gamma}\,:=\,\PO\,(\DO-\tgV)\,\PO\,,
\end{equation}
is semi-bounded on
$\PO\,C_0^\infty(\RR^3,\CC^4)$ if and only
if $\gamma\klg \gcnp $. Thus, for $\gamma\klg \gcnp$, it
has a self-adjoint Friedrichs extension which we again
denote by $\BR{\gamma}$ and which 
actually satisfies Tix' inequality,
$\BR{\gamma}\grg(1-\gamma)\,\PO$, $\gamma\in[0,\gcnp]$; see \cite{Tix1998}.
We exploit the semi-boundedness of $\BR{\gamma}$ in the
proof of Theorem~\ref{thm-sb-np} below.

Finally, we introduce a convention used throughout this review:
We will frequently use the symbols $H^\sharp_V$, $H^{\sharp}_\gamma$, $\gamma_c^{\sharp}$, 
etc. 
when we treat both the relativistic Pauli-Fierz 
and no-pair operators at the same time;
that is, $\sharp$ is $\Pf$ or $\np$.


\subsection{How to deal with the non-local terms}\label{sec-htd}

\noindent
Although general strategies to prove the existence of ground
states have been developed in the framework of non-relativistic
QED \cite{BFS1998b,BFS1999,GLL2001} the application of these
ideas to the models discussed in this review poses a variety
of new mathematical problems. This is mainly due to the
non-locality of the operators $|\DA|$ and $\PA$ appearing in
$H_V^\sharp$. In this respect the the no-pair
operator is considerably more difficult to analyze than the
semi-relativistic Pauli-Fierz operator since also
the projected potential and radiation field energies become
non-local. As a consequence a variety of commutator
estimates involving $|\DA|$, $\PA$, $\Hf$, cut-off
functions etc. is required for a spectral analysis of $H_V^\sharp$.
Most of these commutator estimates are based on the
observations and facts we collect in this subsection.
We shall only present one proof in the present subsection
in order to illustrate some simple ideas and defer
other technical arguments to Section~\ref{sec-commutators}.
We introduce a general hypothesis on the coupling function
which is sometimes used in the sequel:

\begin{hypothesis}\label{hyp-G}
The map
$\RR^3\times(\RR^3\times\ZZ_2)\ni(\V{x}, k)\mapsto \V{G}_{\V{x}}(k)$
is measurable such that $\V{x}\mapsto \V{G}_{\V{x}}(k) $ 
is continuously differentiable, for almost every $k$, and
\begin{equation}\label{G(-k)}
\V{G}_\V{x}(-\V{k},\lambda)\,=\,
\ol{\V{G}}_\V{x}(\V{k},\lambda)\,,\qquad
\V{x}\in\RR^3,\;\textrm{a.e.}\;\V{k}\,,\;\lambda\in\ZZ_2\,.
\end{equation}
There exist
$d_{-1},d_0,d_1,d_2\in(0,\infty)$ satisfying
\begin{equation}\label{def-dell}
2\int\omega(k)^{\ell}\,\|\V{G}(k)\|_\infty^2\,dk\,\klg\,d_\ell^2\,,
\quad
2\int\frac{\|\nabla_\V{x}\wedge\V{G}(k)\|_\infty^2}{\omega(k)}\,dk\,\klg\,d_1^2
\,,
\end{equation}
where $\|\V{G}(k)\|_\infty:=\sup_{\V{x}}|\V{G}_\V{x}(k)|$, etc.
\end{hypothesis}

\noindent
We remark that,
if \eqref{G(-k)} is fulfilled, then $[A^{(j)}(\V{x}),A^{(k)}(\V{y})]=0$,
for all $j,k\in\{1,2,3\}$, $\V{x},\V{y}\in\RR^3$.
For later reference we also recall the following 
well-known relative bounds,
valid for every $\psi\in\dom(\Hf^{1/2})$,
\begin{align}
\|\valpha\cdot a(\V{G})\,\psi\|^2
&\klg d_{-1}^2 \,\|\Hf^{1/2}\psi\|^2,\label{rb-a}
\\
\|\valpha\cdot\ad(\V{G})\,\psi\|^2&\klg
d_{-1}^2\,\|\Hf^{1/2}\psi\|^2+d_0^2\,\|\psi\|^2.\label{rb-ad}
\end{align}
In order to cope with the non-locality of $\PA$ we write
$$
\RA{iy}\,:=\,(\DA-iy)^{-1}\,,\qquad y\in\RR\,,
$$
and use the following representation of the
sign function of $\DA$ as a strongly convergent
principal value (see Lemma~VI.5.6 in \cite{Kato}),
\begin{equation}\label{sgn}
\SA\,\vp\,:=\,\DA\,|\DA|^{-1}\,\vp\,=\,
\lim_{\tau\to\infty}\int_{-\tau}^\tau \RA{iy}\,\vp\,\frac{dy}{\pi}\,,
\qquad \vp\in\HR.
\end{equation}
In addition we observe that
\begin{equation}\label{|DA|PA}
 |\DA|\,=\,\SA\,\DA\,,\qquad \PA\,=\,\frac{1}{2}\,\id+\frac{1}{2}\,\SA\,.
\end{equation}
These formulas reduce computations involving $|\DA|$ or $\PA$
to computations involving $\DA$ and integrals over its resolvent.
To study the exponential localization it is hence useful to
recall that, for all 
$y\in\RR$, $a\in[0,1)$,
and 
$F\in C^\infty(\RR^3_\V{x},\RR)$ having a fixed sign and
satisfying $|\nabla F|\klg a$, we have 
$iy\in\vr(\DA+i\valpha\cdot\nabla F)$,
\begin{equation}\label{exp-marah0}
R_{\V{A}}^F(iy)\,:=\,e^F\,\RA{iy}\,e^{-F}
=(\DA+i\valpha\cdot \nabla F
-iy)^{-1}\!\!\upharpoonright_{\dom(e^{-F})}\,,
\end{equation}
and
\begin{equation}\label{exp-marah1}
\|R_{\V{A}}^F(iy)\|\,\klg\,
\frac{\sqrt{6}}{\sqrt{1+y^2}}\cdot
\frac{1}{1-a^2}
\,.
\end{equation}
For classical vector potentials
this essentially follows from a computation we learned from
\cite{BeGe1987}; 
see also \cite{MatteStockmeyer2008b} where \eqref{exp-marah0} and
\eqref{exp-marah1} are proved in the form stated above.
It is, however, clear that the arguments in \cite{MatteStockmeyer2008b}
work for a quantized vector potential, too.
Moreover, it is easy to verify that
\begin{equation}\label{thomas}
[\RA{iy}\,,\,\chi\,e^F]\,e^{-F}\,=\,
\RA{iy}\,i\valpha\cdot(\nabla\chi+\chi\,\nabla F)\,R_\V{A}^F(iy)\,,
\end{equation}
where $\chi\in C^\infty(\RR^3_\V{x},[0,1])$ is some smooth function
of the electron coordinates and $F$ is as above.
Finally, we note that
\begin{equation}\label{thomas2}
\|i\valpha\cdot(\nabla\chi+\chi\,\nabla F)\|\,\klg\,\|\nabla\chi\|_\infty+a\,,
\end{equation}
since $\|\valpha\cdot \V{v}\|=|\V{v}|$, $\V{v}\in\RR^3$, by the
Clifford algebra relations, and $|\nabla F|\klg a$.
As an example we treat some commutator estimates whose
proofs make use of these
remarks and a few further useful observations.

\begin{lemma}
Assume that $\V{G}_\V{x}$ fulfills Hypothesis~\ref{hyp-G}.
Let $\chi$ and $F$ be as above, assume additionally
that $F$ is bounded, and set $\HT:=\Hf+E$, for
some sufficiently large $E\grg1$ (depending on $d_1$).
Let $V\in L^1_\loc(\RR^3,\RR)$ be relatively form-bounded
with respect to $\sqrt{-\Delta}$.
Then, for all $a_0,\kappa\in[0,1)$, $\nu\grg0$, and $a\in[0,a_0]$,
\begin{align}\label{clelia1}
\big\|\,|\DA|^\kappa\,[\PA\,,\,\chi\,e^F]\,e^{-F}\,\big\|\,
&\klg\,
\const(a_0,\kappa)\cdot(a+\|\nabla\chi\|_\infty)\,,
\\\label{clelia2}
\big\|\,\HT^\nu\,[\PA\,,\,\chi\,e^F]\,e^{-F}\,\HT^{-\nu}\,\big\|\,
&\klg\,
\const(a_0,\nu)\cdot(a+\|\nabla\chi\|_\infty)\,,
\\\label{clelia3}
\big\|\,|V|^{1/2}\,[\PA\,,\,\chi\,e^F]\,e^{-F}\,\HT^{-1/2}\,\big\|\,
&\klg\,\const(a_0,V)\cdot(a+\|\nabla\chi\|_\infty)\,.
\end{align}
\end{lemma}

\noindent
Notice that the $a_0$-dependence of the constants originates from the
singularity at $a=1$ in \eqref{exp-marah1}. Notice also that we may
choose $V=|\V{x}|^{-1}$ in \eqref{clelia3} in view of Kato's
inequality.

Before the proof we further remark that all operators appearing in the norms
in \eqref{clelia1}--\eqref{clelia3} and in similar estimates
below are always well-defined a priori on $\sD$ and have
unique bounded extensions to the whole Hilbert space.
In fact, $\PA\,\sD\subset\dom(\DO)\cap\bigcap_{\nu>0}\dom(\Hf^\nu)$
as we have recalled from \cite{MatteStockmeyer2009a} above already.
To simplify the presentation 
we shall not comment on this anymore from now on.

\smallskip

{\proof}
We use the fact that an operator, $T$, acting in some
Hilbert space is bounded if and only if
$\sup_{\|\vp\|,\|\psi\|=1}|\SPn{\vp}{T\,\psi}|$
is bounded in which case it is equal to $\|T\|$.
Here it is sufficient to take the supremum over all
normalized $\vp$ and $\psi$ from a dense set which is a core for $T$. 
Combining \eqref{sgn}, \eqref{|DA|PA}, and \eqref{thomas} we find,
for all normalized $\vp,\psi\in\sD$,
\begin{align*}
\SPb{&|\DA|^\kappa\,\vp}{[\PA\,,\,\chi\,e^F]\,e^{-F}\,\psi}
\\
&=\lim_{\tau\to\infty}\int_{-\tau}^\tau
\SPb{|\DA|^\kappa\,\vp}{\RA{iy}\,i\valpha\cdot(\nabla\chi+\chi\,\nabla F)
\,R_\V{A}^F(iy)\,\psi}\,\frac{dy}{2\pi}\,.
\end{align*}
On account of 
$\|\,|\DA|^\kappa\,\RA{iy}\|\klg\const(\kappa)(1+y^2)^{-1/2+\kappa/2}$,
\eqref{exp-marah1}, and \eqref{thomas2} we see that the
scalar product under the integral sign defines some Lebesgue
integrable function of $y$ and
\begin{align*}
\big|\SPb{&|\DA|^\kappa\,\vp}{[\PA\,,\,\chi\,e^F]\,e^{-F}\,\psi}\big|
\klg\const(\kappa)\,(\|\nabla\chi\|_\infty+a)
\int_\RR\frac{dy}{(1+y^2)^{1-\kappa/2}}\,,
\end{align*}
where the last integral is finite. Therefore, 
$[\PA\,,\,\chi\,e^F]\,e^{-F}\,\psi$ belongs to the domain
of $(|\DA|^\kappa)^*=|\DA|^\kappa$ and the first bound \eqref{clelia1}
follows.

In order to prove the second bound \eqref{clelia2} we introduce
another little tool which turns out to be useful in our whole
analysis. Namely, if $E\grg1$ is sufficiently large
depending on $d_1$ we can construct 
$\Upsilon^F_\nu(iy)\in\LO(\HR)$
such that
$R_\V{A}^F(iy)\,\HT^{-\nu}=\HT^{-\nu}\,R_\V{A}^F(iy)\,\Upsilon^F_\nu(iy)$,
for every $y\in\RR$, and such that the
norm of $\Upsilon^F_\nu(iy)$ is uniformly bounded
with respect to $y\in\RR$; see Corollary~\ref{cor-T-Xi}
below. (In particular, $R_\V{A}^F(iy)$ maps
$\dom(\Hf^\nu)$ into itself.) Therefore,
\begin{align*}
\big|\SPb{&\Hf^\nu \vp}{
\RA{iy}\,i\valpha\cdot(\nabla\chi+\chi\,\nabla F)
\,R_\V{A}^F(iy)\,\HT^{-\nu}\,\psi}\big|
\\
&=\,
\big|\SPb{\vp}{
\RA{iy}\,\Upsilon^0_\nu(iy)\,i\valpha\cdot(\nabla\chi+\chi\,\nabla F)
\,R_\V{A}^F(iy)\,\Upsilon^F_\nu(iy)\,\psi}\big|
\\
&\klg\,
C\,(\|\nabla\chi\|_\infty+a)(1+y^2)^{-1}\,,
\end{align*}
and it is clear from the argument above how to derive \eqref{clelia2}.

The last bound \eqref{clelia3} follows from the first two
and the inequality $|V|/C\klg|\DA|+\Hf+E$
proved later on in Theorem~\ref{le-sb-PF4}.
{\qed}

\section{ Self-adjointness}\label{sec-sb}

\noindent
As it is obvious from the definitions in the 
preceding section the operators $H^{\sharp}_0$ are positive. 
In this section we present some basic 
relative bounds that allow to define the 
perturbed operators $H^{\sharp}_V$ as self-adjoint Friedrichs extensions. 
As a rule we denote the self-adjoint
extensions of $H^\sharp_V$ or $H^{\sharp}_\gamma$ -- which are only 
defined on $\sD$ and $\PA\,\sD$ so far --
again by the same symbols. For suitable $V$,
we are also able to characterize the 
quadratic form domains of $H^{\sharp}_V$ which turn out to be the
spaces of all vectors with finite kinetic 
and radiation field energy;
see Theorems~\ref{le-sb-PF4} and~\ref{thm-sb-np} below.

Before we present the afore-mentioned results we discuss various 
(essentially well-known) diamagnetic inequalities in QED; 
see Theorem \ref{thm-dia} below. 
Since these estimates are of independent interest 
we decided to present one way to derive them 
(adapted from \cite{Barry1979}) in detail
which has not been worked out in the literature before, as it
seems to us.


\subsection{Diamagnetic inequalities in QED}\label{ssec-dia}

\noindent
In this subsection it is sufficient to assume that 
$$
\V{A}\,=\,\int_{\RR^3}^{\oplus}\V{A}(\V{x})\,d^3\V{x}\,,
\qquad 
\V{A}(\V{x}):= \id_{\CC^4}\otimes
\big(a^\dagger(\V{G}_\V{x})+a(\V{G}_\V{x})\big)\,,
$$
where $\RR^3\times(\RR^3\times\ZZ_2)\ni(\V{x}, k)\mapsto \V{G}_{\V{x}}(k)$
is measurable such that $\V{x}\mapsto \V{G}_{\V{x}}(k) $ 
is continuously differentiable, for almost every $k$, and
\begin{align*}
\int(\omega(k)^{-1}+\omega(k)^2)\sup_{\V{x}}|\V{G}_{\V{x}}(k)|^2
dk&<\infty\,,\\
\int(1+\omega(k)^{-1})\sup_{\V{x}}
|\nabla_{\V{x}}\V{G}_{\V{x}}(k)|^2dk&<\infty\,.
\end{align*}
The following result is probably well-known but the
argument sketched in its proof might be new.

\begin{lemma} Let $\lambda\grg0$.
Under the above condition on $\V{G}_{\V{x}}$ the operator
$(-i\nabla+\V{A})^2+\lambda\,\Hf$ is
essentially self-adjoint on $\core$.  
\end{lemma}

{\proof}
It is a standard exercise to show that 
$\{-i\nabla,\V{A}\}+\V{A}^2$ is a small operator
perturbation of $-\Delta+c\,\Hf$, provided that $c>0$
is chosen sufficiently large depending on $\V{G}_\V{x}$.
In particular, $\cN:=(-i\nabla+\V{A})^2+c\,\Hf+1$
is essentially self-adjoint on any core of
$-\Delta+c\,\Hf$ and, in particular, on $\core$.
In the next step we apply Nelson's commutator theorem with
the closure of $\cN$ starting from $\core$
as test operator to conclude.
{\qed}

\smallskip

\noindent
We denote the closure of $(-i\nabla+\V{A})^2$ starting
from $\core$ by $\tau_{\V{A}}$.
For every $\phi,\psi\in \HR=L^2(\RR^3_\V{x},\CC^4\otimes\Fock)$, 
we write $\RSP{\phi}{\psi}$ for the (partial)
scalar product on $\CC^4\otimes\Fock$ and denote
$\llb\vp\rrb(\V{x}):=\RSP{\vp(\V{x})}{\vp(\V{x})}^{1/2}$. Furthermore,
we set $S_\phi(\V{x}):=\frac{1}{\llb \phi \rrb (\V{x})}
\,\phi(\V{x})$, for $\phi(\V{x})\not= 0$, and $S_\phi(\V{x})=0$, for
$\phi(\V{x})=0$.
\begin{theorem}\label{thm-dia}
(i)  Let $\phi\in\mathcal{D}(\tau_{\V{A}})$. 
Then $\llb \phi \rrb \in H^1(\RR^3)$, and
  \begin{equation}
    \label{dia2}
  \SPn{\eta}{-\Delta\,\llb \phi\rrb}_{L^2(\RR^3)}\,\klg\,\Re\,\int_{\RR^3}
\,\eta(\V{x})\,\RSPb{S_\phi (\V{x})}{(\tau_{\V{A}}\,\phi)(\V{x}))}\,d^3\V{x}\,,
    \end{equation}
for all $\eta\in H^1(\RR^3)$, $\eta\grg 0$. 
In particular, for $\eta=\llb\phi\rrb$,
\begin{equation}
  \label{dia3}
  \SPn{\llb\phi\rrb}{-\Delta\,\llb\phi\rrb}_{L^2(\RR^3)}\,
\klg\, \SPn{\phi}{\tau_{\V{A}}\,\phi}\,.
\end{equation}
(ii)  Let $\phi\in\mathcal{D}(\tau_{\V{A}}^{1/2})$. 
Then $\llb \phi \rrb \in H^{1/2}(\RR^3)$, and
 \begin{equation}
    \label{dia4}
  \SPn{\eta}{\sqrt{-\Delta}\,\llb \phi\rrb}_{L^2(\RR^3)}\,
\klg\,\Re\int_{\RR^3}
\,\eta(\V{x})\,\RSPb{S_\phi (\V{x})}{(\tau_{\V{A}}^{1/2}\phi)(\V{x})}
\,d^3\V{x}\,,
    \end{equation}
for all $\eta\in H^{1/2}(\RR^3)$, $\eta\grg 0$.
In particular, for $\eta=\llb\phi\rrb$,
\begin{equation}
  \label{dia5}
  \SPn{\llb\phi\rrb}{\sqrt{-\Delta}\,\llb\phi\rrb}_{L^2(\RR^3)}\,\klg\, \SPn{\phi}{\tau_{\V{A}}^{1/2}\,\phi}\,.
\end{equation}
(iii) For all $\psi\in\HR$ and $t\in[0,\infty)$, we have, 
almost everywhere on $\RR^3$,
\begin{align}
  \label{dia6}
  \llb\,e^{-t\tau_{\V{A}}}\,\psi\,\rrb\,
&\klg\,e^{-t(-\Delta)}\,\llb \,\psi\,\rrb\,,
\\\label{dia7}
\llb\,e^{-t\tau_{\V{A}}^{1/2}}\,\psi\,\rrb\,
&\klg\,e^{-t\sqrt{-\Delta}}\,\llb \,\psi\,\rrb\,.
\end{align}
\end{theorem}

\begin{remark}{\em
(1) Arguing as in Theorem~7.21 of \cite{LiebLossb} 
with the corresponding changes as in the proof below 
one can easily verify that, for
$\phi\in\core$, we have $\llb\phi\rrb\in H^1(\RR^3)$ and 
$|\nabla\,\llb\,\phi\,\rrb  (\V{x})|\klg
\llb(-i\nabla+  \V{A}) \phi\,\rrb(\V{x})$,
for a.e. $\V{x}\in\RR^3$.

\smallskip

\noindent
(2) In \cite{Hiroshima} diamagnetic inequalities for infra-red regularized 
vector potentials have been proved by means of dressing transformations.
For an alternative proof  using functional integrals see \cite{Hiroshima2}. 
If all components of the vector potential commute,
$[A^{(j)}(\V{x}),A^{(k)}(\V{y})]=0$, 
then one can also reduce the diamagnetic inequalities to classical ones 
by diagonalizing all components $A^{(j)}(\V{x})$
simultaneously; this argument due to J.~Fr\"ohlich is mentioned in \cite{AHS}.  
The proofs given here are variants of the ones presented in 
\cite{ReedSimonII,Barry1979}.}
\end{remark}

{\proof}
Let $\ve>0$.
First, we assume that $\phi\in\core$ and set 
$u_\ve:=\sqrt{\llb\phi\rrb^2+\ve^2}\in C^\infty(\RR^3,\RR)$. 
Since $A^{(j)}(\V{x})$ is symmetric on $\sC_0$,
for every $\V{x}\in\RR^3$, we have 
$\Re \RSP{\phi(\V{x})}{iA^{(j)}(\V{x})\,\phi(\V{x})}=0$, thus
\begin{equation}
  \label{eq:11}
u_\ve\,\nabla\,u_\ve\,=\,\frac12 \,\nabla\,u_\ve^2\,=\,
\Re\RSP{\phi}{\nabla\,\phi}\,=\,  
\Re\RSP{\phi}{(\nabla+i\V{A})\,\phi}\,.
\end{equation}
In particular,
\begin{equation}
  \label{eq:18}
|\nabla\,u_\ve|\,\klg\,
\frac{\llb\,\phi\,\rrb}{u_\ve}\,\llb(\nabla+i\V{A})\,\phi\rrb\,
\klg\,\llb(\nabla+i\V{A})\,\phi\rrb\,.
\end{equation}
Taking the divergence of \eqref{eq:11} we obtain
\begin{align}
  \nonumber
|\nabla\,u_\ve|^2+u_\ve \,\Delta\,u_\ve\,
&=\,\Re\,\RSP{\nabla\phi}{(\nabla+i\V{A})\,\phi}
+\Re\,\RSP{\phi}{\nabla\,(\nabla+i\V{A})\,\phi}
\\
&\label{eq:19}
=\llb(\nabla+i\V{A})\,\phi\rrb^2-\Re\,\RSP{\phi}{\tau_{\V{A}}\,\phi}\,.
\end{align}
Combining this identity with \eqref{eq:18} we arrive at
\begin{equation}
  \label{eq:20}
  -\Delta\,u_\ve\,\klg\,\Re\,\RSP{u_\ve^{-1}\,\phi}{\tau_{\V{A}}\,\phi}\,.
\end{equation}
Now, assume that $\phi\in\mathcal{D}(\tau_{\V{A}})$. 
Since $\tau_{\V{A}}$ is essentially self-adjoint on $\core$ 
we find $\phi_n\in\core$, $n\in\NN$,
such that $\phi_n\to\phi$ and $\tau_{\V{A}}\phi_n\to\tau_{\V{A}}\phi$ in $\HR$.
On account of \eqref{eq:20} we have 
\begin{equation}
  \label{eq:21}
\int_{\RR^3}(-\Delta\,\eta)(\V{x})\,u_\ve^{(n)}(\V{x})\,d^3\V{x}\,
\klg\,\Re\,\SPn{\eta\, (u_\ve^{(n)})^{-1}\,\phi_n}{\tau_{\V{A}} \,\phi_n}\,,
\end{equation}
for all Schwartz functions
$\eta \in \sS(\RR^3)$, $\eta\grg 0$, 
where $u_\ve^{(n)}:=\sqrt{\llb\phi_n\rrb^2+\ve^2}$, $n\in\NN$. 
Passing to appropriate subsequences if necessary
we may assume that $\llb\phi_n\rrb\to\llb\phi\rrb$ and, hence,
$u_\ve^{(n)}\to u_\ve$ almost everywhere. Using that  
$u_\ve^{-1}, (u_\ve^{(n)})^{-1}\klg 1/\ve$, it is easy to see 
that $\eta\, (u_\ve^{(n)})^{-1}\,\phi_n\to \eta\,u_\ve^{-1}\,\phi$ in $\HR$. 
By virtue of the Riesz-Fischer theorem we further 
find a square-integrable majorant for the sequence $(\llb\phi_n\rrb)$. 
We can thus pass to the limit $n\to\infty$ in \eqref{eq:21} to get, 
for all $\eta\in\sS(\RR^3)$, $\eta\grg 0$, and 
$\phi\in\mathcal{D}(\tau_{\V{A}})$,
\begin{equation}
  \label{eq:22}
\int_{\RR^3}\!(-\Delta\eta)(\V{x})\,u_\ve(\V{x})\,d^3\V{x}\klg
\Re\!\int_{\RR^3}\!\!\eta(\V{x})\,
\RSP{u_\ve^{-1}(\V{x})\phi(\V{x})}{(\tau_{\V{A}}\phi)(\V{x})}\,d^3\V{x}.
\end{equation}
Here we may take the limit $\ve\to0$ 
by means of the dominated convergence theorem 
(with the majorant $\eta\,\llb\tau_{\V{A}}\phi\rrb$ on the right hand side) 
to obtain, for all $\eta\in\sS(\RR^3)$, $\eta\grg 0$, and 
$\phi\in\mathcal{D}(\tau_{\V{A}})$,
\begin{equation}
  \label{eq:23}
  \int_{\RR^3}(-\Delta\,\eta)(\V{x})\,\llb\,\phi\,\rrb(\V{x})\,d^3\V{x}
\,\klg\,
\Re\int_{\RR^3}\eta(\V{x})\,\RSP{S_\phi(\V{x})}{(\tau_{\V{A}}\,\phi)(\V{x})}
\,d^3\V{x}\,.
\end{equation}
Adding $\int\eta\lambda\llb\phi\rrb$ with $\lambda>0$ to both sides 
we obtain
\begin{equation*}
  \begin{split}
\int_{\RR^3}[(-\Delta+\lambda)\eta](\V{x})\,\llb\,\phi\,\rrb(\V{x})\,d^3\V{x}
&\klg\, 
\Re \int_{\RR^3}\eta(\V{x})\,
\RSP{S_\phi}{(\tau_{\V{A}}+\lambda)\phi}(\V{x})\,d^3\V{x}
\\
&\klg\,\int_{\RR^3}\eta(\V{x})\,
\llb(\tau_{\V{A}}+\lambda)\,\phi\rrb(\V{x})\,d^3\V{x}\,.
\end{split}
\end{equation*}
Let $0\klg\chi \in C_0^\infty (\RR^3)$ and $\psi\in\HR$. 
Since $(-\Delta+\lambda)^{-1}$, $\lambda>0$, is positivity preserving, 
we may then choose 
$\eta:=(-\Delta+\lambda)^{-1}\chi\in\sS(\RR^3)$ 
and $\phi:=(\tau_{\V{A}}+\lambda)^{-1}\psi\in\mathcal{D}(\tau_{\V{A}})$ 
and arrive at
\begin{equation*}
  \int_{\RR^3}\chi(\V{x})\,\llb\,(\tau_{\V{A}}+\lambda)^{-1}\psi\,\rrb
(\V{x})\,d^3\V{x}\,\klg\,\int_{\RR^3}\chi(\V{x})\,
[\,(-\Delta+\lambda)^{-1}\llb\,\psi\,\rrb\,](\V{x})\,d^3\V{x}\,.
\end{equation*}
Since $\chi\in C_0^\infty (\RR^3)$ is arbitrary we find 
$\llb\,(\tau_{\V{A}}+\lambda)^{-1}\psi\,\rrb
\klg (-\Delta+\lambda)^{-1}\llb\,\psi\,\rrb$, 
almost everywhere on $\RR^3$, and by induction 
(see \cite{Barry1979}, for the same argument) we get, 
for all $n\in\NN$ and $t>0$,
\begin{equation*}
  \llb\,(\tfrac{n}{t})^n \,(\tau_{\V{A}}+\tfrac{n}{t})^{-n}\,\psi\,\rrb 
\,\klg\,(\tfrac{n}{t})^n (-\Delta+\tfrac{n}{t})^{-n} \llb\,\psi\,\rrb\,.
\end{equation*}
Both sides converge almost everywhere along  some subsequence to 
$\llb e^{-t\tau_{\V{A}}}\psi\rrb$ and $e^{-t(-\Delta)}\llb\psi\rrb$ respectively, 
and \eqref{dia6} follows. 
Equation \eqref{dia7} follows from \eqref{dia6}, the spectral calculus, 
the subordination identity
\begin{equation*}
  e^{-t\lambda^{1/2}}\,=\,\int_0^\infty e^{-s-t^2\lambda/(4s)}\,
\frac{ds}{\sqrt{\pi s}}\,,\qquad t,\lambda\grg0\,,
\end{equation*}
and the properties of the Bochner-Lebesgue integral. 
In the remaining part of the proof we derive 
(following again \cite{Barry1979}) \eqref{dia2} and \eqref{dia4} 
at the same time. To this end let $\nu\in \{1/2, 1\}$ 
and $\phi\in\mathcal{D}(\tau_{\V{A}}^\nu)$.

On account of
$\SPn{\phi}{e^{-t\tau_{\V{A}}^\nu}\phi}
\klg\int\llb\phi\rrb\,\llb e^{-t\tau_{\V{A}}^\nu}\phi\rrb$ 
Equations \eqref{dia6} and \eqref{dia7} imply 
$\SPn{\phi}{e^{-t\tau_{\V{A}}^\nu}\phi}
\klg \int \llb\phi\rrb e^{-t(-\Delta)^\nu}\llb\phi\rrb$, thus
\begin{equation*}
  \SPn{\phi}{t^{-1}(1-e^{-t\tau_{\V{A}}^\nu})\phi}\,
\grg\, \int_{\RR^3} t^{-1}(1-e^{-t|\V{\xi}|^{2\nu}})|\llb\phi\rrb\,\hat{}\,|^2
(\V{\xi})\,d^3\V{\xi}\,. 
\end{equation*}
Since $\phi\in \mathcal{D}(\tau_{\V{A}}^\nu)$ the limit $t\to0$ 
exists on the left hand side of the previous inequality. 
By the monotone convergence theorem 
we conclude that the limit $t\to0$ of the right hand side exists, too, and 
\begin{equation*}
  \SPn{\phi}{\tau_{\V{A}}^\nu\phi}\,
\grg\,\int_{\RR^3} |\V{\xi}|^{2\nu}|\llb\phi\rrb\,\hat{}\,|^2(\V{\xi})\,
d^3\V{\xi}\,.
\end{equation*}
Hence $\llb\phi\rrb\in H^{\nu}(\RR^3)$ and \eqref{dia3} and \eqref{dia5}  
hold true. Using this we may take the derivatives at $t=0$ on the left 
and right sides of the following consequence of \eqref{dia6} and \eqref{dia7},
\begin{equation}
  \label{eq:25}
\begin{split}
 \Re \int_{\RR^3} \eta(\V{x})\RSP{S_\phi}{&e^{-t\tau_{\V{A}}^\nu}\,\phi}
(\V{x})\,d^3\V{x}
\klg\,
 \int_{\RR^3}\eta(\V{x})\llb\,e^{-t\tau_{\V{A}}^\nu}\,\phi\,\rrb(\V{x})
\,d^3\V{x}
\\
&\klg\,\int_{\RR^3}\eta(\V{x})\,e^{-t(-\Delta)^\nu}\,\llb\,\phi\,\rrb(\V{x})
\,d^3\V{x}\,,
\end{split}
\end{equation}
to get 
$-\Re \,\SPn{\eta S_\phi}{\tau_{\V{A}}^\nu\,\phi}
\klg\,-\SPn{\eta}{(-\Delta)^{\nu}\llb\,\phi\,\rrb}_{L^2(\RR^3)}$,
for all $\eta\in H^\nu(\RR^3), \eta\grg 0$. 
Here we have also used that all expressions in \eqref{eq:25} are equal to 
$\int \eta \llb\,\phi\,\rrb$ at $t=0$.
{\qed}


\subsection{Semi-boundedness}

\noindent
The following theorem is a slight generalization of a result from
\cite{MatteStockmeyer2009a}. Its proof is based on two basic steps:
The first one follows immediately from the diamagnetic inequalities
by means of which the form bounds the potential satisfies by
assumption can be turned into form bounds with respect to the
{\em scalar} operators $\tau_{\V{A}}^{1/2}$ or $\tau_\V{A}$.
After that we use relative bounds on the magnetic field
to include spin, $\tau_{\V{A}}\klg\DA^2+\Hf^2+\const$.
To complete the square on the right hand side of the previous bound
we employ the inequality \eqref{eq:8} below.
After completing the square we take square roots on both sides
to obtain a bound on $\tau_{\V{A}}^{1/2}$.

\begin{theorem}\label{le-sb-PF4}
 For $\nu\in\{1/2,1\}$, let $V_\nu\in L^1_{\rm loc}(\RR^3, \RR)$ and
assume that there is some $c_\nu\grg 0$
  such that, for every $\vp\in H^\nu(\RR^3)$,
  \begin{equation}\label{hyp-Vnu}
\SPn{\vp}{V_\nu\,\vp}_{L^2(\RR^3)}\,
\klg\,\SPn{\vp}{(-\Delta)^\nu\,\vp}_{L^2(\RR^3)}
+c_\nu\,\|\,\vp\,\|_{L^2(\RR^3)}^2.
  \end{equation}
Then there is some $C\in(0,\infty)$ such that, 
for all $\V{G}_\V{x}$ fulfilling Hypothesis~\ref{hyp-G},
$\phi\in\core$, and $\delta>0$,
  \begin{equation}
    \label{gustav-gen}
    \SPb{\phi}{V_\nu \phi}\,\klg\,\SPb{\phi}{\big(|\DA|+\delta\,
      \Hf+(\delta^{-1}+C\,\delta)
      \,d_1^2\big)^{2\nu}\,\phi}+c_\nu\,\|\phi\|^2.
  \end{equation}
In particular,
\begin{equation}\label{gustav}
\frac{1}{4}\,\big\|\,|\V{x}|^{-1}\,\phi\,\big\|^2\,\klg\,
\big\|\,\big(|\DA|+\delta\,\Hf+(\delta^{-1}+C\,\delta)
\,d_1^2\big)\,\phi\,\big\|^2\,\,,
\end{equation}
for all $\phi\in\core$, and
\begin{equation}\label{gustav2}
\frac{2}{\pi}\frac{1}{|\V{x}|}\,\klg\,|\DA|+\delta\,\Hf
+(\delta^{-1}+C\,\delta)\,d_1^2\,,
\end{equation}
in the sense of quadratic forms on $\core$.
Therefore, $\PF{V_{1/2}}$ and $\PF{\gamma}$, $\gamma\in[0,\gcPF]$,
have self-adjoint Friedrichs extensions -- henceforth again denoted
by the same symbols -- and $\core$ is a form core for these extensions. 
Moreover,
for $a\in[0,1)$ and $\gamma\in[0,\gcPF)$, we know that 
$\form(\PF{aV_{1/2}})=\form(\PF{\gamma})=\form(\PF{0})=\form(|\DO|)\cap \form(\Hf)$.
\end{theorem}

{\proof}
First, we show that  
$\form(\PF{0})=\form(|\DO|)\cap \form(\Hf)$;
see \cite{KMS2009a,Stockmeyer2009}.
The remaining statements on form domains will then be a
consequence of \eqref{gustav-gen} and \eqref{gustav2}.
In fact, this follows from the bounds \cite{MatteStockmeyer2009a}
$$
\big\|\,|\DO|^{1/2}(\SA-\SO)\,\HT^{-1/2}\big\|\klg C\,,\qquad
\big\|\,|\DA|^{1/2}(\SA-\SO)\,\HT^{-1/2}\big\|\klg C\,,
$$
where $\HT:=\Hf+E$ with $E\grg1+(4d_1)^2$.
(These bounds are derived exactly as in Lemma~\ref{veronique} below.)
Together with
$|\DA|-|\DO|=\DO\,(\SA-\SO)+\valpha\cdot\V{A}\,\SA$
and \eqref{rb-a}\&\eqref{rb-ad}
the first bound implies
\begin{align}\nonumber
|\SPb{\vp}{(|\DA|-|\DO|)\,\vp}|
&\klg C'\,\big\|\,|\DO|^{1/2}\vp\big\|\,\|\HT^{1/2}\vp\|
\\\label{pavel}
&\klg C''\SPb{\vp}{(|\DO|+\Hf)\,\vp}\,,
\end{align}
for all $\vp\in\sD$.
Analogously, the second bound implies \eqref{pavel} with
$|\DO|$ replaced by $|\DA|$ on the right hand side.
Consequently, the form norms of $|\DA|+\Hf$ and $|\DO|+\Hf$
are equivalent on $\sD$ which implies 
$\form(\PF{0})=\form(|\DO|)\cap \form(\Hf)$.

All details missing in the proof of 
\eqref{gustav-gen}--\eqref{gustav2} sketched
 below can be found in \cite{MatteStockmeyer2009a}. 
We set 
  $\HT:=\Hf+E$, for $E>0$. Besides some standard arguments the
 main ingredient in this proof is the following bound proven
 in \cite[Lemma~4.1]{MatteStockmeyer2009a}:
 We find some constant, $C>0$, such that, for
 all $E>C\,d_1^2$ and $\phi\in \core$,
\begin{equation}
  \label{eq:8}
  \Re\SPb{|\DA|\,\phi}{\HT\,\phi}\,\grg\,0\,.
\end{equation}
This estimate follows from the following identity 
$
\Re(|\DA|\,\HT)=\HT^{1/2}(|\DA|-\mathcal{T})\HT^{1/2}
$
on $\core$, where 
$$
\mathcal{T}:=\Re\{[|\DA|,\HT^{-1/2}]\HT^{1/2}\}
\klg\ve\,|\DA|+\ve^{-1}\,\const\,d_1/E^{1/2},
$$
for $\ve\in(0,1]$ and $E\grg(4d_1)^2$,
as we shall see at the end of Subsection~\ref{exopu}.
To make use of the bound \eqref{eq:8} we recall that,
since $[A^{(j)}(\V{x}),A^{(k)}(\V{y})]=0$, we have
\begin{equation}\label{ttt1}
\DA^2\,\phi\,=\,\tau_\V{A}\,\phi\,
+\,\V{S}\cdot\V{B}\,\phi\,+\,\phi\,,\qquad \phi\in\core\,,
\end{equation}
where 
the entries of the formal three-vector $\V{S}$
are $S_j=\sigma_j\otimes\id_2$ and
$\V{B}$ is the magnetic field, i.e.
$
\V{S}\cdot\V{B}=
\V{S}\cdot\ad(\nabla_{\V{x}}\wedge\V{G})
+\V{S}\cdot a(\nabla_{\V{x}}\wedge\V{G})
$.
By \eqref{rb-a} with $(\nabla_\V{x}\wedge\V{G},d_1)$ instead of
$(\V{G},d_{-1})$ we have, for all $\delta>0$ and $\phi\in\core$, 
\begin{equation}\label{rb-B}
\big|\SPb{\phi}{\V{S}\cdot\V{B}\,\phi}\big|\,\klg\,
2\,d_1\,\|\phi\|\,\big\|\,\Hf^{1/2}\,\phi\,\big\|\,\klg\,
\delta\,\SPb{\phi}{(\Hf+\delta^{-2}\,d_1^2)\,\phi}\,.
\end{equation}
Choosing $E=(\delta^{-2}+C)\,d_1^2$
we infer from \eqref{eq:8}--\eqref{rb-B},
for all $\phi\in \core$,
\begin{align}
  \SPb{\phi}{\tau_\V{A}\,\phi}
&\klg\,\label{bea1}
\SPb{\DA\,\phi}{\DA\,\phi}+\delta\,\SPb{\phi}{\HT\,\phi}-\|\phi\|^2
\\
&\klg \,\nonumber
\SPb{\DA\,\phi}{\DA\,\phi}+\SPb{\phi}{\delta^2\,\HT^2\,\phi}
+2\Re\,\SPb{|\DA|\,\phi}{\delta\,\HT\,\phi}
\\
&=\,\nonumber
\big\|\,(|\DA|+\delta\,\HT)\,\phi\,\big\|^2.
\end{align}
Furthermore, since the square root is
operator monotone it follows from \eqref{bea1} 
that $\SPb{\phi}{\tau_\V{A}^{1/2}\,\phi}\,\klg
\,\SPb{\phi}{(|\DA|+\delta\,\HT)\,\phi}$.
Using the diamagnetic inequalities \eqref{dia3} and \eqref{dia5}  
we further find, for $\nu\in\{1/2, 1\}$,
\begin{align}\nonumber
  \SPn{\phi}{ V_\nu \,\phi}\,&=\,\SPn{\llb\,\phi\rrb}{
    V_\nu\,\llb\,\phi\rrb}_{L^2(\RR^3)}
\klg\,\SPn{\llb\,\phi\rrb}{(-\Delta)^{\nu}\,\llb\,\phi\,\rrb}_{L^2(\RR^3)}
+c_\nu\,\|\phi\|^2
\\
&\klg\,\SPb{\phi}{\tau_\V{A}^{2\nu}\,\phi}+c_\nu
  \|\phi\|^2,\label{nunu}
\end{align}
and we conclude that \eqref{gustav-gen} holds true. 
Inequalities \eqref{gustav} and \eqref{gustav2} follow from 
\eqref{nunu} together with Hardy's and Kato's inequality, respectively. 
{\qed}

\smallskip

\noindent
Theorem~\ref{le-sb-PF4} has a straightforward extension
to the case of $N$ electrons \cite{Matte2009}. 
We discuss this extension in the next corollary
mainly since its proof gives the opportunity to
introduce some identities and estimates which are used
later on. Let $\HR_N$ and $\sD_N$, $N\in\NN$, be defined in the same way
as $\HR$ and $\sD$ but with the $L^2(\RR^3,\CC^4)=L^2(\RR^3\times\ZZ_4)$ 
replaced by $L^2((\RR^3\times\ZZ_4)^N)$. The spatial coordinates
of the $i$-th electron are denoted by $\V{x}_i\in\RR^3$ and
we designate an operator acting only on $\V{x}_i$, the $i$-th
spinor components,
and on the photon field by a superscript $(i)$.

\begin{corollary}\label{hoe-prop-sb-srPF}
Assume that $N,K\in\NN$, $e>0$, $\gamma_1,\ldots,\gamma_K\in(0,2/\pi]$,
and $\{\V{R}_1,\ldots,\V{R}_K\}\subset\RR^3$. Then
\begin{equation}\label{hoe-manuela-1}
\sum_{i=1}^N|\DA^{(i)}|
-\sum_{i=1}^N\sum_{k=1}^K\frac{\gamma_k}{|\V{x}_i-\V{R}_k|}
+\sum_{i<j}\frac{e^2}{|\V{x}_i-\V{x}_j|}+\,\delta\,\Hf\,>\,-\infty\,,
\end{equation}
for every $\delta>0$, in the sense of quadratic forms on $\sD_N$.
\end{corollary}

{\proof}
In view of \eqref{gustav2} we only have to explain how to
localize the non-local kinetic energy terms. 
To begin with we note the following special cases of
\eqref{clelia1} and \eqref{exp-dc-DA}, respectively:
For every $\chi\in C^\infty(\RR^3_\V{x},[0,1])$, 
\begin{align}\label{hoe-tim1}
\|\,[\chi,\SA]\,\|\,\klg\,\|\nabla\chi\|_\infty\,,
\qquad
\big\|\DA\,\big[\chi\,,\,[\chi,\SA]\,\big]\big\|\,
\klg\,2\,\|\nabla\chi\|_\infty^2\,.
\end{align}
Let $\ball{r}{\V{z}}$ denote the open ball  in $\RR^3$ of radius $r>0$
centered at $\V{z}\in\RR^3$.
We set $\vr:=\min\{|\V{R}_k-\V{R}_\ell|:\,k\not=\ell\}/2$ and 
pick a smooth partition of unity on $\RR^3$, $\{\chi_k\}_{k=0}^K$,
such that $\chi_k\equiv1$ on $\ball{\vr/2}{\V{R}_k}$ and
$\supp(\chi_k)\subset \ball{\vr}{\V{R}_k}$, for $k=1,\ldots,K$,
and such that $\sum_{k=0}^K\chi_k^2=1$.
Combining the following IMS type localization formula,
\begin{equation}\label{hoe-IMS-DA1}
|\DA|\,=\,
\sum_{k=0}^K\Big\{\,\chi_k\,|\DA|\,\chi_k\,+\,\frac{1}{2}\,
\big[\chi_k\,,\,[\chi_k,|\DA|\,]\,\big]\,\Big\}
\quad\textrm{on}\;\core\,,
\end{equation}
and the identities 
\begin{align}\label{hoe-IMS-DA2}
\big[\chi_k\,,\,[\chi_k,|\DA|\,]\,\big]\,&=\,
2\,i\valpha\cdot(\nabla\chi_k)\,[\chi_k,\SA]+
\DA\,\big[\chi_k\,,\,[\chi_k,\SA]\,\big]
\end{align}
and $\|\valpha\cdot\nabla\chi_k\|=|\nabla\chi_k|$ 
with \eqref{hoe-tim1}, we obtain
\begin{equation}\label{hoe-IMS-Da3}
\big\|\,\big[\chi_k\,,\,[\chi_k,|\DA|\,]\,\big]\,\big\|\,
\klg\,4\,\|\nabla\chi_k\|_\infty^2\,,
\end{equation}
for all $k\in\{0,\ldots,K\}$.
Since we are able to localize the kinetic energy terms
and since, by the choice of the partition of unity,
the functions $\RR^3\ni\V{x}\mapsto
|\V{x}-\V{R}_k|^{-1}\,\chi_\ell^2(\V{x})$
are bounded, for $k\in\{1,\ldots,K\}$,
$\ell\in\{0,\ldots,K\}$, $k\not=\ell$,
the bound \eqref{hoe-manuela-1} is now an immediate 
consequence of \eqref{gustav2}.
{\qed}

\smallskip

\noindent
Next, we discuss the
semi-boundedness of the no-pair operator.
The idea is to reduce the stability of the no-pair operator
to the one of the purely electronic
Brown-Ravenhall operator.

\begin{theorem}\label{thm-sb-np}
Assume that $\V{G}_\V{x}$ fulfills Hypothesis~\ref{hyp-G}.
Let $\delta\in (0,1]$ 
and let $V\in L^2_\loc(\RR^3,\RR)$ be form bounded with respect to
$\sqrt{-\Delta}$ and satisfy
\begin{equation}\label{irina}
\SPn{\PO\,\vp}{V\,\PO\,\vp}\klg a\,\SPn{\vp}{\DO\,\PO\,\vp}
+b\,\|\PO\,\vp\|^2,\quad \vp\in C_0^\infty(\RR^3)\,,
\end{equation}
for some $a\in(0,1)$ and $b\grg0$.
Then there exist
  constants $c_V, C\in(0,\infty)$, $C\equiv
  C(\delta,V,d_{-1},d_1)$, such that, for all
  $\vp^+\in\PA\,\core$, $\|\vp^+\|=1$,
\begin{equation}\label{eqqq22V}
\SPn{\vp^+}{(\DA+V+\delta\Hf)\,\vp^+}\,
\grg\,c_{V}\SPn{\vp^+}{|\DO|\,\vp^+}
-C\,,
\end{equation}
and in particular, for every
$\gamma\in[0,\gcnp)$, 
\begin{equation}\label{eqqq22}
\SPn{\vp^+}{(\DA-\tgV+\delta\Hf)\,\vp^+}\,
\grg\,c_{\gamma}\SPn{\vp^+}{|\DO|\,\vp^+}
-C(\delta,\gamma,d_{-1},d_1)\,.
\end{equation}
Therefore, $\NPO{V}$ and 
$\NPO{\gamma}$ have self-adjoint Friedrichs
extensions -- henceforth again denoted by the same symbols --
and $\PA\,\core$ is a form core for these extensions. 
Furthermore, 
$\form(\NPO{V})=\form(\NPO{\gamma})=\form(\NPO{0})
=\form(|\DO|)\cap \form(\Hf)\cap \mathrm{Ran}\PA$.
\end{theorem}

{\proof}
 The statement on the form domains 
follows from Theorem~3.2 of \cite{KMS2009b}.
(See also Section~3.4 of the {\em first} preprint version of
\cite{KMS2009b}
available on the arXiv for an alternative proof.)
The estimate \eqref{eqqq22} is derived in  
\cite{MatteStockmeyer2009a} and we shall outline its proof in what follows.
  
We pick some $\rho>1$ with $\rho\,a<1$ and write,
for $\vp^+\in\PA\,\core$,
\begin{equation}\label{id-br12}
\begin{split}
\SPb{\vp^+}{(\DA+V)\,\vp^+}
&=\,
\rho^{-1}
\SPb{\vp^+}{\PO\,(\DO+\rho\,V)\,\PO\,\vp^+}
\\
& 
\;+\;
(1-\rho^{-1})\,\SPb{\vp^+}{\PO\,\DO\,\vp^+}
\\
& 
\;+\;
\SPb{\vp^+}{\valpha\cdot \V{A}\,\vp^+}
\,
\\
& 
\;+\;
\SPb{\vp^+}{\POm\,(\DO+V)
\,\POm\,\vp^+}
\\
& 
\;+\;
2\,\Re\SPb{\vp^+}{\PO\,V\,\POm\,\vp^+}\,.
\\
\end{split}
\end{equation}
The estimate \eqref{eqqq22} is based on the identity above 
and the following bound on the difference between the spectral 
projections with and without field 
(see Lemma~\ref{veronique} for similar statements): 
For $E\grg1+(4d_1)^2$, there is a constant, $C\equiv C(d_{-1},d_0)>0$, such that
\begin{eqnarray}
\big\|\,|\DO|^{3/4}(\POpm-\PApm)\HT^{-1/2}\big\|
&\klg&\, C\,,\label{peki111}
\end{eqnarray}
where $\HT=\Hf+E$.
On account of $
\POm\,\vp^+\,=\,
(\POm\,-\,\PAm)\,\vp^+$, for $\vp^+\in\PA\,\core$, 
and \eqref{peki111} we have, for every $\ve>0$, 
\begin{align*}
\big\|\,&|\DO|^{1/2}\,\POm\,\vp^+\,\big\|^2
\klg\,
\big\|\,|\DO|^{1/4}\,\POm\,\vp^+\,\big\|\,
\big\|\,|\DO|^{3/4}\,(\POm-\PAm)\,\vp^+\,\big\|
\\
&\klg\,
\frac{1}{2}\,\big\|\,|\DO|^{1/2}\,\POm\,\vp^+\,\big\|^2
+\frac{C(\ve,d_{-1},d_0)}{2}\,\big\|\,\POm\,\vp^+\,\big\|^2
+\frac{\ve}{2}\,\big\|\,\HT^{1/2}\,\vp^+\,\big\|^2,
\end{align*}
that is, 
\begin{equation}
\big\|\,|\DO|^{1/2}\,\POm\,\vp^+\,\big\|^2
\,\klg\,\label{clelia100}
\ve\,\big\|\,\HT^{1/2}\,\vp^+\,\big\|^2
+C(\ve,d_{-1},d_0)\,\big\|\,\POm\,\vp^+\,\big\|^2.
\end{equation}
By virtue of $|V|\klg C\,|\DO|$, 
the previous estimate further implies, for every $\tau>0$, 
\begin{align}
&\nonumber
\big|\SPb{\PO\,\vp^+}{V\,\POm\,\vp^+}\big|
\\
&\klg\label{clelia102}
\tau\,\big\|\,|\DO|^{1/2}\,\PO\,\vp^+\,\big\|^2+
\ve\,\big\|\,\HT^{1/2}\,\vp^+\,\big\|^2
+C_{\ve,\tau}\,
\big\|\,\POm\,\vp^+\,\big\|^2.
\end{align}
Here the second term on the RHS of \eqref{id-br12} can be used to 
control the first term on the RHS of \eqref{clelia102}.
Recalling the definition \eqref{def-Bel} and
applying \eqref{rb-a}, \eqref{clelia100}, \eqref{clelia102}, and \eqref{irina} to
the various terms in 
\eqref{id-br12} we thus find, for every $\delta\in(0,1]$, 
some constant, $C\equiv C(\delta,\rho,d_{-1},d_1)\in(0,\infty)$, such that
\begin{equation*}
\SPn{\vp^+}{(\DA+V+\delta\Hf)\,\vp^+}\grg c_{a,\rho}
\SPn{\vp^+}{\DO\,\PO\,\vp^+}
-C\,\|\vp^+\|^2\,.
\end{equation*}
Using \eqref{clelia100} once more to replace $\DO\,\PO$ by
$|\DO|$ on the right hand side, 
we arrive at the first asserted estimate.
According to the remarks made below \eqref{def-Bel} 
the first estimate applies in particular to the
Coulomb potential, as long as $\gamma\in[0,\gcnp)$.
{\qed}

\smallskip

\noindent
From the previous theorem and our commutator estimates
one can also infer the semi-boundedness of a no-pair
operator for $N\in\NN$ electrons and $K\in\NN$ static
nuclei, analogously to Corollary~\ref{hoe-prop-sb-srPF}, as long as all
Coulomb coupling constants $\gamma_1,\ldots,\gamma_K$ are
less than $\gcnp$; see Proposition~A.2 of \cite{Matte2009}.

Since we are addressing the question of finding distinguished
self-adjoint realizations of $H_\gamma^\sharp$ it is also natural
to state the following theorem whose proof can be found in 
Corollary~3.4 of~\cite{KMS2009b}.

\begin{theorem}\label{Cor-Friedr-Extens}
Let $\gamma\in[0,1/2)$ and assume that $\V{G}_\V{x}$
fulfills Hypothesis~\ref{hyp-G}.
Then $\PF{\gamma}$ and $\NPO{\gamma}$ 
are essentially self-adjoint on $\core$ and
$\PA\,\core$, respectively.
\end{theorem}

\noindent
For sufficiently small values of $|e|$ and/or $\UV$, the essential
self-adjointness of $H^{\rm{PF}}_0$ has been shown earlier
in \cite{MiyaoSpohn2009}.

\section{Bounds on the ionization energy}\label{sec-binding}

\noindent
As a first step towards the proof of the existence of ground states we
need to show that binding occurs in the atomic system defined by
$H^{\sharp}_V$ in the sense that
$\inf\spec[H^{\sharp}_V]<\inf\spec[H^{\sharp}_0]$.  This
information will be exploited mathematically when we apply a bound on
the spatial localization of low-lying spectral subspaces of
$H^{\sharp}_V$ from \cite{MatteStockmeyer2009a}.  The
localization estimate in turn enters into the proof of the existence
of ground states at various places, for instance, into the derivation
of the infra-red estimates and into the compactness argument
given in Subsection~\ref{ssec-compact}.  
Theorem \ref{thm-binding-PF} below is the
main result of this section. 
In its statement we abbreviate
($\sharp\in\{\Pf,\np\}$)
$$
E_{V}^\sharp:=\inf\spec[H^{\sharp}_V]\,,
\quad E_\gamma^\sharp:=\inf\spec[H^{\sharp}_\gamma]\,,\;
\gamma\in(0,\gc^\sharp)\,,
\quad
\Sigma^\sharp\,:=\,\inf\spec[H^{\sharp}_0]\,,
$$
where $V$ satisfies the conditions under which
$H^\sharp_V$ has been defined in the previous section.
To simplify the exposition we only consider the physical
choice of the coupling function $\V{G}_\V{x}$ given in
\eqref{Gphys}, as always for arbitrary values
of $e$ and $\UV$. Our proofs work, however, equally well
for other coupling functions, for instance, for
the infra-red cut-off and discretized coupling
functions introduced in Section~\ref{sec-ex-m},
and we obtain uniform bounds on the binding energies
in these cases. If we consider coupling functions
other than \eqref{Gphys} then the unitary transformation
$U$ introduced below has to be changed accordingly;
see \cite{KMS2009a,KMS2009b}.

\begin{theorem}
\label{thm-binding-PF}
(i)
Let
$V\in L^2_{\rm loc}(\RR^3, \RR)$ 
be form bounded with respect to $\sqrt{-\Delta}$ with form
bound less than or equal to one.
(So $V$ fulfills \eqref{hyp-Vnu} with $\nu=1/2$.)
Define the self-adjoint operator
$h_V:=\sqrt{1-\Delta}+V$ by means of a Friedrichs extension
starting from $C_0^\infty(\RR^3)$ and
assume that $\inf\spec[h_V]$ is an eigenvalue.
Then
\begin{equation}\label{giselle0}
\ThPF-E^{\mathrm{PF}}_V\,\grg\,1-\inf\sigma[h_V]\,.
\end{equation}
In particular, for $\gamma\in[0,\gcPF]$,
\begin{equation}\label{giselle0C}
\ThPF-E_\gamma^{\mathrm{PF}}\,\grg\,
1-\inf\sigma[\sqrt{1-\Delta}-\tgV]\,.
\end{equation}
(ii)
Let $V\klg0$ be relatively form bounded with respect to
$\sqrt{-\Delta}$ and assume that $V$ satisfies \eqref{irina}
with $a\in(0,1)$.
Additionally, assume there exist $r\grg1$, $c>0$, and $\theta\in(0,2)$
such that
$$
V(\V{x})\klg-c\,|\V{x}|^{\theta-2},\qquad|\V{x}|\grg r\,.
$$ 
Then
\begin{equation}\label{giselle}
\Thnp-E_V^{\mathrm{np}}>0\,,
\end{equation}
and in particular, for $\gamma\in(0,\gcnp)$,
$$
\Thnp-E_\gamma^{\mathrm{np}}>0\,.
$$
\end{theorem}

\begin{remark}{\em\label{rem-HiroshimaSasaki}
(i) The bound \eqref{giselle0} has been obtained first 
  in \cite{HiroshimaSasaki2010} (under the assumption  that 
  $\PF{V}$ be essentially self-adjoint which, in the case
  $V=-\tgV$, is true, at least for all $\gamma<1/2$). 
  The result of \cite{HiroshimaSasaki2010} improved a lower bound
  on the binding energy in an earlier preprint version of \cite{KMS2009a}.
  The latter was given in terms of the {\em non-relativistic}
  ground state energy of an electronic Schr\"odinger operator.

\smallskip

\noindent(ii)
In a forthcoming work of the first two authors \cite{KM2012b} is it shown that
the inequalities \eqref{giselle0} and \eqref{giselle0C} are actually
{\em strict}, for all $e,\Lambda>0$.
Moreover, there is a certain class of short-range potentials $V$
such that
$\ThPF-E^\Pf_V>0$ and in particular --
according to the present article -- $E_V^\Pf$ is an eigenvalue of $\PF{V}$
although $\inf\spec[h_V]=1$ and $1$ is not an
eigenvalue of $h_V$. This effect is called {\em enhanced binding}
due to the quantized radiation field and we are again able to prove
its occurrence, for arbitrary large values of $e$ and $\Lambda$.
There are numerous results on enhanced binding in non-relativistic
QED;
up to now complete proofs were, however, available only for small $e$.
The proofs in \cite{KM2012b} extend the ideas and methods underlying
the proof of Theorem~\ref{thm-binding-PF} given below.}
\end{remark}

Because of lack of space we shall only describe the proof 
of Theorem~\ref{thm-binding-PF} for the
semi-relativistic Pauli-Fierz operator 
\cite{KMS2009a} in detail. The proof of \eqref{giselle} follows similar lines
and can be found in \cite{KMS2009b}; see also Remark~\eqref{Rmk-binding-V}
below.

Our proof of \eqref{giselle0} and \eqref{giselle}
is based on a direct fiber decomposition of $\HR$
with respect to fixed values of the total
momentum $\pe\otimes\id+\id\otimes\pf$, where $\pe:=-i\nabla_\V{x}$ and
\begin{equation}
  \label{eq:1} 
\pf\, := \,d \Gamma(\V{k})\,:=\,\big(d\Gamma(k^{(1)})\,,\,
d\Gamma(k^{(2)})\,,\,d\Gamma(k^{(3)})\big)
\end{equation}
is the photon momentum operator.
In fact, 
a conjugation of the Dirac operator with
the unitary operator $e^{i\pf\cdot\V{x}}$
-- which is simply a multiplication
with the phase 
$(\RR^3)^n\ni(\V{k}_1,\ldots,\V{k}_n)\mapsto e^{i(\V{k}_1+\dots+\V{k}_n)\cdot\V{x}}$
in each Fock space sector $\Fock^{(n)}[\HP]$ -- 
yields
$$
e^{i\pf\cdot\V{x}}\,\DA\,e^{-i\pf\cdot\V{x}}
\,=\,\valpha\cdot(\V{p}-\pf+\V{A}(\V{0}))+\beta\,,
$$
and a further conjugation with the Fourier transform, 
$\fourier:L^2(\RR^3_\V{x})\to L^2(\RR^3_{\V{P}})$,
turns the latter expressions into
\begin{equation}\label{tf-Dirac}
(\fourier\otimes\id)\,e^{i\pf\cdot\V{x}}\,\DA\,
e^{-i\pf\cdot\V{x}}\,(\fourier^{-1}\otimes\id)
\,=\,\int_{\RR^3}^\oplus \wh{D}(\V{P})\,d^3\V{P}\,.
\end{equation}
Here the operators
$$
\wh{D}(\V{P})\,:=\,\valpha\cdot(\V{P}-\pf+\V{A}(\V{0}))+\beta\,,
\qquad \V{P}\in\RR^3\,,
$$
acting in $\CC^4\otimes\Fock[\HP]$,
are fiber Hamiltonians of
the transformed Dirac operator in \eqref{tf-Dirac}
with respect to the isomorphism
\begin{equation}
  \label{eq:2} 
\HR\,=\,L^2(\RR^3_\V{P},\CC^4)\otimes\Fock[\HP]\,\cong
\int_{\RR^3}^\oplus\CC^4\otimes\Fock[\HP]\,d^3\V{P} \,.
\end{equation}
(In particular, the transformed Dirac operator in \eqref{tf-Dirac}
again acts in $\HR$,
where the variable in the first tensor factor, $\V{P}$,
is now interpreted as the total momentum of the
combined electron-photon system.)
Accordingly, we have the direct integral representation
(compare, e.g., Theorem~XIII.85 in \cite{ReedSimonIV})
\begin{equation}
  \label{eq:3} 
(\fourier\otimes\id)\,e^{i\pf\cdot\V{x}}\,\PF{0}\,
e^{-i\pf\cdot\V{x}}\,(\fourier^{-1}\otimes\id)
\,=\,
\int_{\RR^3}^\oplus\PF{0}(\V{P})
\,d^3\V{P} \,,
\end{equation}
where
$$
\PF{0}(\V{P})\,:=\,|\wh{D}(\V{P})|+\Hf\,.
$$
Let $\ve>0$. Then
we know that the Lebesgue measure
of the set of all $\V{P}\in\RR^3$ satisfying
$\spec[\PF{0}(\V{P})]\cap(\ThPF-\ve,\ThPF+\ve)\not=\varnothing$
is strictly positive. In particular, 
we find some $\V{P}_\star\in\RR^3$ and some
normalized 
$\vp_\star\in\form(\PF{0}(\V{P}_\star))$
such that
\begin{align}
  \label{eq:6}
    &\SPb{\vp_{\star}}{\PF{0}(\V{P}_\star) \, 
    \vp_{\star}}_{\CC^4\otimes\Fock[\HP]} 
    \,<\, \ThPF + \ve\,.
\end{align}
We define the unitary transformation
\begin{equation}
\label{def-U-binding} 
U\, : =\, e^{i ( \pf - \pp_{\star}) \cdot \mathbf{x}}
\end{equation}
and observe as above that
\begin{equation*}
  U\,\DA\,U^*\,=\,\wh{D}_{\V{p}}(\V{P}_\star)\,:=\,
\valpha\cdot(\pe+\pp_\star-\pf+\V{A}(\V{0}))+\beta\,.
\end{equation*}
It is sufficient to prove the bound \eqref{giselle0}
for the unitarily equivalent operator
\begin{equation}
  \label{eq:9} U H^\Pf_V U^{\ast}\,=\,|\wh{D}_{\V{p}}(\V{P}_\star)|
+V+\Hf\,.
\end{equation}
{\proof}({\it Theorem \ref{thm-binding-PF}: The semi-relativistic Pauli-Fierz case.})
 Let $\ve > 0$ and $\pp_{\star}$ be as in the paragraph
preceding the statement. We abbreviate $\bts : =
  \pp_{\star} - \pf + \V{A}(\V{0})$ and, for $\eta \grg 0$,
\begin{align*}  
R_1({\eta}):=
\big(\pe^2+({\valpha}{\cdot}{\bts})^2+{\eta}+1\big)^{-1},\quad
R_2({\eta}):=\big(({\valpha}{\cdot}(\pe+\bts))^2+{\eta}+1\big)^{-1}.
\end{align*}  
Since the anti-commutator of $\valpha\cdot\pe$ and 
$\valpha\cdot\bts$ is equal to $2\,\pe\cdot\bts$
it holds 
$(\valpha\cdot(\pe+\bts))^2=(\valpha\cdot\bts)^2+2\,\pe\cdot\bts+\pe^2$.
We deduce that, for any $\varphi \in \core$, 
\begin{align}
    -R_2({\eta}){\vp} \nonumber
\,&=\,
-R_2({\eta})\,[\,{\pe}^2+(\valpha\cdot\bts)^2
+1+{\eta}\,]\,R_1({\eta}){\vp}
\\ \label{eq:10} 
&=\,
R_2({\eta})\,[2\,\pe\cdot\bts]\,R_1({\eta}){\vp}-R_1({\eta}){\vp}\,.
\end{align}
We use the following formula, for a self-adjoint operator $T > 0$,
\begin{align}\label{john1}  
{\sqrt{T}}\,{\vp}\,=\,
\int_0^{{\infty}}\Big(1-{\frac{{\eta}}{T+{\eta}}}\Big)
\,{\vp}\,{\frac{d{\eta}}{{\pi}{\sqrt{{\eta}}}}}\:,
\qquad {\vp}\,{\in}\,{\dom}(T)\,,
\end{align}
and the resolvent identity \eqref{eq:10}
to obtain, for any $\vp \in \core$,
\begin{align}
\SPb{\vp&}{\big(\sqrt{( \valpha \cdot(\pe+\nonumber
      \bts))^2 + 1}\,-\, 
      \sqrt{{\pe}^2+( \valpha \cdot \bts)^2 + 1} \,\big)\, \vp}
      \\
      &= \,\nonumber
      \int_0^{\infty} \SPb{\vp}{\big( R_1 (\eta) - R_2 (\eta) \big)
      \, \vp} \sqrt{\eta} \: \frac{d \eta}{\pi}\\
      &= \,\nonumber
      \int_0^{\infty} \SPb{R_2 (\eta)\,\vp}{[2\,\pe\cdot\bts] 
      \,R_1 (\eta)
      \,\vp} \sqrt{\eta} \: \frac{d \eta}{\pi}\\
      &= \,\nonumber
      \int_0^{\infty} \SPb{\vp}{R_1(\eta)\,[2\,\pe\cdot\bts]
      \,R_1 (\eta)
      \, \vp} \sqrt{\eta} \: \frac{d \eta}{\pi}
\\
&\quad\nonumber
- \int_0^{\infty} \SPb{\vp}{
        R_1 (\eta)\, [2\,\pe\cdot\bts]
        \,R_2 (\eta)\,[2\,\pe\cdot\bts]
        \,R_1 (\eta) \, \vp} \sqrt{\eta} \: \frac{d \eta}{\pi}
\\
&\klg\,\label{eq:7000}
      \int_0^{\infty} \SPb{\vp}{R_1 (\eta) \left[2\,\pe\cdot\bts 
      \right] R_1 (\eta)
      \, \vp} \sqrt{\eta} \, \frac{d \eta}{\pi}
      \,.
\end{align}
In the last step we used the positivity of $R_2(\eta)$.
  We consider now 
$\vp : = \vp_1 \otimes \vp_2$ where 
$\vp_1 \in C_0^{\infty}( \RR^3, \RR)$ 
and $\vp_2 \in \CC^4 \otimes \sC_0$ 
with $\| \vp_j \|= 1$, $j=1,2$. 
Writing 
$\Phi_2(\vxi,\eta):=(\vxi^2+(\valpha\cdot\bts)^2+1+\eta)^{-1}
\,\vp_2=\Phi_2(-\vxi,\eta)$
we find that
\begin{align}\nonumber
\SPb{\vp}{&R_1 (\eta) \, \pe\cdot\bts\, R_1 (\eta)\, \vp}  
 \\
& 
= \,\label{eq:14}
   \int_{\RR^3}\vxi\cdot\SPb{\Phi_2(\vxi,\eta)}{
   \bts\,\Phi_2(\vxi,\eta)}
\,|\wh{\vp}_1(\vxi)|^2d^3\vxi\,=\,0\,,
\end{align}
  due to the fact that $\vp_1$ is real and, hence,
  $|\wh{\vp}_1(\vxi)|=|\wh{\vp}_1(-\vxi)|$. 
Furthermore, 
\begin{align}\label{eq:15a}
&|\wh{D}_{\V{p}}(\V{P}_\star)|=\sqrt{( \valpha \cdot(\pe+\bts))^2 + 1}\,,
\quad|\wh{D}(\V{P}_\star)|=\sqrt{( \valpha \cdot\bts)^2 + 1}\,,
\\\label{eq:15b}
&\sqrt{{\pe}^2+( \valpha \cdot \bts)^2 + 1}
\klg\sqrt{( \valpha \cdot\bts)^2 + 1}+\sqrt{\pe^2+1}-1\,,
\end{align}
where we used $[\pe^2,( \valpha \cdot\bts)^2]=0$ in the second line.
Combining \eqref{eq:7000}--\eqref{eq:15b}
we arrive at
  \begin{align*}
    \SPb{\vp}{&U H^\Pf_V U^{\ast} \, \vp}
\klg \, \SPb{\vp_2}{\PF{0} ( \pp_{\star})\, \vp_2}+ 
    \SPb{\vp_1}{h_V \, \vp_1}-1
    \, .
  \end{align*}
  By a limiting argument the previous inequality extends to any 
  real-valued $\vp_1\in \form(h_V)$.
  We choose $\vp_1$ to be
he normalized, strictly positive 
  eigenfunction of 
  $h_V$
  corresponding to the eigenvalue at the bottom of its spectrum
  and $\vp_2 =\vp_{\star}$. 
  By the choice of $\vp_\star$ in \eqref{eq:6}, where
  $\ve>0$ is arbitrary, this proves the assertion.
{\qed}

\begin{remark}{\em\label{Rmk-binding-V}
  As already mentioned the proof of Theorem~\ref{thm-binding-PF} for the
  no-pair operator employs ideas similar to those described above. 
  However, due to the
  more complex structure of the no-pair Hamiltonian the resulting 
  bound \eqref{giselle} is not as
  satisfactory as the one for $\PF{V}$. Again, we
  have the representation $H^{\mathrm{np}}_0\cong
  \int^\oplus_{\RR^3}H^{\mathrm{np}}_0(\V{P})\,d^3\V{P}$ with
$$
H^{\mathrm{np}}_0(\V{P})=
\wh{P}(\V{P})\,\big(\wh{D}(\V{P})+\Hf\big)\,\wh{P}(\V{P})\,,
\qquad \wh{P}(\V{P}):=\id_{[0,\infty)}(\wh{D}(\V{P}))\,.
$$
Given $\ve>0$, we again find $\V{P}_\star\in\RR^3$ 
and some normalized
$\vp_\star=\wh{P}(\V{P})\,\vp_\star\in\form(H^{\mathrm{np}}(\V{P}_\star))$ 
such that
  $\SPn{\vp_{\star}}{H^{\mathrm{np}}_0(\V{P}_\star)\,
    \vp_{\star}}_{\CC^4\otimes\Fock[\HP]}<\Thnp + \ve$. 
Conjugating $H^{\mathrm{np}}_\gamma$ with $U$ defined
as in \eqref{def-U-binding} we obtain
\begin{equation}\label{giselle1}
  U \,H^{\mathrm{np}}_\gamma\,U^\ast\,=\,
\wh{P}_\V{p}(\V{P}_\star)\,\big(\wh{D}_{\V{p}}(\V{P}_\star)
+V\Hf\big)\,\wh{P}_\V{p}(\V{P}_\star)\,,
\end{equation}
where $\wh{P}_\V{p}(\V{P}_\star):=\id_{[0,\infty)}(\wh{D}_{\V{p}}(\V{P}_\star))$.
As a test function we now choose 
$\Phi_R:=\wh{P}_\V{p}(\V{P}_\star)(\chi_R\otimes\vp_\star)$,
where $\chi_R$ is some normalized smooth and non-negative
function supported in $\{R\klg|\V{x}|\klg2R\}$ with $R\grg1$.
It turns out that $\|\Phi_R\|=1+\bigO(1/R^2)$.
We then exploit the fact that the negative contribution
of the potential in \eqref{giselle1} to the expectation
value of $\Phi_R$
decays as $1/R^{2-\vt}$ whereas all other terms yield a contribution 
$\Sigma^\np+\ve+\bigO(1/R^2)$,
as $R$ tends to infinity. 
To show all this it is convenient to work on a non-projected
Hilbert space by adding a suitable ``positronic'' no-pair operator
to $\NPO{\gamma}$
(similarly as in \eqref{nurleen2011} below).
For then one can again make use of the bounds
\eqref{eq:7000}--\eqref{eq:15b} derived in the previous proof;
see Section~V of \cite{KMS2009b}.
(It is entirely obvious that the Coulomb potential can be replaced
by a more general one in Section~V of \cite{KMS2009b}.)}
\end{remark}

\section{Exponential localization}\label{sec-exp}

\noindent
Our next aim is to discuss the exponential 
localization with respect to the electron coordinates
of low-lying spectral subspaces
of the no-pair and semi-relativistic Pauli-Fierz operators.
Since the multiplication with some exponential weight
function, $e^F$, acting on the electron coordinates does
not map the projected Hilbert space $\HR^+_{\V{A}}$ into
itself it is convenient to extend the no-pair operator
to some continuously invertible operator
on the whole Hilbert space $\HR$ in the discussion below.
Therefore, we set
\begin{equation}\label{nurleen2011}
\wh{H}_V^\np\,:=\,\NPO{V
}+H_{0}^{\np,-}\,,\qquad
H_{0}^{\np,-}:=\PAm\,(|\DA|+\Hf)\,\PAm\,.
\end{equation}
In Lemma~\ref{le-H+H-} below we show that 
$\Thnp=\inf\spec[H_{0}^{\np,-}]$
by constructing some
anti-linear map $\tau:\HR\to\HR$ with
$\NPO{0}\,\tau=\tau\,H_{0}^{\np,-}$.
Therefore,
\begin{equation}\label{luise1}
\wh{H}_0^\np\,\grg\,\Thnp\,.
\end{equation}
To unify the notation we further set
$\whPF{V}:=\PF{V}$
and write $\wh{H}_{V}^\sharp$, ${\Sigma}^\sharp$, etc., 
when we treat both 
$\PF{V}$ and $\wh{H}_V^\np$ at the same time.
In the whole section
we assume that $\V{G}_\V{x}$ fulfills Hypothesis~\ref{hyp-G}.

\begin{theorem}\label{el}
(i) Let $V\in L^2_\loc(\RR^3,\RR)$ be relatively form bounded
with respect to $\sqrt{-\Delta}$ with relative form bound
less than or equal to one. Moreover, assume that
\begin{equation}\label{john2}
\exists\,r\grg1\::\quad
\sup_{|\V{x}|\grg r/4}|V(\V{x})|<\infty\quad
\textrm{and}\quad V(\V{x})\,\xrightarrow{\;|\V{x}|\to\infty\;}\,0\,.
\end{equation}
Define
$$
\rho(a):=1-(1-a^2)^{1/2},\quad a\in[0,1)\,,
$$
and let $I\subset (-\infty,{\Sigma}^\Pf)$ 
be some compact interval. Then there exists  $k\in(0,\infty)$,
such that, for all $a\in(0,1)$ satisfying
$\ve:=\Sigma^\Pf-\sup I-\rho(a)\in(0,1]$, we have
\begin{equation}\label{for-exp-decay}
\big\|\,e^{a|{\bf x}|}\,\id_I({H}_{V}^\Pf)\,\big\|\,\klg\,
k\,(\Sigma^\Pf-E^\Pf_V)\,e^{k/\ve}.
\end{equation}
(ii) Assume that $V\in L^2_\loc(\RR^3,\RR)$ satisfies
$H^1(\RR^3)\subset\dom(V)$ (which implies $|V|\klg\const\,|\DO|$)
as well as \eqref{irina} with $a<1$ and \eqref{john2}.
Let $I\subset (-\infty,{\Sigma}^\np)$ 
be some compact interval.
Then we find some $a'>0$ such that 
$\Ran(\id_I(\wh{H}_V^\np))\subset\dom(e^{a'|\V{x}|})$.
In particular, 
\begin{equation}\label{for-exp-decay2}
\big\|\,e^{a'|{\bf x}|}\,\id_I(\NPO{V})\,\big\|_{\LO(\HR^+_{\V{A}},\HR)}
\,\klg\,\const\,.
\end{equation}
If $\V{G}_\V{x}$ is modified, then we get uniform lower bounds on
$a'$ and uniform upper bounds on the constant in \eqref{for-exp-decay2},
provided that we have uniform upper bounds on
$d_{-1},d_1,\Thnp$ and uniform lower bounds on
$\Thnp-\sup I$.
\end{theorem}  

{\proof}
The proof is given in the succeeding three
subsections. 
{\qed}

\noindent
Note that the potential $V$ is {\em not} assumed to be a {\em small}
form perturbation of $\sqrt{-\Delta}$ in Part~(i) of the previous theorem.
In particular, the assumptions of (i) cover the Coulomb potential
$-\gamma/|\V{x}|$ with coupling constants $\gamma\in[0,\gcPF]$
including the critical one. This improves on
\cite{MatteStockmeyer2009a}
where Coulomb potentials have been treated, for subcritical $\gamma$.
By a modification of the arguments of this section
it is actually also possible to prove exponential localization
for the no-pair operator with Coulomb potential in the critical case $\gamma=\gcnp$,
which is not covered by Part~(ii) of the above theorem;
see \cite{KM2011}.

The bound on the decay rate $a$ of Part~(i) has been found first
in \cite{KM2011} (where only the Coulomb potential is treated explicitly). It reduces to the typical
relativistic decay rate known for eigenvectors of electronic Dirac or
square-root operators when $\V{G}_\V{x}$ is set equal to zero.


\subsection{A general strategy to prove
the localization of spectral subspaces} 

\noindent
The general strategy of the proof of Theorem \ref{el} is essentially due
to \cite{BFS1998b}. We shall present a variant of the argument
used in \cite{BFS1998b} in Lemma~\ref{le-ed-BFS} below.
In order to apply this lemma to $H_V^\sharp$ we shall
also benefit from some useful observations made in \cite{Griesemer2004}.
The main advantage of Lemma~\ref{le-ed-BFS} and its earlier
variants is that it allows to study the localization of
{\em spectral subspaces} without any a priori knowledge on
the spectrum. Its proof does not exploit eigenvalue equations
as it is the case in Agmon type arguments nor does
it assume discreteness of the spectrum or the presence of spectral
gaps. This is important for us since the spectra
of both the no-pair and the semi-relativistic Pauli-Fierz operators
will be continuous up to their minima.  
Roughly speaking the proof of Lemma~\ref{le-ed-BFS} rests on a combination 
of the following:
\begin{enumerate}
\item[$\bullet$] The representation \eqref{li1} for the spectral projection
$\id_I(\wh{H}^\sharp_\gamma)$. Here a {\em comparison operator}, $Y$, 
enters into the analysis whose resolvent stays bounded after conjugation
with suitable exponential weights, for all relevant spectral parameters.
\eqref{li1} is valid since its also satisfies $\id_I(Y)=0$.
\item[$\bullet$] The {\em Helffer-Sj\"ostrand formula} (re-derived below
for the convenience of the reader) which is used to represent smoothed
versions of $\id_I(\wh{H}^\sharp_\gamma)$ and $\id_I(Y)$ as integrals
over resolvents.
\item[$\bullet$] The second resolvent identity; in fact, $Y$ 
will be chosen such that $\wh{H}^\sharp_\gamma-Y$ is well-localized
and is hence able to control exponential weights.
\end{enumerate}
In the somewhat technical parts of this section following
after Lemma~\ref{le-ed-BFS} we shall verify the applicability
of Lemma~\ref{le-ed-BFS} to our models by defining and analyzing
suitable comparison operators $Y$.

Let us now introduce some prerequisites for the proof of
Lemma~\ref{le-ed-BFS}.
In order to find a representation of the spectral projection
which is accessible for an analysis we smooth out the
projection and employ the Helffer-Sj\"ostrand formula.
More precisely, let $I\subset(-\infty,{\Sigma}^\sharp)$
be some compact interval.  Then we pick some slightly
larger compact interval $J\subset(-\infty,{\Sigma}^\sharp)$,
$\mr{J}\supset I$, and some $\chi\in C_0^\infty(\RR,[0,1])$
such that $\chi\equiv1$ on $I$ and $\chi\equiv0$ outside
$J$. We pick another cut-off function
$\rho\in C_0^\infty(\RR,[0,1])$ such that $\rho=1$
in a neighborhood of $0$ and $\rho(y)=0$, for $|y|\grg1/2$, 
and extend $\chi$ to
a compactly supported smooth function of the
complex plane setting 
$$
\wt{\chi}(x+iy)\,:=\,\chi(x)+\chi'(x)\,iy\,\rho(y)\,,\qquad
x,y\in\RR\,;
$$
compare, e.g., \cite{Davies,DimassiSjoestrand}. Then we have
$$
2\,\partial_{\ol{z}}\wt{\chi}(z)\,:=\,(\partial_x+i\partial_y)\chi(z)
\,=\,\chi'(x)\,(1-\rho(y)-y\,\rho'(y))
+\chi''(x)\,iy\,\rho(y)\,,
$$
for every $z=x+iy\in\CC$, and the choice of $\rho$ implies
\begin{equation}\label{ben1}
|\partial_{\ol{z}}\wt{\chi}(z)|\,\klg\,C_\chi|\Im z|\,,\qquad z\in\CC\,,
\end{equation}
for some $C_\chi\in(0,\infty)$.
Moreover, the following Helffer-Sj\"ostrand
formula is valid, for every self-adjoint
operator, $X$, on some Hilbert space,
\begin{equation}\label{ben2}
\chi(X)\,=\int_\CC(X-z)^{-1}\,d\mu(z)\,,
\qquad d\mu(z):=-\frac{1}{\pi}\partial_{\ol{z}}\wt{\chi}(z)\,dxdy\,.
\end{equation}
If $X$ were
a complex number \eqref{ben2} were just a special case of Pompeiu's formula
for some path encircling the support of $\chi$.
$X$ can, however, be inserted in that formula by means of
the spectral calculus; see, e.g., \cite{DimassiSjoestrand}.

To facilitate the discussion of operator domains we 
replace $a|\V{x}|$ in \eqref{for-exp-decay} by some
$F:\RR^3_{\V{x}}\to\RR$
satisfying
\begin{equation}\label{hyp-F<>}
F\in C^\infty\cap L^\infty(\RR^3_{\V{x}},\RR)\,,
\quad F=a\,r\;\textrm{on}\;\ball{r/2}{0}\,,\quad
F\grg0\,,\quad|\nabla F|\klg a\,,
\end{equation}
where $r\grg1$ is the parameter appearing in \eqref{john2}.

Finally, we recall our notation $\sharp\in\{\np,\Pf\}$.

\begin{lemma}\label{le-ed-BFS}
Let $I$, $J$, $\wt{\chi}$, and $C_\chi$
be as described above and assume that  
$Y$ is a self-adjoint operator
in $\HR$ with $\dom(Y)=\dom(\wh{H}^\sharp_V)$ and $Y>\sup J$.
Furthermore, let $a>0$ and
assume there exist $C,C'\in(0,\infty)$ such that, for all $F$ 
satisfying \eqref{hyp-F<>},
\begin{equation}\label{hyp-XY}
\big\|\,e^F\,(\wh{H}^\sharp_V-Y)\,\big\|\,\klg\,C\,,\quad
\sup\limits_{z\in J+i\RR}
\big\|\,e^F\,(Y-z)^{-1}\,e^{-F}\,\big\|\,\klg\,C'\,.
\end{equation}
Then $\Ran(\id_I(\wh{H}^\sharp_V))\subset \dom(e^{a|\V{x}|})$ and
\begin{equation}\label{ben3}
\big\|\,e^{a|\V{x}|}\,\id_I(\wh{H}^\sharp_V)\,\big\|\,\klg\,
C\,C'\,C_\chi\,\cL(\supp(\chi'))/\pi\,,
\end{equation}
where $\cL$ denotes the Lebesgue measure on $\RR$. 
\end{lemma}

{\proof}
Since $\id_I=\chi\,\id_I$ and $J\subset\vr(Y)$ we have
\begin{equation}\label{li1}
\id_I(\wh{H}^\sharp_V)
=(\chi(\wh{H}^\sharp_V)-\chi(Y))\,\id_I(\wh{H}^\sharp_V)\,.
\end{equation}
Applying the Helffer-Sj\"ostrand formula \eqref{ben2} 
and the second resolvent identity we infer that
\begin{equation}\label{li2}
e^F\,\id_I(\wh{H}^\sharp_V)\,=
\int_\CC \{e^F\,(Y-z)^{-1}\,e^{-F}\}\{e^F(Y-\wh{H}^\sharp_V)\}
(\wh{H}^\sharp_V-z)^{-1}\,d\mu(z)\,.
\end{equation}
Estimating the norm of these expressions using \eqref{ben1} 
and $\|(\wh{H}^\sharp_V-z)^{-1}\|\klg1/|\Im z|$ we
obtain \eqref{ben3} with $a|\V{x}|$ replaced by $F$.
Then \eqref{ben3} follows by choosing a suitable sequence
of functions $F_n$ satisfying \eqref{hyp-F<>} and
converging monotonically on $\{|\V{x}|\grg2r\}$ to $a|\V{x}|$
and applying the monotone convergence theorem to
$e^{F_n}\,\id_I(\wh{H}^\sharp_V)\,\psi$, for every $\psi\in\HR$.
{\qed}


\subsection{Choice of the comparison operator $Y$}

\noindent
In the next step
we thus have
to find a suitable operator $Y$ fulfilling the
conditions of Lemma~\ref{le-ed-BFS}. 
The hardest problem is to verify the second bound
in \eqref{hyp-XY} and this is actually the main new
mathematical challenge in the study of the exponential
localization in our non-local models.
We defer the discussion of the second bound
in \eqref{hyp-XY} to the next subsection.

In order to construct an operator $Y$ whose spectrum
differs only slightly from the spectrum of the
free operator 
$\wh{H}_0^\sharp$, so that $Y>\sup J$, and which is defined on the
same domain as $\wh{H}_V^\sharp$ we simply add
some bounded term to $\wh{H}_V^\sharp$ which compensates for the 
singularities and wells of the electrostatic potential 
and thus pushes the spectrum up to the
ionization threshold. A similar choice of a comparison operator
has been employed in the non-relativistic setting in \cite{Griesemer2004}.
More precisely,
we first introduce a scaled partition of unity. 
That is, we pick $\chi_{0,R},\chi_{1,R}\in C^\infty(\RR^3,[0,1])$,
$R\grg r$, such that 
$\chi_{0,R}\equiv1$ on $\ball{R}{0}$,
$\chi_{0,R}\equiv0$ on $\RR^3\setminus\ball{2R}{0}$,
$\chi^2_{0,R}+\chi^2_{1,R}=1$, and $\|\nabla\chi_{k,R}\|_\infty\klg c/R$,
$k=0,1$, for some $R$-independent constant $c\in(0,\infty)$.
Then we define $Y$ as follows:

\smallskip

{\it The semi-relativistic Pauli-Fierz operator.} 
We define
\begin{equation}\label{agua}
\YPF{V}:=\PF{V}\,
+(\ThPF-E_{V}^{\mathrm{PF}})\,\chi_{0,R}^2\,,
\end{equation}
for some $R\grg1$ which shall be fixed sufficiently large later on.
Of course, $\PF{V}$ and $\YPF{V}$ have the same
domain and the first bound in \eqref{hyp-XY} which provides the
control on the exponential weights holds trivially,
$$
\|e^F(\PF{V}-\YPF{V})\|\,\klg\,
(\ThPF-E_{V}^{\mathrm{PF}})\,\|e^{F}\chi_{0,R}^2\|_\infty\,\klg\,
(\ThPF-E_{V}^{\mathrm{PF}})\,e^{ar+2aR},
$$
for every $F$ satisfying \eqref{hyp-F<>}.
We shall sketch the proof of the condition $Y>\sup J$
which follows from the next lemma.

\begin{lemma}\label{le-philip}
$\YPF{V}\grg\ThPF-o(R^0)$, $R\to\infty$, in the sense of quadratic forms
on $\sD$, where the little-$o$-symbol depends only on $V$ and $\chi_{0,1}$.
\end{lemma}

{\proof}
We employ the localization formula \eqref{hoe-IMS-DA1} with $K=1$,
the error estimate \eqref{hoe-IMS-Da3} , and $\PF{V}\grg E_V^{\mathrm{PF}}$, 
$\PF{0}\grg\ThPF$, to get
\begin{align*}
\YPF{V}\,&\grg\,\chi_{0,R}\,\PF{V}\,\chi_{0,R}
+\chi_{1,R}\,\PF{0}\,\chi_{1,R}
\\
&\quad+(\ThPF-E_V^{\mathrm{PF}})\,\chi_{0,R}^2
+\,\chi_{1,R}^2\,V-\bigO(1/R^2)\,\grg\,\ThPF-o(R^0)\,. 
\end{align*}
We also used that $\sup_{|\V{x}|\klg R}|V(\V{x})|=o(R^0)$ and
$\chi_{0,R}^2+\chi_{1,R}^2=1$.
{\qed}

{\it The no-pair operator.}  
Since $|\DA|=\PA\,\DA+\PAm\,|\DA|$ we have
$$
\wh{H}_V^\np\,=\,|\DA|+\PA\,V\,\PA+\Hf^{\mathrm{diag}},
\quad\Hf^{\mathrm{diag}}:=
\PA\,\Hf\,\PA+\PAm\,\Hf\,\PAm,
$$
on $\sD$. We write
$\wh{H}_{V,R}^\np:=\wh{H}_V^\np-(1/R)\,\Hf^{\mathrm{diag}}$,
for some $R>1$, so that
$E_{V,R}^\np:=\inf\spec[\wh{H}_{V,R}^\np\,\PA]>-\infty$
by \eqref{eqqq22V}, and
define
\begin{equation}\label{agua2}
\Ynp{V}=\,\wh{H}_V^\np+
(\Thnp-E_{V,R}^\np)\,\chi_{0,R}\,\PA\,
\chi_{0,R}\,.
\end{equation}
Again it is clear that $\wh{H}_V^\np$ and $\Ynp{V}$
are self-adjoint on the same domain.
Also the first bound in \eqref{hyp-XY} again follows trivially,
$$
\|e^F(\wh{H}_V^\np-\Ynp{V})\|\,\klg\,
(\Thnp-E_{V,R}^\np)\,e^{ar+2aR},
$$
for every $F$ satisfying \eqref{hyp-F<>}.
Besides the second bound in \eqref{hyp-XY} it remains
to derive the following lemma.

\begin{lemma}
$\Ynp{V}\grg\Thnp-\Thnp/R-o(R^0)-k\,d_1^2/R^2$ as quadratic forms
on $\sD$, where $k\in(0,\infty)$ and the Little-$o$-symbol depend only
on $V$ and $\chi_{0,1}$.
\end{lemma}  

{\proof}
Again we
use an IMS type localization formula
to infer that
\begin{align}\nonumber
&\Ynp{V}\grg\,\chi_{0,R}\,\wh{H}_{V,R}^\np\,\chi_{0,R}
+\chi_{1,R}\,\wh{H}_{0,R}^\np\,\chi_{1,R}
+(\Thnp-E_{V,R}^\np)\,\chi_{0,R}\,\PA\,\chi_{0,R}
\\
&+\tfrac{1}{R}\,\Hf^{\mathrm{diag}}\label{febri}
+\chi_{1,R}\PA\,V\,\PA\,\chi_{1,R}
+\frac{1}{2}\sum_{k=0,1}
\big[\chi_{k,R}\,,\,[\chi_{k,R},\wh{H}_{V,R}^\np]\big]\,.
\end{align}
As a consequence of \eqref{clelia2}, \eqref{clelia3}, \eqref{hoe-IMS-Da3},
\eqref{exp-dc-HT}, \eqref{exp-dc-VC},
and $\Hf\klg2\Hf^{\mathrm{diag}}$
the double commutator in the last line is bounded from below
by $-(k/R^2)(\Hf^{\mathrm{diag}}+d_1^2+1)$, where $k\in(0,\infty)$
depends only on $V$ and $\chi_{0,1}$.
To control this error we use the term 
$\tfrac{1}{R}\,\Hf^{\mathrm{diag}}$ in \eqref{febri}.
Furthermore, we
put $\mu_R:=\chi_{0,R/2}$, so that $\chi_{1,R}\,\mu_R=0$.
Then $(1-\mu_R^2)\,V=o(R^0)$ and, by \eqref{clelia3},
\begin{align*}
-\chi_{1,R}\PA\,V\,\PA\,
\chi_{1,R}&\klg-\chi_{1,R}[\PA,\mu_R]\,V\,[\mu_R,\PA]\,
\chi_{1,R}+o(R^0)\,\chi_{1,R}^2
\\
&\klg o(R^0)\,\chi_{1,R}^2
+\bigO(1/R^2)\,(\Hf^{\mathrm{diag}}+d_1^2+1)\,,
\end{align*}
so that the second term in the last line of \eqref{febri}
can again be controlled by the first one.
Using these remarks, 
$\wh{H}_{V,R}^\np\grg E_{V,R}^\np\,\PA+(1-1/R)\,\Thnp\,\PAm$,
and
$\wh{H}_{0,R}^\np\grg(1-1/R)\,\Thnp$ (by \eqref{luise1}), 
we arrive at the assertion.
{\qed}


\subsection{Conjugation of $Y$ with exponential weights}

\noindent
In order to prove Theorem~\ref{el} it only remains to
verify the second bound in \eqref{hyp-XY}.
The following lemma \cite{MatteStockmeyer2009a}
gives a criterion for this condition to hold true.

\begin{lemma}\label{le-MS-Y}
Let $Y$ be a non-negative operator in $\HR$
which 
admits $\core$ as a form core.
Set $b:=\inf\spec(Y)$ and
let $J\subset(-\infty,b)$ be some compact interval.
Let $a\in(0,1)$ and assume that, for all
$F$ satisfying \eqref{hyp-F<>},
we have $e^{\pm F}\,\form(Y)\subset\form(Y)$.
(Notice that $e^{\pm F}$ maps $\core$ into itself.)
Assume further
that there exist constants $c(a),f(a),g(a),h(a)\in[0,\infty)$
such that $c(a)<1/2$ and 
$\delta:=b-\max J-b\,g(a)-h(a)>0$ and, for all
$F$ satisfying \eqref{hyp-F<>} and $\vp\in\core$,
\begin{align}
\big|\SPb{\vp}{(e^{F}\,Y\,e^{-F}-Y)\,\vp}\big|\,&\klg\,\label{eq-Y2a}
c(a)\,\SPn{\vp}{Y\,\vp}+f(a)\,\|\vp\|^2,
\\ \label{eq-Y2}
\Re\SPb{\vp}{e^F\,Y\,e^{-F}\,\vp}
\,&\grg\,(1-g(a))\,\SPb{\vp}{Y\,\vp}-h(a)\,\|\vp\|^2.
\end{align}
Then we have, for all 
$F$ satisfying \eqref{hyp-F<>},
\begin{equation}\label{eq-Y3}
\sup_{z\in J+i\RR}
\big\|\,e^F\,(Y-z)^{-1}\,e^{-F}\,\big\|\,\klg\,
\delta^{-1}.
\end{equation}
\end{lemma}

{\proof} We only sketch the proof and refer to Lemma~5.2 of
\cite{MatteStockmeyer2009a} for the details.
The assumptions $e^{\pm F}\,\form(Y)\subset\form(Y)$ and \eqref{eq-Y2a}
ensure that the closure, $Y_F$, of 
$(e^F\,Y\,e^{-F})\!\!\upharpoonright_\sD$ agrees with the
closed operator $e^F\,Y\,e^{-F}$.
The bound \eqref{eq-Y2} shows that the numerical range
of $Y_F$ is contained in the half space $\{z\in\CC:\,\Re z\grg\sup J+\delta\}$.
Moreover, we can argue that, for $z\in J+i\RR$, the deficiency
of $Y_F-z$ is zero and, hence, the norm of 
$(Y_F-z)^{-1}=e^F(Y-z)e^{-F}$ can be estimated by one over 
the distance of $z$ to the numerical range of $Y_F$.
{\qed}

\smallskip

{\it The semi-relativistic Pauli-Fierz operator.} 
Next, we apply Lemma~\ref{le-MS-Y} to $\PF{V}$.
In order to verify condition \eqref{eq-Y2} with a
good bound on the exponential decay rates we apply
the following technical lemma from \cite{KM2011}:

\begin{lemma}\label{le-retno}
For all $a\in(0,1)$, $F$ satisfying \eqref{hyp-F<>}, and $\vp\in\core$,
\begin{align}\nonumber
\Re\SPb{\vp}{e^F\,|\DA|\,e^{-F}\vp}
&\grg
\SPb{\vp}{(\DA^2-|\nabla F|^2)^{1/2}\,\vp}
\\\label{retno1}
&\grg \SPb{\vp}{(|\DA|-\rho(a))\,\vp}\,.
\end{align}
\end{lemma}

{\proof}
For every $\vp\in\core$, we infer from \eqref{john1} that
\begin{align*}
\SPb{\vp}{\big(e^F\,|\DA|\,e^{-F}-(\DA^2-|\nabla F|^2)^{1/2}\big)\,\vp}
=\int_0^{\infty}J[\vp;\eta]\,\frac{\eta^{1/2}d\eta}{\pi}\,,
\end{align*}
with 
\begin{align*}
J[\vp;\eta]&:=\SPb{\vp}{\big(\sR_F(\eta)-e^F\sR_0(\eta)\,e^{-F}\big)\,\vp}\,,
\\
\sR_G(\eta)&:=(\DA^2-|\nabla G|^2+\eta)^{-1},\quad G\in\{0,F\}\,.
\end{align*}
Now, let $\phi:=e^F(\DA^2+\eta)\,e^{-F}\psi$, for some $\psi\in\core$.
Then
\begin{align*}
\Re\SPb{\phi}{e^F\sR_0(\eta)\,e^{-F}\phi}
&=
\Re\SPb{e^F(\DA^2+\eta)\,e^{-F}\psi}{\psi}
\\
&=\SPb{(\DA^2-|\nabla F|^2+\eta)\,\psi}{\psi}
\\
&\grg(1-a^2+\eta)\,\|\psi\|^2\grg0\,.
\end{align*}
Since $\DA^2$ is essentially self-adjoint on
$\core$ and multiplication with $e^{-F}$
maps $\core$ bijectively onto itself, we know that
$(\DA^2+\eta)\,e^{-F}\core$ is dense in $\HR$.
Since $F$ is bounded we conclude that the previous
estimates hold, for all $\phi$ in some dense domain,
whence $\Re[e^F\sR_0(\eta)\,e^{-F}]\grg0$ as a quadratic form
on $\HR$. Next, we set
$Q:=(\valpha\cdot\nabla F)\,\DA+\DA\,(\valpha\cdot\nabla F)$
and let 
$$
\vp:=(\DA^2-|\nabla F|^2+\eta)\,\psi
=e^{\pm F}(\DA^2+\eta)\,e^{\mp F}\psi\mp iQ\,\psi\,, 
$$
for $\psi\in\core$. Then
\begin{align*}
J[\vp;\eta]&=
i\SPb{e^{-F}\sR_0(\eta)\,e^{F}\vp}{Q\,\psi}
=i\SPn{\psi}{Q\,\psi}
+\SPb{Q\,\psi}{e^F\sR_0(\eta)\,e^{-F}Q\,\psi}\,.
\end{align*}
Here $\DA^2-|\nabla F|^2$ is essentially self-adjoint
on $\core$ and $Q$ is symmetric on the same domain.
Hence, $\Re J[\vp;\eta]\grg0$, for all $\vp$ in a dense set, 
thus for all $\vp\in\HR$, and we arrive at the first inequality in
\eqref{retno1}.
Since the square root is operator monotone, $|\nabla F|\klg a$,
and $|\DA|\grg1$, we further have
$$
(\DA^2-|\nabla F|^2)^{1/2}\grg|\DA|+(\DA^2-a^2)^{1/2}-|\DA|
\grg|\DA|-\rho(a)\,.
$$
{\qed}

In what follows we abbreviate
$$
\cK_F\,:=\,[\PA\,,\,e^F]\,e^{-F},
$$
and recall from \eqref{clelia1} 
that $\||\DA|^{1/2}\,\cK_F\|=\bigO_{a_0}(a)$, 
for all $F$ satisfying \eqref{hyp-F<>} with $0<a\klg a_0<1$.

\begin{lemma}\label{exp-le-O(a)-PF}
For all $0<a\klg a_0<1$ and $F$ satisfying \eqref{hyp-F<>}, 
\begin{equation}\label{exp-eq-O(a)-PF}
\Re\SPb{\vp}{e^F\,\YPF{V}\,e^{-F}\,\vp}
\grg\SPn{\vp}{\YPF{V}\,\vp}-\rho(a)\,\|\vp\|^2,\quad \vp\in\core.
\end{equation}
Moreover, for all $\ve>0$ and $a_0\in(0,1)$, there is some constant, 
$C(a_0,\ve,V)\in(0,\infty)$,
such that, for all $F$ satisfying \eqref{hyp-F<>} with $a\in[0,a_0]$ and $\vp\in\core$,
\begin{equation}\label{exp-iris-PF}
\big|\SPb{\vp}{(e^F\,\YPF{V}\,e^{-F}-\YPF{V})\,\vp}\big|
\,\klg\,
\ve\,\SPb{\vp}{\YPF{V}\,\vp}\,+\,C(a_0,\ve,V)\,\|\vp\|^2\,.
\end{equation}
\end{lemma}

{\proof}
\eqref{exp-eq-O(a)-PF} follows immediately from \eqref{retno1}.
To derive \eqref{exp-iris-PF} we write
$$
e^F\,|\DA|\,e^{-F}-|\DA|\,=\,-2\,\DA\,\cK_F+i\valpha\cdot(\nabla F)\,
e^F\,\SA\,e^{-F}
$$
on $\core$
and make a little observation. Since $F\equiv a\,r$ on $\ball{r/2}{0}$
we find some $\mu\in C^\infty(\RR^3_\V{x},[0,1])$ such that
$\mu=0$ on $\ol{\cB}_{r/4}(0)$ and $\nabla F=\mu\,\nabla F$.
For $\vp,\psi\in\core$, we thus have
(recall \eqref{sgn} and \eqref{thomas})
\begin{align*}
\big|\SPb{\DA\,\vp}{\cK_F\,\psi}\big|
\,&\klg
\int_\RR\big|\SPb{\vp}{\DA\,\RA{iy}\,\mu\,\,i\valpha\cdot
\nabla F\,R_\V{A}^F(iy)\,\psi}\big|
\,\frac{dy}{2\pi}\,.
\end{align*}
Here we can write
$$
\DA\,\RA{iy}\,\mu=\mu\,|\DA|\,\SA\,\RA{iy}+[\DA,\mu]\,\RA{iy}
+\DA\,\RA{iy}\,[\mu,\DA]\,\RA{iy}\,,
$$ 
where $\|[\DA,\mu]\|\klg\|\nabla\mu\|_\infty=\bigO(1)$.
On account of 
$$
\|\,|\DA|^{1/2}\,\RA{iy}\|\klg\bigO(1)\,(1+y^2)^{-1/4}
$$
and $\|\DA\,\RA{iy}\|=\bigO(1)$
it is now straightforward to verify that
\begin{align*}
\big|\SPb{&\DA\,\vp}{\cK_F\,\psi}\big|
\,\klg\,\bigO_{a_0}(a)\,\big\{\big\|\,|\DA|^{1/2}\,\mu\,\vp\,\big\|+\|\vp\|\big\}
\,\|\psi\|\,.
\end{align*}
Some elementary estimates using
$\|\nabla F\|_\infty\klg a$, $\|e^F\,\SA\,e^{-F}\|=\bigO_{a_0}(1)$,
and the previous bound
now show that 
\begin{align}\nonumber
\big|\SPb{\vp}{(e^F\,&|\DA|\,e^{-F}-|\DA|)\,\vp}\big|
\\
&\klg\,\label{vera1}
\ve_1\,\SPb{\mu\,\vp}{|\DA|\,\mu\,\vp}
+(\ve_1^{-1}\,\bigO_{a_0}(a^2)+\bigO_{a_0}(a))\,\|\vp\|^2
\\\nonumber
&\klg\,
\ve_1\,\bigO(1)\,\SPb{\mu\,\vp}{\YPF{\gamma}\,\mu\,\vp}
+\const(a_0,\ve_1)\,\|\vp\|^2,
\end{align}
for every $\ve_1\in(0,1]$. In the second step we used that $\mu\,V$
is bounded because $\mu=0$ on $\ol{\cB}_{r/4}(0)$. Since we may assume that there is some 
$\wt{\mu}\in C_0^\infty(\RR^3,[0,1])$ such that
$\mu^2+\wt{\mu}^2=1$ we can employ an IMS localization formula
as in the proof of Lemma~\ref{le-philip} to show that
$\mu\,\YPF{V}\,\mu\klg\YPF{V}+\bigO(1)$ on $\sD$.
Altogether
this proves \eqref{exp-iris-PF}.
{\qed}

\begin{lemma}\label{exp-le-fdom-PF}
There exist constants, $c_1,c_2\in(0,\infty)$, such that,
for all $a\in(0,1/2]$ and $\pm F$ satisfying \eqref{hyp-F<>},
and $\vp\in\core$,
\begin{equation}\label{vera2}
\SPb{e^{F}\,\vp}{\YPF{V}\,e^F\,\vp}\,\klg\,c_1\,\|e^F\|^2\,
\SPb{\vp}{\YPF{V}\,\vp}+
c_2\,\|e^F\|^2\,\|\vp\|^2.
\end{equation}
In particular, $e^F\,\form(\YPF{V})\subset\form(\YPF{V})$.
\end{lemma}

{\proof}
We pick a smooth partition of unity with respect to
the electron coordinates, $\mu_0^2+\mu_1^2=1$, where
$\supp(\mu_0)\subset\ball{r/2}{0}$ and $\mu_0=1$ on $\ol{\cB}_{r/4}(0)$.
Then $\SPb{e^{F}\,\vp}{\YPF{V}\,e^F\,\vp}
=\sum_{i=0,1}\SPb{\mu_i\,e^{F}\,\vp}{\YPF{V}\,\mu_i\,e^F\,\vp}+R_\vp$,
where $|R_\vp|\klg\bigO(1)\|e^F\|^2\,\|\vp\|^2$.
(This holds in particular for $F=0$, of course.)
Therefore, it is sufficient to prove the bound \eqref{vera2}
with $\vp=\mu_i\,\psi$, $i=0,1$, $\psi\in\sD$.
For $\vp=\mu_0\,\psi$, the bound holds, however, true trivially,
for all $c_1,c_2\grg1$, since $F=1$
on the support of $\mu_0$.

Let us assume that $\vp=\mu_1\,\psi$, for some $\psi\in\sD$, in the rest of
this proof.
Of course,
$\|\chi_{0,R}\,e^F\,\vp\|^2\klg\|e^F\|^2\,\|\chi_{0,R}\,\vp\|^2$ and,
since $\Hf$ and $e^F$ commute, 
$\|\Hf^{1/2}\,e^F\,\vp\|^2\klg\|e^F\|^2\,\|\Hf^{1/2}\,\vp\|^2$.
Furthermore, 
$|\SPn{\vp}{V\,\vp}|\klg\bigO(1)\|\vp\|^2$,
since $V$ is bounded on $\supp(\mu_1)$.
To conclude we write 
$|\DA|=\PA\,\DA-\PAm\,\DA$ and employ the following bound
derived in 
\cite[Equation~(4.24) and the succeeding paragraphs]{MatteStockmeyer2008a},
\begin{equation}\label{exp-heidi99}
\SPb{\vp}{e^F\,\PApm\,(\pm\DA)\,e^F\,\vp}\klg c_3\,
\|e^F\|^2\,\SPn{\vp}{\PApm\,(\pm\DA)\,\vp}+c_4\,\|e^F\|^2\,
\|\vp\|^2,
\end{equation}
for every $\vp\in\core$.
We actually derived this bound in \cite{MatteStockmeyer2008a}
for classical vector potentials.
The proof works, however, also for the quantized vector potential
without any change. Moreover, we only treated the choice
of the plus sign in \eqref{exp-heidi99}. But again an obvious
modification of the 
proof in \cite{MatteStockmeyer2008a} shows that \eqref{exp-heidi99}
is still valid when we choose the minus sign.
{\qed}

\smallskip

{\em The no-pair operator.}
The following lemma implies that the conditions
\eqref{eq-Y2a} and \eqref{eq-Y2} are fulfilled in
the case of the no-pair operator, too.

\begin{lemma} There exist
$c\equiv c(V)\in(0,\infty)$ and
$c'\equiv c'(V,d_{-1},d_1,\Thnp-E^\np_{V,R})\in(0,\infty)$
such that, for all $F$
satisfying \eqref{hyp-F<>},
\begin{align*}
\big|\SPb{\vp}{(e^F\,\Ynp{V}\,e^{-F}-\Ynp{V})\,\vp}\big|
\,&\klg\,\bigO(a)\,\SPb{\vp}{(c\,\Ynp{V}+c')\,\vp}\,,
\quad \vp\in\sD.
\end{align*}
\end{lemma}

{\proof}
On account of \eqref{vera1} and 
$\|e^F\,\chi_{0,R}\,\PA\,\chi_{0,R}\,e^{-F}-\chi_{0,R}\,\PA\,\chi_{0,R}\|
\klg\|\cK_F\|=\bigO_{a_0}(a)$ it suffices to consider
$$
\triangle^\pm(T)\,:=\,
e^F\,\PApm \,T\,\PApm\,e^{-F}-\PApm \,T\,\PApm
\,=\,
2\Re\big[\PApm \,T\,\delta P\big]+\delta P\,T\,\delta P,
$$
where $\delta P:=e^F\,\PApm \,e^{-F}-\PApm$ and
$T$ is $\Hf$ or $V$. Clearly,
$$
|\SPn{\vp}{\triangle^\pm(T)\,\vp}|\,\klg\,
\ve\,\SPn{\vp}{\PApm \,T\,\PApm\,\vp}+(1+\ve^{-1})\,
\big\|\,|T|^{1/2}\,\delta P\,\vp\,\big\|^2,
$$
for all $\ve>0$ and $\vp\in \sD$.
Since, by \eqref{clelia2} and \eqref{clelia3},
$\|\,|T|^{1/2}\,\delta P\,\vp\|^2\klg\bigO(a^2)\,
\SPn{\vp}{(\Hf+(4d_1)^2+1)\,\vp}
\klg\bigO(a^2)\,\SPn{\vp}{(\Hf^{\mathrm{diag}}+d_1^2+1)\,\vp}$,
we may choose $\ve$ proportional to $a$ and use
\eqref{eqqq22V} to conclude.
{\qed}

\smallskip

\noindent
To complete the proof of Theorem~\ref{el} also in the case of
the no-pair operator we note that the bound \eqref{vera2}
still holds true when $\YPF{V}$ is replaced by
$\Ynp{0}$. To this end we only have to observe in addition
to the remarks in the proof of Lemma~\ref{exp-le-fdom-PF}
that 
$\|\HT^{1/2}\,\PApm\,e^F\,\vp\,\|\klg\bigO(1)\,\|e^F\|\,\|\HT^{1/2}\,\vp\|$.
This follows, however, immediately from \eqref{eva99}
which implies
$
\|\HT^{1/2}\,\PApm\,e^F\,\vp\,\|\klg
(1+\|\cC_{1/2}\|/2)\,\|e^F\,\HT^{1/2}\,\vp\|
$. Thus,
$e^{\pm F}\form(\Ynp{0})\subset\form(\Ynp{0})$. 
By Theorem~\ref{thm-sb-np} and the assumptions on $V$
we know that $\form(\Ynp{V})=\form(\Ynp{0})$
and we conclude.


\section{Existence of ground states with mass}\label{sec-ex-m}

\noindent
In this section we present an intermediate step of
the proof of
the existence of ground states for $\Hs{V}$.
Namely, 
we prove the existence of ground state eigenvectors, $\phi^\sharp_m$, for modified Hamiltonians,
$\Hs{V,m}$, which are defined by means of an infra-red cut-off coupling
function.
The infra-red cut-off parameter, $m>0$,
is referred to as the photon mass.
Later on in Section~\ref{ExistenceGroundStates}
we shall remove the infra-red cut-off by showing that
every sequence, $\{\phi^\sharp_{m_j}\}_j$, $m_j\searrow0$,
contains a strongly convergent subsequence whose limit
turns out to be a ground state eigenvector of $\Hs{V}$.
The compactness argument used to show this in Section~\ref{ExistenceGroundStates}
requires the infra-red bounds derived before in
Section~\ref{InfraredBounds}.

In the present section the existence of $\phi^\sharp_m$ is shown by
discretizing the photon degrees of freedom.
After the infra-red cut-off operators $\Hs{V,m}$ have been
defined in Subsection~\ref{ssec-mass} we construct
discretized versions of them, denoted by
$\Hs{V,m,\ve}$, in Subsection~\ref{ssec-NPe}.
We collect some technical estimates needed to compare
the original, infra-red cut-off, and discretized operators
in Subsections~\ref{ssec-veronique} and~\ref{ssec-hoe}.
As another preparation we study the continuity of the
ground state energy and ionization threshold with respect
to the parameters $m$ and $\ve$ in Subsection~\ref{ssec-cont}.
The main result of this section,
Theorem~\ref{prop-gs-m} on the existence of $\phi^\sharp_m$,
is stated and proved in Subsection~\ref{ssec-ex-m}
and we refer the reader to that subsection for
some brief remarks on its proof.
Many arguments of this section 
(in particular those in Subsections~\ref{ssec-hoe} and~\ref{ssec-cont}) are
alternatives to the corresponding ones in \cite{KMS2009a,KMS2009b}.

In the whole section $\V{G}_\V{x}$
is the coupling function given by \eqref{Gphys}.
To clarify which properties of $V$ are exploited we introduce
the following hypothesis. It is fulfilled by the Coulomb
potential in the subcritical cases:

\begin{hypothesis}\label{hyp-V1}
In the case $\sharp=\Pf$
the potential $V\in L^2_\loc(\RR^3,\RR)$ is relatively form-bounded with respect
to $\sqrt{-\Delta}$ with relative form bound strictly less
than one.
In the case $\sharp=\np$ the potential $V\in L^2_\loc(\RR^3,\RR)$
satisfies $H^1(\RR^3)\subset\dom(V)$ and 
\eqref{irina} with $a<1$.
\end{hypothesis}

We shall strengthen the assumptions on $V$ later on in order to
apply the localization estimates of Section~\ref{sec-exp}.


\subsection{Operators with photon mass}\label{ssec-mass}

\noindent
For every $m>0$, the infra-red cut-off coupling
function is given as
\begin{equation}\label{def-Gmx}
\V{G}_{\V{x},m}(k)\,:=\,-e\,
\frac{\id_{\{m\klg|\V{k}|\klg\UV\}}}{2\pi\sqrt{|\V{k}|}}\,e^{-i\V{k}\cdot\V{x}}
\,\veps(k)\,,
\end{equation}
for all $\V{x}\in\RR^3$
and almost every $k=(\V{k},\lambda)\in\RR^3\times\ZZ_2$.
To compare $\V{G}_{\V{x},m}$ with $\V{G}_\V{x}$ defined in \eqref{Gphys}
we introduce the parameter
\begin{equation}\label{def-trianglem}
\triangle^2(m)\,:=\,\int\big(\omega(k)+\omega(k)^{-1}\big)\,
\sup_{\V{x}}|\V{G}_\V{x}(k)-\V{G}_{\V{x},m}(k)|^2\,dk\,.
\end{equation}
Of course, 
$\triangle^2(m)=(e^2/2\pi^2)\int_{|\V{k}|<m}(1+|\V{k}|^{-2})d^3\V{k}\to0$,
as $m\searrow0$. For $m>0$,
we further define the infra-red cut-off vector potential,
$$
\V{A}_m:=\ad(\V{G}_m)+a(\V{G}_m)\,,\quad
a^\sharp(\V{G}_m):=\int_{\RR^3}^\oplus\id_{\CC^4}\otimes a^\sharp(\V{G}_{\V{x},m})
\,d^3\V{x}\,,
$$
and the infra-red regularized Hamiltonians
\begin{align}
\PF{V,m}&\,:= \,|D_{\mathbf{A}_m}|+V+\Hf\,,
\\
\NPO{V,m}&\,:=\, P^+_{\V{A}_m}\,(D_{\V{A}_m}+V+\Hf)\,P^+_{\V{A}_m}\,.
\end{align}
We define 
these operators as self-adjoint Friedrichs extensions
starting from $\core$. 
The ground state energies and ionization thresholds, 
for positive photon mass $m>0$, are denoted by
$$
E_{V,m}^\sharp:=\inf\spec[\Hs{V,m}]\,,
\qquad\Sigma_{m}^\sharp:=\inf\spec[\Hs{0,m}]\,.
$$ 
As a first step we introduce a truncated Fock space where
the radiation field energy $\Hf$ is bounded from below
by $m>0$ on the orthogonal complement of the vacuum sector.
Namely, we split the one-photon Hilbert space
into two mutually orthogonal subspaces
$$
\HP=\HP_m^>\oplus\HP_m^<, \qquad \HP_m^{>}:=
L^2(\cA_m\times\ZZ_2)\,,\qquad\cA_m:=\{|\V{k}|\grg m\}\,.
$$
It is well-known that $\Fock[\HP]\,=\,\Fock[\HP_m^>]\otimes\Fock[\HP_m^<]$.
We observe that $\V{A}_{m}$ creates and annihilates
photon states in $\HP_m^>$ only and $\Hf$ leaves the Fock space factors associated
with the subspaces $\HP_m^{\lessgtr}$ invariant.
We shall designate operators acting in the Fock space
factors $\Fock[\HP_m^{>}]$ or $\Fock[\HP_m^{<}]$ by a superscript
$>$ or $<$, respectively. 
Under the isomorphism
\begin{equation}\label{isom-m<>}
\HR\,\cong\,\big(L^2(\RR^3,\CC^4)\otimes
\Fock[\HP_m^>]\big)\otimes\Fock[\HP_m^<]\,=:\,\HR_m^>\otimes \Fock[\HP_m^<],
\end{equation}
we then have 
$D_{{\V{A}}_{m}}\cong D_{\V{A}^>_{m}}\otimes\id$,
$|D_{{\V{A}}_{m}}|\cong |D_{\V{A}^>_{m}}|\otimes\id$, 
$P^+_{\V{A}_{m}}\cong P^+_{\V{A}_{m}^>}\otimes\id$,
and
$\Hf=\Hf^>\otimes\id+\id\otimes\Hf^<$ with
$H^>_{f,m}:=d\Gamma(\omega\!\!\upharpoonright_{\cA_m\times\ZZ_2})$,
$H^<_{f,m}:=d\Gamma(\omega\!\!\upharpoonright_{\cA_m^c\times\ZZ_2})$.
As a consequence,
the semi-relativistic Pauli-Fierz and no-pair operators
decompose under the isomorphism \eqref{isom-m<>}
as
\begin{align}\label{DefOpm}
\PF{V,m}\,&=\,\ol{\PF{V,m,0}\otimes\id+\id\otimes \Hf^<}\,,
\\ \nonumber
\NPO{V,m}\,&=\,\ol{\NPO{V,m,0}\otimes\id
+P^+_{\V{A}_{m}^>}\otimes H_{\mathrm{f}}^<}\,,
\end{align}
where
\begin{align*}
\PF{V,m,0}\,&:=\,|D_{\V{A}^>_{m}}|+V+\Hf^>\,,
\\
\NPO{V,m,0}\,&:=\,P^+_{\V{A}_{m}^>}\,(D_{\V{A}_{m}^>}
+V+\Hf^>)\,P^+_{\V{A}_{m}^>}\,.
\end{align*}
The latter operators act in the Hilbert spaces
$\HR_m^>$ and $P^+_{\V{A}_m^>}\HR_m^>$, respectively.
The following lemma \cite{KMS2009a,KMS2009b} shows 
in particular that it 
suffices to prove the existence of ground states in
these truncated Hilbert spaces.

\begin{lemma}\label{le-E-Eg}
Assume that $V$ fulfills Hypothesis~\ref{hyp-V1}. Then,
for all $m>0$,
$$
E^{\sharp}_{V,m}=\inf \spec[\Hs{V,m,0}],\qquad
\Sigma^{\sharp}_{m}= \inf \spec[\Hs{0,m,0}]\,.
$$
Moreover, if 
$E^{\sharp}_{V,m}$ is an eigenvalue of $\Hs{V,m,0}$,
then it is an eigenvalue of
$\Hs{V,m}$, too.
\end{lemma}

{\proof}
It is clear that $\Hs{V,m,0}\otimes\id\klg\Hs{V,m}$,
whence $\inf \spec[\Hs{V,m,0}]\klg\inf \spec[\Hs{V,m}]$.
Next, we pick a minimizing sequence of
normalized vectors $\psi_n^>\in\form(\Hs{V,m,0})$,
$\SPn{\psi_n^>}{\Hs{V,m,0}\,\psi_n^>}\to\inf \spec[\Hs{V,m,0}]$.
Setting $\psi_n:=\psi_n^>\otimes\Omega^>$, where $\Omega^>$ is 
the vacuum vector in $\Fock[\HP_m^>]$, we observe that
$$
\SPn{\psi_n}{\Hs{V,m}\,\psi_n}=
\SPn{\psi_n^>}{\Hs{V,m,0}\,\psi_n^>}\,,
$$ 
thus
$\inf \spec[\Hs{V,m,0}]\grg\inf \spec[\Hs{V,m}]$.
Likewise, if
$\phi_{m}^\sharp\in \HR_m^>$ is a ground state eigenvector
of $\Hs{V,m,0}$, 
then $\phi_{m}^\sharp\otimes\Omega^>$ is a ground state eigenvector
of $\Hs{V,m}$. 
{\qed}

\smallskip

\noindent 
In order to show the existence of a ground state for $\Hs{V,m,0}$
in the next step it is sufficient
to show that the spectrum of $\Hs{V,m,0}$ is discrete in a neighborhood
of $E_{V,m}^\sharp$. 
($E_{V,m}^\sharp$ is contained in the essential
spectrum of $\Hs{V,m}$ on the contrary.)
A general strategy to achieve this would be the following.
We could seek for a 
self-adjoint operator, $A$, 
satisfying $-\infty<A\klg\Hs{V,m,0}$ and having
discrete spectrum in
$(-\infty,\,E_{V,m}^\sharp+c]$, 
for some $c>0$. If such an operator $A$ exists then also $\Hs{V,m,0}$
has discrete spectrum in $(-\infty,E_{V,m}^\sharp +c]$.
We need, however, a modification of this strategy.
Let $\chi$ denote the spectral projection of $\Hs{V,m,0}$ 
corresponding to some half-line $(-\infty,\,E_{V,m}^\sharp+c]$,
$c>0$. Then we seek for 
a self-adjoint auxiliary operator $A$ such that
\begin{equation}
\chi\,A\,\chi\, \klg \,
\chi\,\{ \Hs{V,m,0}-E_{V,m}^\sharp-2c \} \,\chi \,\klg\,-c\,\chi
\end{equation} 
and $\operatorname{Tr}\{\chi\, A \,\chi\}>-\infty$, 
where $\mathrm{Tr}$ denotes the trace.
For in this case we have
$\operatorname{Tr}\{\chi\}<\infty$.
The latter strategy is advantageous since 
we only have to compare $A$ and $\Hs{V,m,0}$
on the range of $\chi$ whose elements are exponentially localized
by Theorem~\ref{el}, provided that 
$c>0$ is appropriately chosen.
A suitable comparison operator $A$ is constructed by means of a
discretization of the photon momenta in the next subsection.


\subsection{Discretization of the photon momenta}\label{ssec-NPe}

\noindent
On $\HR_m^>$ we introduce a discretization in the photon momenta:
For every $\ve>0$,
we decompose $\cA_m=\{|\V{k}|\grg m\}$ as
$$
\cA_m=\!\!\bigcup_{\vnu\in(\ve\ZZ)^3}\!Q_m^\ve(\vnu)\,,
\quad Q_m^{\ve}(\vnu):=\big(\vnu+[-\ve/2\,,\,\ve/2)^3\big)
\cap\cA_m\,,\;\,
\vnu\in(\ve\ZZ)^3.
$$
Of course, for every $\V{k}\in\cA_m$, we find a unique
vector, $\wt{\vnu}_\ve(\V{k})\in(\ve\ZZ)^3$, such that
$\V{k}\in Q_m^\ve(\wt{\vnu}_\ve(\V{k}))$.
To each $\vnu\in(\ve\ZZ)^3$ with $Q_m^{\ve}(\vnu)\not=\varnothing$
we further associate
some $\boldsymbol{\varkappa}_{m,\ve}(\vnu)\in \ol{Q^\ve_m(\vnu)}$ such that
$$
|\boldsymbol{\varkappa}_{m,\ve}(\vnu)|\,=\,\inf_{\V{k}\in Q_m^\ve(\vnu)}|\V{k}|\,.
$$
In this way we obtain a map
\begin{equation}\label{def-nu}
\vnu_\ve\::\;\cA_m\times\ZZ_2\,\longrightarrow\,\RR^3\,,
\qquad
k=(\V{k},\lambda)\,\longmapsto\,\vnu_\ve(k)\,:=\,
\boldsymbol{\varkappa}_{m,\ve}\big(\wt{\vnu}_\ve(\V{k})\big)\,.
\end{equation}
It is evident that the vectors $\boldsymbol{\varkappa}_{m,\ve}(\vnu)$
can be chosen such that
\begin{equation}\label{xenia}
\vnu_\ve(-\V{k},\lambda)\,=\,-\vnu_\ve(\V{k},\lambda)\,,
\qquad\textrm{for almost every}
\;\V{k}\in\cA_m\,.
\end{equation}
The set of Lebesgue measure zero where 
the identity \eqref{xenia} might not hold
is contained in the union of all planes which are perpendicular
to some coordinate axis and contain points of the lattice $(\ve\,\ZZ)^3$.
We define the $\ve$-average of a locally integrable
function, $f$, on $\cA_m\times\ZZ_2$ by
\begin{equation}\label{Pdisc}
[P_\ve f](k)\,:=\,
\frac{1}{|Q_m^\ve(\wt{\vnu}_\ve(\V{k}))|}
\int\limits_{Q_m^\ve(\wt{\vnu}_\ve(\V{k}))}f(\V{p},\lambda)
\,d^3\V{p}\,,
\end{equation}
and introduce the following discretized coupling function,
\begin{equation*}
\V{G}_{\V{x},m,\ve}(k)\,:=\,
-(e/2\pi)\,
e^{-i\vnu_\ve(k)\cdot\V{x}}\,P_\ve\big[\id_{\cA_m} \,\omega^{-1/2}\,\veps\big](k)\,,
\end{equation*}
for all $\V{x}\in\RR^3$ and almost every $k=(\V{k},\lambda)\in\cA_m\times\ZZ_2$.
In order to compare $\V{G}_{\V{x},m}$ with $\V{G}_{\V{x},m,\ve}$
we put
\begin{equation}\label{def-triangleeps}
\triangle_*^2(a,m,\ve):=\!\!\!\!
\int\limits_{\cA_m\times\ZZ_2}\!\!\!\!\!\big(\omega(k)+\omega^{-1}(k)\big)
\,\sup_{\V{x}}\big\{e^{-a|\V{x}|}
|\V{G}_{\V{x},m}(k)-\V{G}_{\V{x},m,\ve}(k)|^2\big\}dk,
\end{equation}
for $a,m,\ve>0$. It is elementary to verify that
$\triangle_*(a,m,\ve)\to0$, as $\ve\searrow0$, for fixed
$a,m>0$. Notice that $\triangle_*(a,m,\ve)$ did not
converge to zero if we chose $a=0$. The fact that we need
some weight function in $\V{x}$ to control the difference
between $\V{G}_{\V{x},m}$ and $\V{G}_{\V{x},m,\ve}$ is one of the
reasons why a localization estimate for spectral subspaces
is required to prove the existence of ground states.
The discretized vector potential is now given as
\begin{equation*}
\V{A}_{m,\ve}:=
\ad(\V{G}_{m,\ve})+a(\V{G}_{m,\ve}),\quad
a^\sharp(\V{G}_{m,\ve}):=\int_{\RR^3}^\oplus\id_{\CC^4}\otimes
a^\sharp(\V{G}_{\V{x},m,\ve})\,d^3\V{x}\,.
\end{equation*}
The reason why we choose vectors $\vnu_\ve$ fulfilling \eqref{xenia}
is its consequence
$$
\V{G}_{\V{x},m,\ve}(-\V{k},\lambda)=\ol{\V{G}}_{\V{x},m,\ve}(\V{k},\lambda)\,.
$$
The latter identity ensures that different components,
$\V{A}_{m,\ve}^{(i)}(\V{x}),\V{A}_{m,\ve}^{(j)}(\V{y})$,
$\V{x},\V{y}\in\RR^3$, $i,j\in\{1,2,3\}$, of the 
discretized vector potential still commute.
We have used this property in Section~\ref{sec-sb}.
The dispersion relation is discretized as
\begin{equation*}
\omega_\ve(k)
\,:=\,\inf\big\{\,|\V{p}|\,:\:\V{p}\in Q_m^\ve(\wt{\vnu}_\ve(\V{k}))\,\big\}
\,,\qquad 
k=(\V{k},\lambda)\in\cA_m\times\ZZ_2\,.
\end{equation*}
Then $|\vnu_\ve|\klg\omega_\ve$ on $\cA_m\times\ZZ_2$ and
\begin{align}
m\,\klg&\,
\omega_\ve\,\klg\,\omega\;\;\textrm{on}\;\;\cA_m\times\ZZ_2\,,
\qquad \label{omega-omegaeps2}
H_{\mathrm{f},m,\ve}\,:=\,d\Gamma(\omega_\ve)\,\klg\,\Hfmg\,.
\end{align}
Here the operators $H_{\mathrm{f},m,\ve}$ and $\Hfmg$ are acting in
$\Fock[\HP_m^>]$. Finally, we define discretized Hamiltonians, 
$\Hs{V,m,\ve}$, acting in $L^2(\RR^2,\CC^4)\otimes \Fock[\HP_m^>]$,
\begin{align*}
\PF{{V},m,\ve}\,
&:=\,|D_{\V{A}_{m,\ve}}|+V+H_{\mathrm{f},m,\ve}
\,,
\\
\NPO{V,m,\ve}\,&:=\,
P^+_{\V{A}_{m,\ve}}\,(D_{\V{A}_{m,\ve}}+V
+H_{\mathrm{f},m,\ve})\,P^+_{\V{A}_{m,\ve}}\,.
\end{align*}


\subsection{Comparison of operators with different coupling functions
}
\label{ssec-veronique}

\noindent
In order to compare the various modified operators
we derive some bounds on differences of
projections whose proofs  are essentially consequences of the ideas collected
in Subsection~\ref{sec-htd} and the bounds
\begin{align}\label{veronique9}
\big\|\,\valpha\cdot(\V{A}-\V{A}_m)\,\HT^{-1/2}\,\big\|\,&=\,
\bigO\big(\triangle(m)\big)\,,
\\
\label{veronique10}
\big\|\,\valpha\cdot
(\V{A}_m^>-\V{A}_{m,\ve})\,
\HT^{-1/2}\,e^{-a|\V{x}|}\,\big\|\,&=\,\bigO\big(\triangle_*(a,m,\ve)\big)\,.
\end{align}
Here we use the notation \eqref{def-trianglem} and \eqref{def-triangleeps}.

\begin{lemma}\label{veronique}
Let $V$ be a symmetric multiplication operator in $L^2(\RR^3)$
which is relatively form bounded with respect to $\sqrt{-\Delta}$.

\smallskip

\noindent
(i) Set $\HT=\Hf+E$, for some sufficiently
large $E\grg1$ depending on $e$ and $\UV$, and let $\nu\grg0$. Then,
as $m\searrow0$,
\begin{align}
\big\|\,|\DA|^{1/2}\,(P_{\V{A}}^\pm-P_{\V{A}_m}^\pm)\,\HT^{-1/2}\,\big\|\,&\klg\,
\bigO\big(\triangle(m)\big)\,,\label{veronique1}
\\
\big\|\,|D_{\V{A}_m}|^{1/2}\,(P_{\V{A}}^\pm-P_{\V{A}_m}^\pm)\,\HT^{-1/2}\,\big\|\,
&\klg\,
\bigO\big(\triangle(m)\big)\,,\label{veronique2}
\\
\big\|\,\HT^{\nu}\,(P_{\V{A}}^\pm-P_{\V{A}_m}^\pm)\,\HT^{-\nu-1/2}\,\big\|\,&\klg\,
\bigO\big(\triangle(m)\big)\,,\label{veronique3}
\\
\big\|\,|V|^{1/2}\,(P_{\V{A}}^\pm-P_{\V{A}_m}^\pm)\,\HT^{-1}\,\big\|\,&\klg\,
\bigO\big(\triangle(m)\big)\,.\label{veronique4}
\end{align}
\noindent(ii) Let $\HT$ be $\Hfmg+E$ or $\Hfme+E$, for some sufficiently
large $E\grg1$ depending on $e$ and $\UV$, and let $\nu\grg0$ and $a_0\in(0,1)$. Then,
for every $a\in(0,a_0]$
and $F\in C^\infty(\RR^3_\V{x},[0,\infty))$
satisfying $|\nabla F|\klg a$, $F(\V{x})\grg a|\V{x}|$, 
for all $\V{x}\in\RR^3$, and
$F(\V{x})=a|\V{x}|$,
for large $|\V{x}|$, and for all sufficiently small $m,\ve>0$,
\begin{align}
\big\|\,|D_{\V{A}_{m}^>}|^{1/2}\,(P_{\V{A}_m^>}^\pm-P_{\V{A}_{m,\ve}}^\pm)\,\HT^{-1/2}
\,e^{-F}\,\big\|\,&\klg\,
\bigO\big(\triangle_*(a,m,\ve)\big)\,,\label{veronique5}
\\
\big\|\,|D_{\V{A}_{m,\ve}}|^{1/2}\,(P_{\V{A}_m^>}^\pm-P_{\V{A}_{m,\ve}}^\pm)\,\HT^{-1/2}
\,e^{-F}\,\big\|\,&\klg\,
\bigO\big(\triangle_*(a,m,\ve)\big)\,,\label{veronique6}
\\
\big\|\,\HT^{\nu}\,(P_{\V{A}_{m}^>}^\pm-P_{\V{A}_{m,\ve}}^\pm)\,\HT^{-\nu-1/2}
\,e^{-F}\,\big\|
\,&\klg\,
\bigO\big(\triangle_*(a,m,\ve)\big)\,,\label{veronique7}
\\
\big\|\,|V|^{1/2}\,(P_{\V{A}_{m}^>}^\pm-P_{\V{A}_{m,\ve}}^\pm)\,\HT^{-1}\,e^{-F}\,\big\|
\,&\klg\,
\bigO\big(\triangle_*(a,m,\ve)\big)\,.\label{veronique8}
\end{align}
\end{lemma}

{\proof}
By the assumption on $V$ and Theorem~\ref{le-sb-PF4}
we have $|V|\klg C\,(|\DA|+\HT)$.
Therefore, the bounds
 \eqref{veronique4} and \eqref{veronique8}  
are consequences of \eqref{veronique1}\&\eqref{veronique3} 
and \eqref{veronique5}\&\eqref{veronique7}, respectively.  
In order to prove the
remaining estimates we pick two vector potentials, $\V{A}_1$
and $\V{A}_2$, such that the set 
$\{\V{A}_1,\V{A}_2\}$ equals either $\{\V{A},\V{A}_m\}$ 
(in which case $F:=0$ in what follows) or $\{\V{A}_m^>,\V{A}_{m,\ve}\}$.  
For $j=1,2$ and $y\in\RR$, we set $R_{\V{A}_j}(iy):=(D_{\V{A}_j}-iy)^{-1}$ and
$R_{{\V{A}_j}}^F(iy):=(D_{\V{A}_j}+i\valpha\cdot\nabla F-iy)^{-1}$.
(Recall \eqref{exp-marah0} and \eqref{exp-marah1}.) 
Then we have the following resolvent identity, 
for $\mu\grg1/2$ and $y\in\RR$, 
\begin{align*}
  (R_{\V{A}_1}(iy)&-R_{\V{A}_2}(iy))\,e^{-F}\HT^{-\mu}\\
 & =\,R_{\V{A}_1}(iy)\,\valpha\cdot(\V{A}_2-\V{A}_1)\,\HT^{-\mu}e^{-F}
R_{{\V{A}_2}}^F(iy)\,\Upsilon_{\mu}^F(iy)\,,
\end{align*}
where $\Upsilon_{\mu}^F(iy)$ is the bounded operator 
defined in \eqref{def-THT} below (with $\V{A}$ replaced by $\V{A}_1$). 
Using \eqref{sgn} we find, for all $\phi$ and $f$ in the Hilbert space,
\begin{align}\label{aa-bb}
  &\big|\SPb{f}{(P_{\V{A}_1}^\pm-P_{\V{A}_2}^\pm)\,\HT^{-\mu}
\,e^{-F}\,\phi}\big|
\\
&\nonumber\klg
\int_\RR\big|
\SPb{f}{R_{\V{A}_1}(iy)\,\valpha\cdot(\V{A}_2-\V{A}_1)\,\HT^{-\mu}e^{-F}
R_{{\V{A}_2}}^F(iy)\Upsilon_{\mu}^F(iy)\,\phi}\big|\,\frac{dy}{\pi}\,.
\end{align}
In the case $\mu =1/2$, 
we choose $f=|D_{\V{A}_1}|^{1/2}\,\psi$, $\psi\in\dom(|D_{\V{A}_1}|^{1/2})$, 
and observe that the integrand in \eqref{aa-bb} is then bounded by 
  \begin{equation*}
\bigO_*\,\||D_{\V{A}_1}|^{1/2}R_{\V{A}_1}(iy)\|
\,\|R_{{\V{A}_2},L}(iy)\|\,\|\Upsilon_{1/2,L}(iy)\|\,\|\phi\|\,\|\psi\|,
\end{equation*}
which is integrable due to \eqref{exp-marah1},
\eqref{veronique9}, \eqref{veronique10}, and the spectral theorem.
Here and below $\bigO_*=\bigO(\triangle(m))$ or 
$\bigO_*=\bigO(\triangle_*(a,m,\ve))$
depending on the choice of $\V{A}_j$. 
This concludes the proof of \eqref{veronique1}, \eqref{veronique2},
\eqref{veronique5}, and \eqref{veronique6}. 

In order to prove 
\eqref{veronique3} and \eqref{veronique7} 
we infer from 
Lemma~\ref{le-tim} that the commutator  
$T_\nu:=\HT^{\nu}\,[\valpha\cdot(\V{A}_2-\V{A}_1)\,e^{-F}\,,\,\HT^{-\nu}]$ 
extends to a bounded operator with $\|T_\nu\|=\bigO_*$. 
Together with \eqref{veronique9}, \eqref{veronique10}, and
\eqref{eva3b} this implies, for $\nu\grg0$ and $\psi,\phi\in\dom(\Hf^\nu)$, 
\begin{align*}
&\big|\SPb{\HT^{\nu}\psi}{R_{\V{A}_1}(iy)\,
\valpha\cdot(\V{A}_2-\V{A}_1)\,\HT^{-\nu-1/2}e^{-F}
R_{{\V{A}_2}}^F(iy)\,\Upsilon_{\nu+1/2}^F(iy)\,\phi}\big|
\\
&=\big|\SPb{\psi}{R_{\V{A}_1}(iy)\,\Upsilon_{{\nu}}^F(iy)\,
\HT^{\nu}\,\valpha\cdot(\V{A}_2-\V{A}_1)\,\HT^{-\nu-1/2}e^{-F} \times
\\
&\qquad\qquad\qquad\qquad
\qquad\qquad\qquad\qquad\times
R_{{\V{A}_2}}^F(iy)\,\Upsilon_{\nu+1/2}^F(iy)\,\phi}\big|
\\
&\klg
\bigO_*\,
\|R_{\V{A}_1}(iy)\|\,\|\Upsilon_{{\nu}}^F(iy)\|\,\|R_{{\V{A}_2}}^F(iy)\|\,
\|\Upsilon_{\nu+1/2}^F(iy)\|\,\|\psi\|\,\|\phi\|
\\
&\klg
\bigO_*\,(1+y^2)^{-1}\,\|\psi\|\,\|\phi\|\,.
\end{align*}
Therefore, \eqref{veronique3} and 
\eqref{veronique7} follow from \eqref{aa-bb} 
upon choosing $\mu=\nu+1/2$ and  $f=\HT^{\nu}\psi$.
 {\qed}

\subsection{Higher order estimates and their consequences}\label{ssec-hoe}

\noindent
As a preparation
for the proof of the existence of ground states for
$\Hs{V,m,0}$ we derive bounds on certain expectation values
of the difference $\Hs{V,m,0}-\Hs{V,m,\ve}$ in this
subsection. Moreover, we compare the ground state energies
and ionization thresholds of $\Hs{V,m,0}$ with those
of $\Hs{V,m,\ve}$ defined by
$$
E_{V,m,\ve}^\sharp:=\inf\spec[\Hs{V,m,\ve}]\,,\qquad
\Sigma_{m,\ve}^\sharp:=\inf\spec[\Hs{0,m,\ve}]\,,
$$
for $m,\ve>0$. It is not possible to
compare the no-pair operators $\NPO{V,m,0}$
and $\NPO{V,m,\ve}$ in a quadratic form sense.
For some error terms in the difference of these two operators
``have the size of $|\V{x}|\,\Hf^{3/2}$'' and can only be controlled
when we take expectations with respect to states in
some low-lying exponentially localized spectral subspace.
To control the higher power $\Hf^{3/2}$ of the radiation field
energy we need, however, yet another non-trivial
ingredient, namely the higher order estimates of the 
next theorem. 
Because of lack of space we cannot comment on the 
proof of Theorem~\ref{prop-hoe} and refer to \cite{Matte2009} instead.
We remark that, for the semi-relativistic Pauli-Fierz operator,
higher order estimates have been obtained earlier in
\cite{FGS2001}. Their proof given in \cite{Matte2009} is, however,
different and more model-independent so that the no-pair
operator can also be treated along the same
lines in \cite{Matte2009}.
In the case of the no-pair operator only the
Coulomb potential is considered in \cite{Matte2009}.
An inspection of the proofs in \cite{Matte2009}
shows, however, that they immediately extend to
all potentials satisfying Hypothesis~\ref{hyp-V1}.

\begin{theorem}\label{prop-hoe}
Let $e\in\RR$, $\UV\in(0,\infty)$, and assume that $V$
fulfills Hypothesis~\ref{hyp-V1}. Then 
$\dom((\Hs{V,m,\ve})^{n/2})\subset\dom(\Hf^{n/2})$, for 
every $n\in\NN$, and there exist constants,
$\ve_0,m_0,C\in(0,\infty)$,
such that, for all $\ve\in[0,\ve_0]$ and $m\in(0,m_0]$,
\begin{align*}
\big\|\Hfme^{n/2}\!\!\upharpoonright_{P^+_{\V{A}_{m,\ve}}\HR_m^>}\!
\big(\NPO{V,m,\ve}-(E^\np_{V,m,\ve}-1)
P^+_{\V{A}_{m,\ve}}\big)^{-n/2}\big\|
& \klg
C(1+|E^\np_{V,m,\ve}|)^{2n},
\\
\big\|\Hfme^{n/2}
\big(\PF{V,m,\ve}-(E^\Pf_{V,m,\ve}-1)
\big)^{-n/2}\big\|
& \klg
C(1+|E^\Pf_{V,m,\ve}|)^{2n}.
\end{align*}
If $V=0$, then $E^\sharp_{V,m,\ve}$ has to be
replaced by $\Sigma^\sharp_{m,\ve}$ in these bounds.
Analogous bounds hold for $\Hs{V}$.
\end{theorem}

\begin{lemma}\label{MainLem0}
Let $e\in\RR$, $\UV\in(0,\infty)$, and assume that $V$
fulfills Hypothesis~\ref{hyp-V1}.
Then we find some $m_0>0$ such that the following holds:
 
\smallskip

\noindent(i)
For all $m\in(0,m_0]$
and
$\psi^+\in\Ran(\id_{(-\infty,\Sigma^\sharp+1]}(\Hs{V}))$, 
\begin{align}
\big|\SPb{\psi^+}{&\Hs{V,m}\,\psi^+}-
\SPb{\psi^+}{\Hs{V}\,\psi^+}\big|\nonumber
\,\klg\,\const(\Sigma^\sharp,|E_V^\sharp|)\,o(m^0)\,
\|\psi^+\|^2.
\end{align}
\noindent(ii)
For all $m\in(0,m_0]$
and
$\psi^+\in\Ran(\id_{(-\infty,\Sigma_m^\sharp+1]}(\Hs{V,m}))$, 
\begin{align}
\big|\SPb{\psi^+}{&\Hs{V,m}\,\psi^+}-
\SPb{\psi^+}{\Hs{V}\,\psi^+}\big|\nonumber
\,\klg\,\const(\Sigma_m^\sharp,|E_{V,m}^\sharp|)\,o(m^0)\,
\|\psi^+\|^2.
\end{align}
\end{lemma}

{\proof}
We treat only the no-pair operator explicitly.
On account of the formula
$|\DA|=\PA\,\DA-\PAm\,\DA$ it will then be clear
how to obtain the result also for the
semi-relativistic Pauli-Fierz operator.
We remark only once that, for instance, 
the inclusion $P^+_{\V{A}_m}\,\Ran(\id_{(-\infty,\Sigma^\sharp+1]}(\Hs{V}))
\subset\form(\Hs{V,m})$ follows from the characterization of the
form domains in Theorems~\ref{le-sb-PF4} and~\ref{thm-sb-np}.
In the rest of this section we shall use similar remarks
without further notice to simplify the exposition.
Let
$\delta P:=P^+_{\V{A}}-P^+_{\V{A}_m}$ and
$\cM:=\{D_{\V{A}},V,\Hf\}$.
Then we have
\begin{align}\label{Difference}
\NPO{V}-\NPO{{V},m}
&=
P^+_{{\V{A}_m}}\valpha\cdot(\V{A}-{\V{A}_m})P^+_{{\V{A}_m}}\!
+\!\!\sum_{T\in\mathcal{M}}\!\big\{
2\Re\big[P^+_{\V{A}}\,T\,\delta P\big]-
\delta P\,T\,\delta P\big\}
\end{align}
in the sense of quadratic forms on $\form(|\DO|)\cap\form(\Hf)$.
Now, let 
$$
\psi^+\in\Ran(\id_{(-\infty,\Sigma^\sharp+1]}(\Hs{V}))\,.
$$
From \eqref{gustav-gen}, \eqref{veronique9}, 
\eqref{veronique1}--\eqref{veronique4}, and \eqref{Difference} 
we readily infer that
\begin{align}\nonumber
\big|\SPb{\psi^+}{(\NPO{V}-\NPO{{V},m})\,\psi^+}\big|
&\klg
\bigO\big(\triangle(m)\big)\,
\big\{\|\,|\DA|^{1/2}\,\psi^+\|^2+\|\HT\,\psi^+\|^2\big\}.
\end{align}
Here $\|\,|\DA|^{1/2}\,\psi^+\|\klg\bigO(1)\,\|\psi^+\|$
since $\psi^+$ belongs to the spectral subspace
$\Ran(\id_{(-\infty,\Sigma^\sharp+1]}(\Hs{V}))$
and the term containing $V$ is a small form perturbation of $\Hs{V}$
by Hypothesis~\ref{hyp-V1} and Theorem~\ref{le-sb-PF4}
or Theorem~\ref{thm-sb-np}, respectively. Moreover,
$\|\HT\,\psi^+\|\klg\const(\Sigma,|E_V|)\,\|\psi^+\|$
because of the higher order estimates.
We can argue analogously if $\psi^+$ belongs to a spectral
subspace of $\Hs{V,m}$. In this case the right hand
side of the estimate depends on $\Sigma_m$ and $|E_{V,m}|$
since we apply higher order estimates for $\Hs{V,m}$.
{\qed}

\smallskip

\noindent
We recall once more that some $\V{x}$-dependent weight
is required to control the difference between
$\V{A}_{m}^>$ and $\V{A}_{m,\ve}$.
If we consider only vectors $\psi^+$ in a spectral subspace
corresponding to sufficiently low energies of
$\Hs{V,m,\ve}$, then we can borrow
this weight from the exponential localization of $\psi^+$.
At this point we have to introduce further conditions on the
potential in order to guarantee that {\em there are} non-trivial
spectral subspaces below the ionization threshold.

\begin{hypothesis}\label{hyp-V2}
$V$ satisfies \eqref{john2} and
there exist $c,m_\star,\ve_\star>0$ such that, for all
$m\in(0,m_\star]$ and all $\ve\in(0,\ve_\star]$, 
\begin{align}\label{bert}
\Sigma^\sharp-E_V^\sharp\grg c\,,\quad
\Sigma^\sharp_m-E_{V,m}^\sharp\grg c\,,\quad
\Sigma^\sharp_{m,\ve}-E_{V,m,\ve}^\sharp\grg c\,.
\end{align}
\end{hypothesis}

Examples of potentials fulfilling the previous hypothesis
have been found in Theorem~\ref{thm-binding-PF}.
In fact, as already pointed out there,
the proof of Theorem~\ref{thm-binding-PF} works also
for the discretized operators when the unitary transformation
employed in Section~\ref{sec-binding} is modified suitably;
see \cite{KMS2009a,KMS2009b} for details.
 There are, however, potentials fulfilling Hypotheses~\ref{hyp-V1}
and~\ref{hyp-V2} which are not covered by
Theorem~\ref{thm-binding-PF},
for instance, those mentioned in Remark~\ref{rem-HiroshimaSasaki}.
This is the reason why we work with the very implicit 
Hypothesis~\ref{hyp-V2} in what follows.

\begin{lemma}\label{MainLem}
Let $e\in\RR$, $\UV\in(0,\infty)$, and assume that $V$ fulfills
Hypotheses~\ref{hyp-V1} and~\ref{hyp-V2}.
Then there exist $\ve_0,m_0>0$ such that the following holds,
for all $\ve\in(0,\ve_0]$ and $m\in(0,m_0]$:

\smallskip

\noindent(i)
For all
$\lambda\in(E^{\sharp}_{V,m},\Sigma^{\sharp}_{m})$
and 
$\psi^+\in\Ran(\id_{(-\infty,\lambda]}(\Hs{V,m}))$,
\begin{align}
\big|\SPb{\psi^+&}{\Hs{V,m,\ve}\,\psi^+}-
\SPb{\psi^+}{\Hs{V,m,0}\,\psi^+}\big|\nonumber
\\
&\klg\,\const\big(\Sigma^{\sharp}_{m},|E^{\sharp}_{V,m}|,
(\Sigma^{\sharp}_{m}-\lambda)^{-1}\big)\,
o(\ve^0)\,
\|\psi^+\|^2.\label{marianne1}
\end{align}
\noindent(ii)
For all
$\lambda\in(E^{\sharp}_{V,m,\ve},\Sigma^{\sharp}_{m,\ve})$
and
$\psi^+\in\Ran(\id_{(-\infty,\lambda]}(\Hs{V,m,\ve}))$,
\begin{align}
\big|\SPb{\psi^+&}{\Hs{V,m,\ve}\,\psi^+}-
\SPb{\psi^+}{\Hs{V,m,0}\,\psi^+}\big|\nonumber
\\
&\klg\,\const\big(\Sigma^{\sharp}_{m,\ve},|E^{\sharp}_{V,m,\ve}|,
(\Sigma^{\sharp}_{m,\ve}-\lambda)^{-1}\big)\,
o(\ve^0)\,
\|\psi^+\|^2.\label{marianne2}
\end{align}
\end{lemma}

{\proof}
Again we only treat the no-pair operator since the proofs
for the semi-relativistic Pauli-Fierz operator will then be
obvious.

(i): We have a formula for
$\NPO{V,m,0}-\NPO{V,m,\ve}$ similar to \eqref{Difference}
with $(\V{A},\V{A}_m,\Hf)$ replaced by
$(\V{A}_{m}^>,\V{A}_{m,\ve},\Hfmg)$ where one additional
term has to be added, namely 
$P^+_{{\V{A}_{m,\ve}}}\,d\Gamma(\omega-\omega_\ve)\,P^+_{{\V{A}_{m,\ve}}}$.
Using this formula, \eqref{veronique10},
\eqref{veronique5}--\eqref{veronique8},
and 
$|\omega-\omega_\ve|\klg\sqrt{3}\,\ve\klg(\sqrt{3}\,\ve/m)\,\omega_\ve
\klg(\sqrt{3}\,\ve/m)\,\omega$, for $|\V{k}|\grg m$,
which yields
$$
\big|\SPb{\psi^+}{
P^+_{{\V{A}_{m,\ve}}}\,
d\Gamma(\omega-\omega_\ve)\,P^+_{{\V{A}_{m,\ve}}}\,\psi^+}\big|
\,\klg\,o(\ve^0)\,\big\|\HT^{1/2}\,\psi^+\big\|^2,
$$
we arrive at
\begin{align*}
\big|\SPb{\psi^+}{&(\NPO{V,m,0}-\NPO{{V},m,\ve})\,\psi^+}\big|
\\
&\klg\,\big(
\bigO(\triangle_*(a,m,\ve))+o(\ve^0)\big)\,
\big\{\|\,|D_{\V{A}_{m}^>}|^{1/2}\,\psi^+\|^2+\|e^F\,\HT\,\psi^+\|^2\big\},
\end{align*}
for $\psi^+\in\Ran(\id_{(-\infty,\lambda]}(\Hs{V,m}))$.
Here we further have
$$
2\|e^F\,\HT\,\psi^+\|^2\klg\|e^{2F}\,\psi^+\|^2+\|\HT^2\,\psi^+\|^2,
$$ where the norms on the right side can be controlled
by our exponential localization and higher order estimates,
respectively.
Part~(ii) is derived analogously.
The dependence on $\lambda$, the ionization thresholds, and the ground state
energies of the constants on the right hand sides of \eqref{marianne1} and 
\eqref{marianne2} stems from the constants in
the exponential localization and higher order estimates. 
{\qed}


\subsection{Continuity of the ionization thresholds and ground state energies}
\label{ssec-cont}

\noindent
To make use of the bounds of Lemmata~\ref{MainLem0} and~\ref{MainLem}
we still have to verify that the functions
$m\mapsto\Sigma_m^\sharp$, $m\mapsto E_{V,m}^\sharp$,
and $\ve\mapsto\Sigma_{m,\ve}^\sharp$,
$\ve\mapsto E_{V,m,\ve}^\sharp$ are
(semi-)continuous at $0$.
The continuity of $\ve\mapsto E_{V,m,\ve}^\sharp$ will also
enter more directly into the proof of the existence
of ground states in the next subsection.

\begin{corollary}\label{cor-EmE}
Assume that $V$ fulfills Hypothesis~\ref{hyp-V1}. Then it follows that
$\lim_{m\to0}E_{V,m}^\sharp=E_V^\sharp$
and $\lim_{m\to0}\Sigma_m^\sharp=\Sigma^\sharp$.
\end{corollary}

{\proof}
Using Lemma~\ref{MainLem0}(i) it is not difficult to derive
the bounds $\Sigma_m^\sharp\klg\Sigma^\sharp+o(m^0)$.
(Given $\ve>0$ we pick some
$\psi^+\in \Ran(\id_{[\Sigma^\sharp,\Sigma^\sharp+\ve)}(H_0^\sharp))$,
$\|\psi^+\|=1$, and plug
it into the quadratic form of $\Hs{0,m}$. In the case of the no-pair
operator we also have to observe that $\|P^+_{\V{A}_m}\,\psi^+\|\to1$,
$m\searrow0$.)
Since this gives an upper bound on $\Sigma_m^\sharp$
which is uniform, for small $m$,
we can then control the
constants on the right hand side of the estimate in
Lemma~\ref{MainLem0}(ii) to get $\Sigma^\sharp\klg\Sigma_m^\sharp+o(m^0)$
by a similar argument. 
Since the results of Section~\ref{sec-sb} provide uniform
lower bounds on $E_{V,m}$ we can now employ
Lemma~\ref{MainLem0} in a similar fashion to show that
$\lim_{m\to0}E_{V,m}^\sharp=E_V^\sharp$.
{\qed}

\smallskip

\noindent
In order to compare the ionization thresholds 
$\Sigma^\sharp_m$ and $\Sigma^\sharp_{m,\ve}$ we need
a different argument since $\V{x}$-dependent weights are
required to control the difference between
$\V{A}_m^>$ and $\V{A}_{m,\ve}$ but the 
spectral subspaces of the free operators
$\Hs{0,m,\ve}$ are not localized.
Here the essential
self-adjointness of $\Hs{0,m,\ve}$ asserted in Theorem~\ref{Cor-Friedr-Extens} 
is helpful. To be able to work in one
fixed Hilbert space we set
\begin{equation}\label{nurleen2012}
\wh{H}^\np_{0,m,\ve}\,:=\,\NPO{0,m,\ve}+
P_{\V{A}_{m,\ve}}^-\,(|D_{\V{A}_{m,\ve}}|+\Hfme)\,P_{\V{A}_{m,\ve}}^-\,.
\end{equation}
Then Lemma~\ref{le-H+H-} below implies that
$\Sigma_{m,\ve}^\np=\inf\spec[\NPO{0,m,\ve}]=\inf\spec[\wh{H}^\np_{0,m,\ve}]$.

\begin{lemma}\label{le-res-conv}
$\PF{0,m,\ve}\to\PF{0,m,0}$ and $\wh{H}^\np_{0,m,\ve}\to\wh{H}^\np_{0,m,0}$
in the strong
resolvent sense, as $\ve\searrow0$. In particular,
$\limsup_{\ve\searrow0}\Sigma_{m,\ve}^\sharp\klg\Sigma_m^\sharp$.
\end{lemma}

{\proof}
Since all involved operators are essentially self-adjoint
on $\core$ it suffices to show that
$\PF{0,m,\ve}\,\vp\to\PF{0,m,0}\,\vp$ 
and $\wh{H}^\np_{0,m,\ve}\,\vp\to\wh{H}^\np_{0,m,0}\,\vp$,
for every fixed $\vp\in\core$. Since $\vp$ has only
finitely many non-vanishing Fock space components and
the latter are compactly supported it is clear that
$\Hfme\,\vp\to\Hfmg\,\vp$.
Furthermore, we write
$$
|D_{\V{A}_m^>}|-|D_{\V{A}_{m,\ve}}|\,=\,S_{\V{A}_m^>}\,
\valpha\cdot(\V{A}_m^>-\V{A}_{m,\ve})
+(S_{\V{A}_m^>}-S_{\V{A}_{m,\ve}})\,D_{\V{A}_{m,\ve}}\,,
$$
where $S:=D\,|D|^{-1}$ denotes the sign function,
which permits to get
\begin{align*}
&\big\|\,|D_{\V{A}_m^>}|\,\vp-|D_{\V{A}_{m,\ve}}|\,\vp\,\big\|
\,\klg\,
\big\|e^{-F}\valpha\cdot(\V{A}_m^>-\V{A}_{m,\ve})\,\HT^{-1/2}\big\|
\,\|e^{F}\,\HT^{1/2}\,\vp\|
\\
&+
\big\|(S_{\V{A}_m^>}-S_{\V{A}_{m,\ve}})\,\HT^{-1/2}e^{-F}\big\|\,
\big\{\|e^F\HT^{1/2}\DO\,\vp\|+
\|\HT^{1/2}\,\valpha\cdot\V{A}_{m,\ve}\,e^F\vp\|\big\},
\end{align*}
where $\HT:=\Hfmg+E$ and $E\equiv E(e,\UV)$ is sufficiently
large, independently of $\ve$.
Using \eqref{veronique10}, \eqref{veronique5},
and Lemma~\ref{le-tim} it is now
easy to see that $\PF{0,m,\ve}\,\vp\to\PF{0,m,0}\,\vp$.

To show that also $\wh{H}^\np_{0,m,\ve}\,\vp\to\wh{H}^\np_{0,m,0}\,\vp$
it  remains to observe that
\begin{align}\nonumber
&\big\|\PAmgpm\,\Hf^>\,\PAmgpm\,\vp-\PAepm\,\Hf^>\,\PAepm\,\vp\big\|
\\
&\klg\,\nonumber
\big\|(\PAmgpm-\PAepm)\,\HT^{-1/2}\,e^{-F}\,\big\|\,
\|e^F\,\HT^{3/2}\,\PAmgpm\,\HT^{-3/2}\,e^{-F}\|\,\|e^F\,\HT^{3/2}\,\vp\|
\\
&\quad+\,\label{viviane1}
\big\|\Hf\,(\PAmgpm-\PAepm)\,\HT^{-3/2}\,e^{-F}\,\big\|\,
\|e^F\,\HT^{3/2}\,\vp\|\xrightarrow{\;\ve\searrow0\;}\,0\,,
\end{align}
and
\begin{align}
\big\|\PAepm&(\Hf^>-\Hfme)\,\PAepm\vp\big\|
\klg\label{viviane2}
\frac{\sqrt{3}\,\ve}{m}\,\big\|\Hf^>\PAepm\HT^{-1}\big\|\,
\|\HT\,\vp\|\xrightarrow{\;\ve\searrow0\;}0\,.
\end{align}
In fact, \eqref{viviane1} is valid
because of \eqref{veronique5}, \eqref{veronique7}, and 
since $e^F\HT^{3/2}\PAmgpm\HT^{-3/2}e^{-F}$ is well-defined
and bounded as a consequence of \eqref{clelia1} and Lemma~\ref{le-sgn}.
Moreover, \eqref{viviane2} holds true since 
$|\omega-\omega_\ve|\klg\sqrt{3}\,\ve$ and the photon momenta
are $\grg m$ in modulus on $\HR_m^>$, and since
$\|\Hf^>\,\PAepm\,\HT^{-1}\|$ is bounded uniformly
in $\ve>0$ by Lemma~\ref{le-sgn}.
{\qed}

\begin{corollary}\label{MainCor}
Let $m>0$ be sufficiently small and
assume that $V$ fulfills Hypotheses~\ref{hyp-V1}
and~\ref{hyp-V2}. Then
$\lim_{\ve\to0}E^{\sharp}_{V,m,\ve}=E^{\sharp}_{V,m}$.
\end{corollary}

{\proof}
Since Lemma~\ref{le-res-conv} provides upper bounds
on $\Sigma_{m,\ve}^\sharp$ and we have lower bounds
on the spectra of $\Hs{V,m,\ve}$ which are uniform
in $m$ and $\ve$,
we can apply Lemma~\ref{MainLem}
and some straightforward variational arguments to prove
the assertion. 
We only point out one subtlety: In order to show that
$E_{V,m}\klg E_{V,m,\ve}+o(\ve^0)$ we have to pick a test
function in $\Ran(\id_{(-\infty,\lambda]}(\Hs{V,m,\ve}))$,
for some $\lambda\in(E_{V,m,\ve}\,,\,\Sigma^\sharp_{m,\ve})$.
To ensure that this is possible and in order to
have an $\ve$-independent bound on the numbers 
$(\Sigma^\sharp_{m,\ve}-\lambda)^{-1}$ entering into the
constant in \eqref{marianne2} we need the lower bound \eqref{bert} on the binding
energy $\Sigma^\sharp_{m,\ve}-E_{V,m,\ve}\grg c$ which
does not depend on $\ve$. 
Without an estimate on the binding energy we still got 
$E_{V,m,\ve}\klg E_{V,m}+o(\ve^0)$, but
we would not have a useful lower bound on $\Sigma_{m,\ve}^\sharp$.
{\qed}


\subsection{Proofs of the existence of ground states with mass}
\label{ssec-ex-m}

\noindent
The next theorem asserting compactness of spectral
projections of $H_{V,m,0}^\sharp$ associated with sufficiently low energies
is the final result of this section. As in \cite{BFS1998b}
it is proved by estimating the trace of the spectral projection
from above by the trace of some finite rank operator,
namely the one in \eqref{jacob}.
This finite rank operator is constructed by means
of a suitable restriction of the discretized field energy
with discrete spectrum and a harmonic oscillator potential
which compactifies the electronic part of the operator.
In order to sneak in the harmonic oscillator potential
in the proofs below we exploit the exponential localization
of low-lying spectral subspaces once more.
The latter idea stems from \cite{KMS2009a,KMS2009b}
and replaces an argument in \cite{BFS1998b} that works
only for small $e$ and/or $\UV$.

\begin{theorem}\label{prop-gs-m}
Let $e\in\RR$, $\UV\in(0,\infty)$, and 
assume that $V$ fulfills Hypotheses~\ref{hyp-V1} and~\ref{hyp-V2}.
Define 
$\chi:=\mathbbm{1}_{(-\infty\,,\,E_{V,m}^\sharp+m/4]}(H_{V,m,0}^\sharp)$
 and assume
that $m>0$ is sufficiently small.
Then $\mathrm{Tr}\{\chi\}<\infty$.
In particular, $E_{V,m}^\sharp$ 
is an eigenvalue of both $H_{V,m,0}^\sharp$ and 
$H_{V,m}^\sharp$.
\end{theorem}

\smallskip

\noindent
In the proof of the preceding theorem, which is carried through
separately for $\sharp=\Pf$ and $\sharp=\np$ below,
we shall employ an orthogonal splitting of $\HP_m^>$ into
subspaces of discrete and fluctuating
photon states,
\begin{equation*}
\HP_m^d\,:=\,P_\ve\,\HP_m^>\,,\qquad \HP_m^f\,:=\,\HP_m^>\ominus\HP_m^d\,.
\end{equation*}
Here $P_\ve$ is defined in \eqref{Pdisc}.
The splitting $\HP_m^>=\HP_m^d\oplus\HP_m^f$
gives rise to an isomorphism
\begin{equation}\label{SplittingDiscFluc}
L^2(\RR^3,\CC^4)\otimes\Fock[\HP_m^>]\,\cong\,
\big(\,L^2(\RR^3,\CC^4)\otimes\Fock[\HP_m^d]\,\big)\,\otimes\,\Fock[\HP_m^f]\,,
\end{equation}
and we observe that the Dirac operator and the field energy
decompose under the above isomorphism as 
\begin{equation}\label{valerie0}
D_{\V{A}_{m,\ve}}\,\cong\,D_{\V{A}^d_{m,\ve}}\otimes\id^f\,,\qquad
H_{\mathrm{f},m,\ve}\,
=\,H_{\mathrm{f},m,\ve}^d\otimes\id^f+\id^d\otimes H_{\mathrm{f},m,\ve}^f\,.
\end{equation}
Here and in the following we designate 
operators acting in the Fock space factors 
$\Fock[\HP_m^\ell]$, $\ell\in\{d,f\}$, by 
the corresponding superscript 
$d$ or $f$.
In fact, the discretized vector potential $\V{A}_{m,\ve}$
acts on the various $n$-particle sectors in $\Fock[\HP_m^>]$
by tensor-multiplying or taking scalar products with elements
from $\HP_m^d$ (apart from symmetrization and a normalization constant). 

For $\ell\in\{d,f\}$,
we denote the identity on $\Fock[\HP_m^\ell]$ by $\id^\ell$ and
the projection onto the vacuum sector in
$\Fock[\HP_m^\ell]$ by $P_{\Omega^\ell}$ and
write 
$P_{\Omega^\ell}^\bot:=\id^\ell-P_{\Omega^\ell}$.
The identity on $L^2(\RR^3_\V{x},\CC^4)$ is denoted
as $\id^\el$.

\smallskip

{\em The semi-relativistic Pauli-Fierz operator.}

{\proof} We prove Theorem~\ref{prop-gs-m} with $\sharp=\Pf$.
On account of Lemma~\ref{MainLem}(i) we have
$\chi\,\PF{V,m,0}\,\chi\grg\chi\,\PF{V,m,\ve}\,\chi-o(\ve^0)\,\chi$.
Using \eqref{valerie0} and 
$H_{\mathrm{f},m,\ve}^f\,P_{\Omega^f}=0$,
we then obtain
\begin{align}
\nonumber
\chi\,\big\{&\,\PF{V,m,0}-E_{V,m}^\Pf-m/2\,\big\}\,\chi+o(\ve^0)\,\chi
\\
&\grg\,\label{valerie1}
\chi\,\big\{\,\big[
\,|D_{\V{A}_{m,\ve}^d}|+V
+H_{\mathrm{f},m,\ve}^d-E_{V,m}^\Pf-m/2\,
\big]\otimes P_{\Omega^f}\,\big\}\,\chi
\\
& 
\quad+\,\label{valerie2}
\chi\,\big\{\,\big[\,
|D_{\V{A}_{m,\ve}^d}|+V+H_{\mathrm{f},m,\ve}^d
-E_{V,m,\ve}^\Pf
\,\big]\otimes P_{\Omega^f}^\bot\,\big\}\,\chi
\\
& 
\quad+\,\label{valerie3}
\chi\,\big\{\,\id^\el\otimes\id^d\otimes(H_{\mathrm{f},m,\ve}^f
-E_{V,m}^\Pf+E_{V,m,\ve}^\Pf-m/2)\,
P_{\Omega^f}^\bot\,\big\}\,\chi\,.
\end{align}
In view of \eqref{valerie0} we have
$E_{V,m,\ve}^\Pf=\inf\sigma[\,|D_{\V{A}_{m,\ve}^d}|+V
+H_{\mathrm{f},m,\ve}^d]$.
Therefore, the expression in \eqref{valerie2} is a 
non-negative quadratic form.
For sufficiently small $\ve>0$, the expression in \eqref{valerie3} 
is a non-negative quadratic form, too, 
because of 
$H_{\mathrm{f},m,\ve}^f\,P_{\Omega^f}^\bot \grg m\,P_{\Omega^f}^\bot$
and Corollary~\ref{MainCor}. 
In order to treat the remaining term in \eqref{valerie1}
we write
\begin{align} \nonumber
\big[\,&|D_{\V{A}_{m,\ve}^d}|+V
+H_{\mathrm{f},m,\ve}^d\,\big]\otimes P_{\Omega^f}
\\ \nonumber
&=\,
(\id\otimes P_{\Omega^f})\,
\big\{\,|D_{\V{A}_{m,\ve}}|+V+H_{\mathrm{f},m,\ve}\,\big\}\,
(\id\otimes P_{\Omega^f})
\\ \nonumber
&\grg\,
(\id\otimes P_{\Omega^f})\,
\big\{\ve\,|\DO|+\ve\,H_{\mathrm{f},m,\ve}-\const(\ve,m,V,e,\UV)\big\}\,
(\id\otimes P_{\Omega^f})\,.
\end{align}
In the second step we assumed that
$\ve>0$ is small enough.
Altogether, we arrive at
\begin{align}\nonumber
\chi\,\big\{\,&\PF{V,m,0}-E_{V,m}^\Pf-m/2\,\big\}\,\chi\,
+o(\ve^0)\,\chi+\ve\,\chi\,|\V{x}|^2\,\chi
\\
&\grg\,\nonumber
\chi\,\big\{\,\big(\,\ve\,|D_{\V{0}}|+\ve\,|\V{x}|^2+\ve\,H_{\mathrm{f},m,\ve}^d
-\const\,\big)
\otimes P_{\Omega^f}\,\big\}\,\chi
\\ \label{GS-Mass-Eq3}
&\grg\,
\chi\,\big\{\,\big[\,\ve\,|D_{\V{0}}|+\ve\,|\V{x}|^2
+\ve\,H_{\mathrm{f},m,\ve}^d-\const\,\big]_-\otimes P_{\Omega^f}\,\big\}\,\chi
\,,
\end{align}
where $[\cdots]_-\klg0$ denotes the negative part.
The crucial point about the previous estimate is
that both 
$|D_{\V{0}}|+|\V{x}|^2$ and $H_{\mathrm{f},m,\ve}^d$ have a purely discrete
spectrum as operators in the electron and discrete photon Hilbert
spaces. Besides $P_{\Omega^f}$ has rank one, of course. 
($\omega_\ve$ has a discrete spectrum 
{\em as an operator in $\HP_m^d=P_\ve\,\HP$} because the eigenspace in
$\HP_m^d$
corresponding to some value attained by $\omega_\ve$ is finite-dimensional.
Using $\omega_\ve\grg m>0$ it is 
then easy to see that the spectrum of its second quantization,
$H_{\mathrm{f},m,\ve}^d=d\Gamma(\omega_\ve\!\!\upharpoonright_{\HP_m^d})$, 
is discrete, too.)
In particular, we observe that
\begin{equation}\label{jacob}
W^-_{m,\ve}\,:=\,
\big[\,\ve\,|D_{\V{0}}|+\ve\,|\V{x}|^2
+\ve\,H_{\mathrm{f},m,\ve}^d-\const\,\big]_-\otimes P_{\Omega^f}
\end{equation}
is a finite rank operator, for every sufficiently small $\ve>0$,
no matter how large the value of the $(\ve,m,V,e,\UV)$-dependent
constant is.
As a simple consequence of the exponential localization 
we further know that $\chi\,|\V{x}|^2\,\chi$ is bounded.	
Using $\chi\,\big\{\,\PF{V,m,0}-E_{V,m}^\Pf-m/2\,\big\}\,\chi
\klg -(m/4)\, \chi$ we obtain the bound  
$
(o(\ve^0)-m/4)\, \operatorname{Tr}\{\chi\}
\grg \operatorname{Tr}\{\chi W^-_{m,\ve}\chi\}>-\infty
$ from \eqref{GS-Mass-Eq3}. 
Fixing $\ve>0$ sufficiently small we conclude 
$\operatorname{Tr}\{\chi\}<\infty$.
{\qed}

\smallskip

{\em The no-pair operator.}

{\proof} 
We prove Theorem~\ref{prop-gs-m} with $\sharp=\np$.
On account of Lemma~\ref{MainLem}(i) we again have
$\chi\,\NPO{V,m,0}\,\chi
\grg\chi\,\NPO{V,m,\ve}\chi-o(\ve^0)\,\chi$.
In view of \eqref{valerie0} we have
$P_{\V{A}_{m,\ve}}^+=P^+_{\V{A}_{m,\ve}^d}\otimes\id^f$
with
$P^+_{\V{A}_{m,\ve}^d}:=\id_{[0,\infty)}(D_{\V{A}_{m,\ve}^d})$
and
we observe that $\NPO{V,m,\ve}$ decomposes
under the isomorphism \eqref{SplittingDiscFluc} as
\begin{align}\label{DecompDisFluc}
\NPO{V,m,\ve}\,&=\,\ol{X_\ve^d\otimes \id^f+ 
P_{\V{A}^d_{m,\ve}}^+\otimes H_{\mathrm{f},m,\ve}^f}\,,
\\\nonumber
X_\ve^d\,&:=\,P_{\V{A}^d_{m,\ve}}^+\,(D_{\V{A}_{m,\ve}^d}+V+ H_{\mathrm{f},m,\ve}^d)
\,P_{\V{A}_{m,\ve}^d}^+\,.
\end{align}
Writing $\id^{\el}\otimes\id^d= P_{\V{A}^d_{m,\ve}}^++ P_{\V{A}^d_{m,\ve}}^-$
and $\mathbbm{1}^f=P_{\Omega^f}+ P_{\Omega^f}^\bot$ and using
\eqref{DecompDisFluc} and $H_{\mathrm{f},m,\ve}^f\,P_{\Omega^f}=0$, we obtain
\begin{align}\nonumber
\chi&\,\big\{\,\NPO{V,m,0}-E^{\np}_{V,m}-m/2\,\big\}\,
\chi+o(\ve^0)\,\chi
\\
&\grg\,\label{valerie1np}
\chi\,\big\{\,\big(X_\ve^d-(E^\np_{V,m}+m/2)\,P_{\V{A}^d_{m,\ve}}^+\big)\otimes 
P_{\Omega^f}\,\big\}\,\chi
\\ \label{valerie2np}
&\quad+\, \chi\,\big\{\,\big(X_\ve^d-E^\np_{V,m,\ve}
\,P_{\V{A}^d_{m,\ve}}^+\big)\otimes 
P^\bot_{\Omega^f}\,\big\}\,\chi
\\ \label{valerie3np}
&\quad+\, \chi\,\big\{\,P_{\V{A}^d_{m,\ve}}^+\otimes 
\big(E^\np_{V,m,\ve}-E^\np_{V,m}-m/2 +H_{\mathrm{f},m,\ve}^f\big)
P^\bot_{\Omega^f}\big\}\,\chi
\\ \label{valerie4np}
&\quad-\,(E^\np_{V,m}+m/2)\, \chi\,\{P_{\V{A}_{m,\ve}^d}^-\otimes\id^f\}
\,\chi\,.
\end{align}
Next, we observe that
$E^\np_{V,m,\ve}= \inf\sigma[X_\ve^d]$ and proceed as in the proof
of Theorem~\ref{prop-gs-m} with $\sharp=\Pf$: 
We omit the expression in \eqref{valerie2np} 
which is a non-negative quadratic form. 
For sufficiently small $\ve>0$,
the term in \eqref{valerie3np} is non-negative also,
since $H_{\mathrm{f},m,\ve}^f P^\bot_{\Omega^f}\grg m\,P^\bot_{\Omega^f}$ and  
$\lim_{\ve \to 0}E^\np_{V,m,\ve}=E^\np_{V,m}$ by Corollary \ref{MainCor}.
The term in \eqref{valerie4np}, is some
$o(\ve^0)$ on account of 
$\chi\,\{P_{\V{A}_{m,\ve}^d}^-\otimes\id^f\}\,\chi
=\chi\,(P_{\V{A}_{m}^>}^+-P_{\V{A}_{m,\ve}}^+)\,\chi$,
\eqref{veronique5}, and the boundedness of $e^F\,\HT^{1/2}\,\chi$.
(The latter follows from the exponential
localization and higher order estimates.)
Putting all these remarks together
and applying \eqref{eqqq22} in the last step
we obtain
\begin{align*}
\chi\,\big\{&\NPO{V,m,0}-E^{\np}_{V,m}-m/2\big\}\,\chi+o(\ve^0)\,\chi
\\
&\grg\,
\chi\,\big\{\,\big(X_\ve^d-(E^\np_{V,m}+m/2)\,P_{\V{A}^d_{m,\ve}}^+\big)\otimes 
P_{\Omega^f}\,\big\}\,\chi
\\
&=\,\chi\,(\id\otimes P_{\Omega^f})\,
\big\{\NPO{V,m,\ve}-(E^\np_{V,m}+m/2)\,P_{\V{A}_{m,\ve}}^+\big\}
\,(\id\otimes P_{\Omega^f})\,\chi
\\
&\grg\,
\chi\,P_{\V{A}_{m,\ve}}^+
\big\{\,\big[\ve\,|\DO|+\ve|\V{x}|^2+\ve\,H^d_{f,m,\ve}
-\const\big]_-
\otimes P_{\Omega^f}\big\}
\,P_{\V{A}_{m,\ve}}^+\,\chi
\\
&\qquad-\ve\,\chi\,P_{\V{A}_{m,\ve}}^+\,
\{|\V{x}|^2\otimes P_{\Omega^f}\}\,P_{\V{A}_{m,\ve}}^+\,\chi\,,
\end{align*}
where the constant in the penultimate line again depends on
$\ve$, $m$, $V$, $e$, and $\UV$, and the
operator in the last line is bounded due to the localization
of $\chi$.
We may thus conclude as in the end
of the proof of Theorem~\ref{prop-gs-m} with $\sharp=\Pf$.
{\qed}



\section{Infra-red bounds}\label{InfraredBounds}

\noindent
The final step in the proof of the existence of ground
states is a compactness argument showing
that a sequence of normalized
ground state eigenfunctions, $\phi_{m_j}$, $m_j\searrow0$,
of operators with photon masses $m_j$ contains a
strongly convergent subsequence whose limit turns out to be
a ground state eigenfunction for the original operator.
This compactness argument is explained in the
subsequent Section~\ref{ExistenceGroundStates}.
As a preparation we now discuss the infra-red bounds
stated in the following proposition. They are proved in 
\cite{KMS2009a} for
the semi-relativistic Pauli-Fierz operator and in \cite{KMS2009b}
for the no-pair operator starting from a 
suitable representation of $a(k)\,\phi_m$.
In order not to lengthen the present exposition too much
we only outline the proof of the soft photon bound for
the semi-relativistic Pauli-Fierz operator 
in Subsection~\ref{ssec-spb-PF}.
We recall the notation
$$
(a(k)\,\psi)^{(n)}(k_1,\dots,k_n)\,=\,
(n+1)^{1/2}\,\psi^{(n+1)}(k,k_1,\dots,k_n)\,,\quad n\in\NN_0\,,
$$
almost everywhere, where $\psi=(\psi^{(n)})_{n=0}^\infty\in\Fock[\HP]$,
and $a(k)\,\Omega=0$.

\begin{proposition}\label{prop-spb}
Let $e\in\RR$, $\UV\in(0,\infty)$, and 
assume that $V$ fulfills Hypotheses~\ref{hyp-V1} and~\ref{hyp-V2}..
Then there is a constant, $C>0$,
such that, for all sufficiently small $m>0$ and every normalized
ground state eigenvector, $\phi_{m}^\sharp$, of $\Hs{V,m}$,
we have the following soft photon bound,
\begin{equation}\label{eq-spb}
\big\|\,a(k)\,\phi_{m}^\sharp\,\big\|^2\,\klg\,\id_{\{m\klg|\V{k}|\klg\UV\}}
\:\frac{C}{|\V{k}|}\,,
\end{equation}
for almost every $k=(\V{k},\lambda)\in\RR^3\times\ZZ_2$,
as well as the following photon derivative bound,
\begin{equation}\label{pdb-kp}
\big\|\,a(k)\,\phi_{m}^\sharp-a(p)\,\phi_{m}^\sharp\,\big\|
\,\klg\,
C\,|\V{k}-\V{p}|\,
\Big(\frac{1}{|\V{k}|^{1/2}|\V{k}_\bot|}\,+\,
\frac{1}{|\V{p}|^{1/2}|\V{p}_\bot|}\Big)\,,
\end{equation}
for almost every $k=(\V{k},\lambda),p=(\V{p},\mu)\in\RR^3\times\ZZ_2$
with $m<|\V{k}|<\UV$ and $m<|\V{p}|<\UV$. 
\end{proposition}

\smallskip

\noindent
We remark that the photon derivative
bound \eqref{pdb-kp} is actually the only place in the whole article
where the special choice of the polarization vectors \eqref{pol-vec}
enters into the analysis.

\subsection{The gauge transformed operator}

\noindent
In order to derive the infra-red bounds \eqref{eq-spb}
and \eqref{pdb-kp} by the method outlined in
Subsection~\ref{ssec-spb-PF} it is necessary to pass to a suitable
gauge \cite{BFS1999,GLL2001}. For otherwise we
would end up with a more singular infra-red behavior
of their right hand sides.
To define an appropriate operator-valued gauge transformation 
\cite{GLL2001} we recall that,
for $i,j\in\{1,2,3\}$, the components $A^{(i)}_m(\V{x})$ and $A^{(j)}_m(\V{y})$
of the magnetic vector potential at $\V{x},\V{y}\in\RR^3$
commute in the sense that all their spectral projections
commute; see, e.g., Theorem~X.43 of
\cite{ReedSimonII}. 
Therefore, it makes sense to define
\begin{equation*}
U:=\int_{\RR^3}^\oplus \id_{\CC^4}\otimes U_\V{x}\,d^3\V{x}\,,\quad
U_\V{x}:=\prod_{j=1}^3 e^{ix_jA_m^{(j)}(\V{0})},\;\;
\V{x}=(x_1,x_2,x_3)\in\RR^3.
\end{equation*}
Then
the gauge transformed vector potential is given by
\begin{equation*}
\wt{\V{A}}_m(\V{x})\,:=\,\V{A}_m(\V{x})-\V{A}_m(\V{0})
\,=\,\ad(\wt{\V{G}}_{m})+a(\wt{\V{G}}_{m})\,,
\end{equation*}
where $a^\sharp(\wt{\V{G}}_{m})=\int_{\RR^3}^\oplus\id_{\CC^4}\otimes a^\sharp(
\wt{\V{G}}_{\V{x},m})\,d^3\V{x}$, and
\begin{equation*}
\wt{\V{G}}_{\V{x},m}(k)\,:=\,-e\,
\frac{\id_{\{m\klg|\V{k}|\klg\UV\}}}{2\pi\sqrt{|\V{k}|}}\,
(e^{i\V{k}\cdot\V{x}}-1)\,
\veps(k)\,=\,(e^{i\V{k}\cdot\V{x}}-1)
\V{G}_{\V{0},m}(k)\,,
\end{equation*}
for all $\V{x}\in\RR^3$
and almost every $k=(\V{k},\lambda)\in\RR^3\times\ZZ_2$.
Here $\V{G}_{\V{x},m}$ is defined in \eqref{def-Gmx}.
In fact, using $[U\,,\,\valpha\cdot\V{A}_m]=0$ we deduce that
\begin{equation*}
U\, D_{\V{A}_m}\,U^*\,=\,D_{\wt{\V{A}}_m},
\qquad U\,P^+_{\V{A}_m}\,U^*\,=\,P^+_{\wt{\V{A}}_m},\qquad
U\,|D_{\V{A}_m}|\,U^*\,=\,|D_{\wt{\V{A}}_m}|\,.
\end{equation*}
The crucial point observed in \cite{BFS1999} is that 
the transformed vector potential $\wt{\V{A}}_m$
has a better infra-red behavior than $\V{A}_m$
in view of the estimate
\begin{equation}\label{laura1}
|\wt{\V{G}}_{\V{x},m}(k)|\,\klg\,
|\V{k}|\,|\V{x}|\,|\V{G}_{\V{0},m}(k)|\,.
\end{equation}
In particular, infra-red divergent (for $m\searrow0$)
integrals appearing in the derivation of the soft photon bound
are avoided when we work with $\wt{\V{A}}_m$ instead of $\V{A}_m$.
We further set
\begin{align}\label{Htilde}
\Hft\,&:=\,U\, \Hf \,U^*\,=\,\Hf+
i\V{x}\cdot\big( a(\omega\, \V{G}_{\V{0},m})-
\ad(\omega\, \V{G}_{\V{0},m})\big)
\\ \nonumber
&\qquad\qquad\qquad\qquad
+2\,
\SPn{\omega\,\V{x}\cdot\V{G}_{\V{0},m}}{\V{x}\cdot\V{G}_{\V{0},m}}\,,
\\\nonumber
\wt{H}^{\Pf}_{V,m}\,&:=\,U\,\PF{V,m}\,U^*
\,=\,
|D_{\wt{\V{A}}_m}|+V+\Hft\,,
\\
\wt{H}^{\np}_{V,m}\,&:=\,U\,\NPO{V,m}\,U^*
\,=\,\nonumber
P^+_{\wt{\V{A}}_m}\,(D_{\wt{\V{A}}_m}+V+\Hft)\,P^+_{\wt{\V{A}}_m}\,,
\\
\wt{\phi}_{m}^\sharp\,&:=\,U\,\phi_{m}^\sharp\,,
\end{align}
so that $\wt{\phi}_{m}^\sharp$ is a ground state eigenfunction of 
$\wt{H}^{\sharp}_{V,m}$. 
One can easily show that, if the infra-red
bounds \eqref{eq-spb} and \eqref{pdb-kp} hold true
with $\phi_m^\sharp$ replaced by $\wt{\phi}_{m}^\sharp$,
then they are valid for $\phi_m^\sharp$ as well
(with a different constant, of course).


\subsection{Soft photon bound for the 
semi-relativistic Pauli-Fierz operator}\label{ssec-spb-PF}

\noindent
To simplify the notation we write $\wt{\phi}_m$ instead of
$\wt{\phi}_m^\Pf$ in this subsection.
A brief calculation yields
\begin{align}
0&\,\klg\, \SPb{(\,\wt{H}^{\Pf}_{V,m}\,-\, E^{\Pf}_{V,m}\,)
\, a(k)\,\wt{\phi}_m}{ a(k)\,\wt{\phi}_m}
\nonumber\,=\,
 \SPb{[\,\wt{H}^{\Pf}_{V,m},\, a(k)\,]\,\wt{\phi}_m}{ a(k)\,\wt{\phi}_m}
\\ \nonumber
&\,=\, \SPb{[\,S_{\wt{\V{A}}_m},\, a(k)\,]\,D_{\wt{\V{A}}_m}\,\wt{\phi}_m}{ 
a(k)\,\wt{\phi}_m}
+\SPb{S_{\wt{\V{A}}_m}\,[\,D_{\wt{\V{A}}_m},\, a(k)\,]
\,\wt{\phi}_m}{ a(k)\,\wt{\phi}_m}
\\
&\qquad\label{MainInequality}
+\SPb{[\,\Hft,\, a(k)\,]\,\wt{\phi}_m}{ a(k)\,\wt{\phi}_m}.
\end{align}
Combining the commutation relations
\begin{gather*}
[\, a(k)\,,\, \ad(f)\,]=f(k),\quad [\, a(k)\,,\, a(f)\,]=0,\quad
[\, a(k)\,,\, \Hf\,]=\omega(k)\,a(k)\,.
\end{gather*}
with the formula \eqref{Htilde} for $\Hft$ we obtain
$$
[\,\Hft\,,\,a(k)\,]= -\omega(k)\,a(k)+
i\omega(k)\,\V{x}\cdot\V{G}_{\V{0},m}(k)\,.
$$
Furthermore, we get
$
[D_{\wt{\V{A}}_m},\, a(k)]=-\valpha \cdot \wt{\V{G}}_{\V{x},m}(k) 
$
and
\begin{align}\label{eq77}
[\,S_{\wt{\V{A}}_m}\,,\, a(k)\,]\,=\, 
  \int_{-\infty}^\infty R_{\wt{\V{A}}_{m}}(iy)
  \,[\,a(k)\,,\,D_{\wt{\V{A}}_m}\, ]\,R_{\wt{\V{A}}_{m}}(iy) \,\frac{dy}{\pi},
\end{align}
where $R_{\wt{\V{A}}_{m}}(iy)=(D_{\wt{\V{A}}_{m}}-iy)^{-1}$.
Next, we insert \eqref{eq77} into \eqref{MainInequality},
move the term containing
$\omega(k)\,a(k)$ to the left hand side, and divide by $\omega(k)$.
Furthermore, we introduce a weight function,
$F(\V{x})= a\, |\V{x}|^2/\sqrt{1+|\V{x}|^2}$, for some small
$a>0$. Abbreviating
\begin{gather*}
D_{\wt{\V{A}}_{m}}^F:=e^F\,D_{\wt{\V{A}}_{m}}\,e^{-F}=
D_{\wt{\V{A}}_{m}}+i\valpha\cdot\nabla F,\quad
R_{\wt{\V{A}}_{m}}^F(iy):=(D_{\wt{\V{A}}_{m}}^F-iy)^{-1},
\end{gather*}
we obtain the following result,
\begin{align} \nonumber
\|\,a(k)\,\wt{\phi}_m\,\|^2
&\,\klg\,  \int_{-\infty}^\infty 
\SPb{
R_{\wt{\V{A}}_{m}}(iy)\,\big\{
\valpha\cdot\wt{\V{G}}_{\V{x},m}\,e^{-F}\,|\V{k}|^{-1}\big\}\,\times
\\ \nonumber
&\qquad\qquad\quad
\times\,R_{\wt{\V{A}}_{m}}^F(iy)\,
D_{\wt{\V{A}}_{m}}^F\,e^F\,\wt{\phi}_m}{a(k)\,\wt{\phi}_m}\,
\frac{dy}{\pi}
\\ \nonumber
&\quad-\, \SPb{S_{\wt{\V{A}}_m}\,
\big\{\valpha\cdot\wt{\V{G}}_{\V{x},m}\,e^{-F}\,|\V{k}|^{-1}\big\}
\,e^F\,\wt{\phi}_m}{ a(k)\,\wt{\phi}_m}
\\ 
&\quad
-i\V{G}_{\V{0},m}(k)\cdot\SPb{(\V{x}\,e^{-F})\,e^F\,\wt{\phi}_m}{
a(k)\,\wt{\phi}_m}\,.\label{MainInequalityII}
\end{align}
The purpose of the exponentials $e^{-F}$ introduced above is to
control the factor $|\V{x}|$ coming from \eqref{laura1}.
In fact,
\begin{equation*}
\|\valpha\cdot\wt{\V{G}}_{\V{x},m}\,e^{-F}\|
\klg\sqrt{2}\,\sup_{\V{x}}|\wt{\V{G}}_{\V{x},m}\,e^{-F(\V{x})}|
\klg\bigO(1)\,
|\V{k}|^{1/2}\,\id_{\{m\klg|\V{k}|\klg\UV\}}\,.
\end{equation*}
Using this it is easy to see that the sum of the last two
expressions in \eqref{MainInequalityII} is not greater than 
$
\bigO(1)\,|\V{k}|^{-1}\,\id_{\{m\klg|\V{k}|\klg\UV\}}\,\|e^F\,\wt{\phi}_m\|^2 
+\|a(k)\,\wt{\phi}_m \|^2/2
$. 
By virtue of the second resolvent identity we further get
\begin{align*}
R_{\wt{\V{A}}_{m}}^F(iy)&\,D_{\wt{\V{A}}_{m}}^F
=
\big(1-R_{\wt{\V{A}}_{m}}^F(iy)\,
(i\valpha\cdot\nabla F)\big)
 R_{\wt{\V{A}}_{m}}(iy)\,(D_{\wt{\V{A}}_{m}}
+i\valpha\cdot\nabla F)
\,.
\end{align*}
Since we have
$\|R_{\wt{\V{A}}_{m}}^F(iy)\|,\|R_{\wt{\V{A}}_{m}}(iy)\|
\klg\bigO(1)\,(1+|y|)^{-1}$ by \eqref{exp-marah1}
and 
$$
\|R_{\wt{\V{A}}_{m}}(iy)\,|D_{\wt{\V{A}}_{m}}|^{1/2}\|\klg\bigO(1)
(1+|y|)^{-1/2},
$$
we thus obtain
\begin{align*}
\big\|\,R_{\wt{\V{A}}_{m}}^F(iy)&\,D_{\wt{\V{A}}_{m}}^F\,e^F\,\wt{\phi}_m\,\big\|
\,\klg \,\bigO(1)\,
(1+|y|)^{-1/2}\,\|\,|D_{\wt{\V{A}}_{m}}|^{1/2}\,e^F\,\wt{\phi}_m\|\,.
\end{align*}
Therefore, the integral in \eqref{MainInequalityII} converges absolutely 
and we arrive at
\begin{align*}
\|a(k)\,\wt{\phi}_m\|^2
\klg&\, \bigO(1)\,|\V{k}|^{-1}\id_{\{m\klg|\V{k}|\klg\UV\}}
\,\|\,|D_{\wt{\V{A}}_{m}}|^{1/2}\,e^F\,\wt{\phi}_m\|^2+
3\,\|a(k)\,\wt{\phi}_m \|^2/4.
\end{align*}
If $a>0$ is small enough, then
$\|\,|D_{\wt{\V{A}}_{m}}|^{1/2}\,e^F\,\wt{\phi}_m\|$ 
is bounded uniformly in $m>0$,  
due to a strengthened version of our exponential localization
estimates; see Lemma~5.4 of \cite{KMS2009a} whose
proof works for every potential $V$ fulfilling
Hypothesis~\ref{hyp-V1}.
Together with the last remark in the preceding subsection
this yields the soft photon bound.

The above calculations illustrate the importance of the formal
gauge invariance of our models. In fact,
without the gauge invariance we could perform these
calculations only for $\V{G}_{\V{x},m}$ 
instead of $\wt{\V{G}}_{\V{x},m}$.
In this case we would, however, end up with an upper bound
on $\|a(k)\,{\phi}_m\|^2$ of the form
$
\bigO(1)\,|\V{k}|^{-3}\id_{\{m\klg|\V{k}|\klg\UV\}}
$, which is not integrable near zero and, hence, is
not suitable for the arguments presented in the following
section.
%
%
\section{Existence of ground states}\label{ExistenceGroundStates}
\noindent
We have now collected all prerequisites to show that
$\inf\spec[\Hs{V}]$ is an eigenvalue of $\Hs{V}$.
The final compactness argument is presented in the first
subsection below. 
The main idea behind it is borrowed from \cite{GLL2001}:
Namely, to show that $\{\phi^\sharp_{m_j}\}_j$, $m_j\searrow0$,
contains strongly convergent subsequences we may restrict
our attention to finitely many Fock space sectors and to
a compact subset of the $\V{k}$-space. In fact, this is
a consequence of the soft photon bounds. 
On account of the exponential localization estimates 
it further suffices to consider compact subsets of the $\V{x}$-space.
Moreover,
the photon derivative bounds and the localization in energy
lead to bounds on the (half-)derivatives of the vectors
$\phi^\sharp_{m_j}$ w.r.t. $\V{x}$ and $\V{k}$ on compact sets and 
in the finitely many Fock space sectors.
The idea proposed in \cite{GLL2001} is to exploit
such information by applying suitable compact embedding
theorems.
Essentially, we only have to replace
the Rellich-Kondrashov theorem applied there by a
suitable embedding theorem for spaces of functions with fractional
derivatives. (In the non-relativistic case the ground states 
$\phi_{m}^\sharp$
possess weak derivatives with respect to the electron coordinates,
whereas in our case we only have Inequality~\eqref{tina} below.
For a variant of the argument that avoids the Nikol$'$ski{\u\i} spaces
introduced below by switching the roles of the electronic position and
momentum spaces see \cite{KM2011}.)

In the second subsection we discuss
the degeneracy of the ground state energies
by applying Kramers' degeneracy theorem
similarly as in \cite{MiyaoSpohn2009}.

In the whole section the coupling function $\V{G}_{\V{x}}$
is the physical one defined in \eqref{Gphys}.


\subsection{Ground states without photon mass}\label{ssec-compact}

\noindent
We shall apply the following elementary fact:

Let $S$ be some non-negative operator acting in some Hilbert
space, $\sX$. Let $\{\eta_j\}_{j\in\NN}$ 
be some sequence in $\sX$, 
converging weakly
to some $\eta\in\sX$
such that $\eta_j\in\form(S)$ and
$\SPn{\eta_j}{S\,\eta_j}=\|S^{1/2}\,\eta_j\|\to0$.
Then $\eta$ belongs to the domain of $S$ and $S\,\eta=0$.

In fact, the linear functional
$f(\phi)=\SPn{\eta}{S^{1/2}\,\phi}$, 
$\phi\in\mathcal{D}(S^{1/2})$, satisfies
\begin{align*}
f(\phi)&=\lim_{j\to \infty}\SPn{\eta_j}{S^{1/2}\,\phi}
=\lim_{j\to \infty}\SPn{S^{1/2}\,\eta_j}{\phi}=0\,,
\end{align*}
and the self-adjointness of $S^{1/2}$ implies
$\eta\in\mathcal{D}(S^{1/2})$ and $S^{1/2}\,\eta=0$.

We remind the reader of the notation $\sharp\in\{\np,\Pf\}$.

\begin{theorem}\label{thm-ex}
Let $e\in\RR$, $\UV>0$, and assume that $V$ fulfills
Hypotheses~\ref{hyp-V1} and~\ref{hyp-V2}.
Then $E_{V}^\sharp$ is an eigenvalue of $\Hs{V}$.
\end{theorem}

{\proof}
For every sufficiently small $m>0$, there is some
normalized
ground state eigenfunction, $\phi_m^\sharp$, of $\Hs{V,m}$.
We thus find some sequence $m_j\searrow0$ such that
$\{\phi_{m_j}^\sharp\}_{j\in\NN}$ converges weakly
to some $\phi^\sharp\in\HR$.
According to Theorem~\ref{le-sb-PF4} the form domains of 
$\PF{V}$ and $\PF{V,m}$, $m>0$,
coincide whence $\phi^\Pf_{m_j}\in\form(\PF{V})$.
By the characterization of the form domain of the
no-pair operator in Theorem~\ref{thm-sb-np}
we further know that $\PA\,\phi^\np_{m_j}\in\form(\NPO{V})$.
It is trivial that $\PA\,\phi^\np_{m_j}$ converges
weakly to $\PA\,\phi^\np$ in $\PA\,\HR$.
Since $N_j:=(\PA-P^-_{\V{A}_{m_j}})\,\HT^{-1/2}$
converges to zero in norm, where $\HT=\Hf+E$, for some
sufficiently large $E\grg1$, we further have
$\SPn{\psi}{\PAm\,\phi^\np}=
\lim_{j\to\infty}\SPn{N_j\,\HT^{1/2}\,\psi}{\phi^\np_{m_j}}=0$,
for every $\psi\in\core$, whence $\PAm\,\phi^\np=0$, or,
$\PA\,\phi^\np=\phi^\np$.
Hence, 
by the remark preceding the statement it suffices to show that
$\phi^\sharp\not=0$ and
$\SPn{\phi_{m_j}^\sharp}{(\Hs{V}-E^\sharp_V)\,\phi_{m_j}^\sharp}\to0$.

The latter condition immediately follows from
$\lim_{m\searrow0}E_{V,m}=E_V$ and 
Lemma~\ref{MainLem0}(ii) which together imply
\begin{align*}
&\SPn{\phi_{m_j}^\sharp}{(\Hs{V}-E^{\sharp}_V)\,\phi_{m_j}^\sharp}=
\SPn{\phi_{m_j}^\sharp}{(\Hs{V,m_j}-E^{\sharp}_{V,m_j})\,
\phi_{m_j}^\sharp}+o(m_j^0)=o(m_j^0)\,.
\end{align*}
In order to verify that $\phi^\sharp\not=0$ we adapt the ideas
of \cite{GLL2001} sketched in the first paragraph
of this section. 

We write
$\phi_{m}^\sharp=
(\phi_{m,\sharp}^{(n)})_{n=0}^\infty\in\bigoplus_{n=0}^\infty\Fock^{(n)}[\HP]$
in what follows. Let $\ve>0$. By virtue of the soft photon bound
we find $n_0\in\NN$ and $C\in(0,\infty)$ such that, for
all sufficiently small $m>0$,
\begin{equation}\label{sabine1}
\sum_{n=n_0}^\infty\|\phi_{m,\sharp}^{(n)}\|^2
\klg\frac{1}{n_0}\sum_{n=0}^\infty n\,\|\phi_{m,\sharp}^{(n)}\|^2
=
\frac{1}{n_0}\int\|a(k)\,\phi_{m}^\sharp\|^2\,dk\,
\klg
\frac{C}{n_0}<\frac{\ve}{2}\,.
\end{equation}
By the exponential localization estimates
of Theorem~\ref{el},  which hold
uniformly for small $m>0$, we further find some $R>0$ such that
\begin{equation}\label{sabine2}
\int_{|\V{x}|\grg R/2}
\|\phi_{m}^\sharp\|_{\CC^4\otimes\Fock}^2(\V{x})\,d^3\V{x}\,<\,\frac{\ve}{2}\,.
\end{equation}
In addition,
the soft photon bound ensures that
$\phi_{m,\sharp}^{(n)}(\V{x},\vs,k_1,\dots,k_n)=0$, for almost every
$(\V{x},\vs,k_1,\dots,k_n)\in\RR^3\times\{1,2,3,4\}\times(\RR^3\times\ZZ_2)^n$,
$k_j=(\V{k}_j,\lambda_j)$,
such that $|\V{k}_j|>\Lambda$, for some $j\in\{1,\dots,n\}$.
(Here and henceforth $\vs$ labels the four spinor components.)
For $0<n<n_0$ and some fixed
$\ul{\theta}
=(\vs,\lambda_1,\dots,\lambda_n)\in\{1,2,3,4\}\times\ZZ_2^n$
we set 
$$
\phi_{m,\ul{\theta},\sharp}^{(n)}(\V{x},\V{k}_1,\dots,\V{k}_n)
\,:=\,\phi_{m,\sharp}^{(n)}(\V{x},\vs,\V{k}_1,\lambda_1,\ldots,\V{k}_n,\lambda_n)
$$
and similarly for $\phi^\sharp=(\phi_{\sharp}^{(n)})_{n=0}^\infty$.
Moreover, we set, for every $\delta\grg0$, 
$$
Q_{n,\delta}:=
\big\{\,(\V{x},\V{k}_1,\dots,\V{k}_n)\::\;
|\V{x}|< R-\delta\,,
\,\delta<|\V{k}_j|<\Lambda-\delta\,,\;j=1,\ldots,n\,\big\}\,.
$$
Fixing some small $\delta>0$ we
pick some cut-off function $\chi\in C_0^\infty(\RR^{3(n+1)},[0,1])$ such that
$\chi\equiv1$ on $Q_{n,2\delta}$ and $\supp(\chi)\subset Q_{n,\delta}$
and define 
$\psi^{(n)}_{m,\ul{\theta},\sharp}:=\chi\,\phi^{(n)}_{m,\ul{\theta},\sharp}$.
As a next step the photon derivative bound is used to
show that $\{\psi_{m,\ul{\theta},\sharp}^{(n)}\}_{m\in(0,\delta]}$
is a bounded family in the anisotropic Nikol$'$ski{\u\i} space\footnote{
For $r_1,\ldots,r_d\in[0,1]$, $q_1,\ldots,q_d\grg1$, we have
$H^{(r_1,\ldots,r_d)}_{q_1,\ldots,q_d}(\RR^d)
:=\bigcap_{i=1}^d H^{r_i}_{q_ix_i}(\RR^d)$.
For $r_i\in[0,1)$, a measurable function
$f:\RR^d\to\CC$ belongs to the class $H^{r_i}_{q_ix_i}(\RR^d)$,
if $f\in L^{q_i}(\RR^d)$ and there is some $M\in(0,\infty)$
such that
\begin{equation}\label{Nik1}
\|f(\cdot+h\,{\sf e}_i)-f\|_{L^{q_i}(\RR^d)}
\,\klg\,M\,|h|^{r_i}\,,\qquad h\in\RR\,,
\end{equation}
where ${\sf e}_i$ is the $i$-th canonical unit vector in $\RR^d$.
If $r_i=1$, then \eqref{Nik1} is replaced by
\begin{equation}\label{Nik2}
\|f(\cdot+h\,{\sf e}_i)-2f+f(\cdot-h\,{\sf e}_i)\|_{L^{q_i}(\RR^d)}\,
\klg\,M\,|h|\,,\qquad h\in\RR\,.
\end{equation}
$H^{(r_1,\ldots,r_d)}_{q_1,\ldots,q_d}(\RR^d)$ is a Banach space
with norm
$$
\|f\|^{(r_1,\ldots,r_d)}_{q_1,\ldots,q_d}\,:=\,
\max_{1\klg i\klg d}\|f\|_{L^{q_i}(\RR^d)}\,+\,
\max_{1\klg i\klg d} M_i\,,
$$ 
where $M_i$ is the infimum of all constants $M>0$
satisfying \eqref{Nik1} or \eqref{Nik2}, respectively.
Finally, we abbreviate 
$H^{(r_1,\ldots,r_d)}_q(\RR^d):=H^{(r_1,\ldots,r_d)}_{q,\ldots,q}(\RR^d)$.
}
$H^{\V{s}}_{\V{q}}(\RR^{3(n+1)})$, 
where $\V{s}=(1/2,1/2,1/2,1,\dots,1)$ and $\V{q}=(2,2,2,p,\dots,p)$
with
$p\in[1,2)$.
In fact, employing the H\"older
inequality (w.r.t. $d^3\V{x}\,d^{3(n-1)}\V{K}$) and
the photon derivative bound \eqref{pdb-kp}, 
we obtain as in \cite{GLL2001}, for 
$p\in[1,2)$, $m\in(0,\delta]$, and $\V{h}\in\RR^3$,
\begin{align*}
&
\int\limits_{{Q_{n,\delta}\cap\atop\{\delta<|\V{k}+\V{h}|<\UV\}}}\!\!\!\!\!\!\!
\big|\phi_{m,\ul{\theta},\sharp}^{(n)}(\V{x},\V{k}+\V{h},\V{K})
-\phi_{m,\ul{\theta},\sharp}^{(n)}(\V{x},\V{k},\V{K})\big|^p
\,d^3\V{x}\,d^3\V{k}\,d^{3(n-1)}\V{K}
\\
&\klg\,
C
\sum_{\lambda\in\ZZ_2}\int\limits_{{m<|\V{k}|<\UV,\atop m<|\V{k}+\V{h}|<\UV}}
\big\|\,a(\V{k}+\V{h},\lambda)\,\phi_{m}^\sharp
-a(\V{k},\lambda)\,\phi_{m}^\sharp\,\big\|^p
d^3\V{k}
\,\klg\,C'\,|\V{h}|^p,
\end{align*}
where the constants $C,C'\in(0,\infty)$ do not depend on $m\in(0,\delta]$.
Since $\phi_{m,\sharp}^{(n)}$ is symmetric in the photon variables the
previous estimate implies \cite[\textsection4.8]{Nikolskii}
that the weak first order partial derivatives of $\phi_{m,\ul{\theta}}^{(n)}$ 
with respect to its last $3n$ variables exist on 
$Q_{n,\delta}$ and that
$$
\|\phi_{m,\ul{\theta},\sharp}^{(n)}\|_{W^{\V{r}}_p(Q_{n,\delta})}^p
\,:=\,
\|\phi_{m,\ul{\theta},\sharp}^{(n)}\|^p_{L^p(Q_{n,\delta})}+
\sum_{j=1}^n\sum_{i=1}^3
\|\partial_{k_j^{(i)}}\phi_{m,\ul{\theta},\sharp}^{(n)}\|^p_{L^p(Q_{n,\delta})}
\,\klg\,C'',
$$
for $m\in(0,\delta]$ and some $m$-independent $C''\in(0,\infty)$,
with $\V{r}:=(0,0,0,1,\dots,1)$.
The previous estimate implies 
$\|\psi_{m,\ul{\theta},\sharp}^{(n)}\|_{W_p^\V{r}(\RR^{3(n+1)})}\klg C'''$,
for some $C'''\in(0,\infty)$ which does not depend on $m\in(0,\delta]$. 
Moreover, the anisotropic Sobolev space $W^{\V{r}}_p(\RR^{3(n+1)})$
is continuously embedded into $H^{\V{r}}_p(\RR^{3(n+1)})$; see, e.g.,
\cite[\textsection6.2]{Nikolskii}.
By Theorems~\ref{le-sb-PF4} and~\ref{thm-sb-np} we get, for $n\in\NN$, 
\begin{equation}\label{tina}
c^{-1}\,\SPn{\phi_{m,\sharp}^{(n)}}{|\DO|\,\phi_{m,\sharp}^{(n)}}\klg
\SPn{\phi_{m,\sharp}}{\Hs{V,m}\,\phi_{m,\sharp}}+c=
E^\sharp_{V,m}+c\klg E_V^\sharp+2c\,,
\end{equation}
for some $m$-independent $c\in(0,\infty)$.
Therefore, $\{\phi_{m,\ul{\theta},\sharp}^{(n)}\}_{m\in(0,\delta]}$ and, hence, 
$\{\psi_{m,\ul{\theta},\sharp}^{(n)}\}_{m\in(0,\delta]}$
are bounded families in the Bessel potential,
or, Liouville
space $L^{\V{r}'}_2(\RR^{3(n+1)})$, $\V{r}':=(1/2,1/2,1/2,0,\dots,0)$,
where the fractional derivatives are defined by means of the Fourier
transform. The embedding $L^{\V{r}'}_2(\RR^{3(n+1)})\to H^{\V{r}'}_2(\RR^{3(n+1)})$
is continuous, too; see
\textsection9.3 in \cite{Nikolskii}.
Altogether it follows that $\{\psi_{m,\ul{\theta},\sharp}^{(n)}\}_{m\in(0,\delta]}$
is a bounded family in $H^\V{s}_\V{q}(\RR^{3(n+1)})$.
Now,
we may apply 
Theorem~3.2 in \cite{Nikolskii1958}. The latter ensures
that $\{\psi_{m,\ul{\theta},\sharp}^{(n)}\}_{m\in(0,\delta]}$
contains a sequence which is strongly convergent
in $L^2(Q_{n,2\delta})$ provided $1-3n(p^{-1}-2^{-1})>0$.
Of course, we can choose $p<2$ large enough such that
the latter condition is fulfilled, for
all $n=1,\ldots,n_0-1$.
By finitely many repeated selections of subsequences 
we may hence assume without loss of generality
that $\{\phi_{m_j,\ul{\theta},\sharp}^{(n)}\}_{j\in\NN}$ converges strongly
in $L^2(Q_{n,2\delta})$ to its weak limit
$\phi^{(n)}_{\ul{\theta},\sharp}$, for $0\klg n<n_0$.
In particular, by the choice of $n_0$ and $R$ 
in \eqref{sabine1} and \eqref{sabine2},
\begin{equation*}
\|\phi^{\sharp}\|^2\grg
\lim_{j\to\infty}
\sum_{n=0}^{n_0-1}\sum_{\ul{\theta}}
\|\phi^{(n)}_{m_j,\ul{\theta},\sharp}\|_{L^2(Q_{n,2\delta})}^2
\grg\lim_{j\to\infty}\|\phi_{m_j}^\sharp\|^2-\ve-o(\delta^0)\,,
\end{equation*}
where we use the soft photon bound to estimate
\begin{align*}
&\sum_{n=1}^{n_0-1}\sum_{\ul{\theta}}
\big\|\,
\phi^{(n)}_{m_j,\ul{\theta},\sharp}\,\id{\{\exists\,i:\,
|\V{k}_i|\klg2\delta\,\vee\,|\V{k}_i|\grg\UV-2\delta\}}
\,\big\|^2
\\
&\klg\,
\sum_{\lambda\in\ZZ_2}\int\limits_{{\{|\V{k}|\klg2\delta\}\cup\atop
\{|\V{k}|\grg\UV-2\delta\}}}
\!\!\!\!\|a(\V{k},\lambda)\,\phi_{m_j}^\sharp\|^2\,d^3\V{k}
\,=\,o(\delta^0)\,,\qquad\delta\searrow0\,.
\end{align*}
Hence, $\|\phi^\sharp\|^2\grg1-\ve-o(\delta^0)$, 
where $\delta>0$ and $\ve>0$ are arbitrary, that is, 
$\|\phi^{\sharp}\|=1$. (In particular,
$\phi_{m_j}^\sharp\to\phi^{\sharp}$ strongly in $\HR$.)
{\qed}


\subsection{Ground state degeneracy}

\noindent
Suppose that $V(\V{x})=V(-\V{x})$.
As already mentioned in the introduction 
it is remarked in \cite{MiyaoSpohn2009}
that every (speculative) eigenvalue
of $\PF{V}$ and, in particular, its
ground state energy is evenly degenerate
in this case. The authors prove this statement
by constructing some anti-linear involution
commuting with $\PF{V}$ and applying
Kramers' degeneracy theorem.
We shall do the same for the no-pair operator
in the next theorem which originates from \cite{KMS2009b}.

\begin{theorem}
Let $e\in\RR$, $\UV>0$, assume that
$V$ fulfills Hypotheses~\ref{hyp-V1}, and assume in addition
that $V(\V{x})=V(-\V{x})$, for almost every $\V{x}$.
If the ground state energy $E_{V}^\np$ is an eigenvalue of $H_V^\np$,
then it is evenly degenerate.
\end{theorem}

{\proof}
Similarly
as in \cite{MiyaoSpohn2009} we introduce the anti-linear operator 
$$
\vt\,:=\,J\,\alpha_2\,C\,R
\,=\,
-\alpha_2\,J\,C\,R
\,,\qquad 
J\,:=\,\begin{pmatrix}0&\id_2\\-\id_2&0\end{pmatrix}\,,
$$
where $C:\HR\to\HR$ denotes complex conjugation,
$C\,\psi:=\ol{\psi}$, $\psi\in\HR$, and 
$R:\HR\to\HR$ is the parity transformation
$(R\,\psi)(\V{x}):=\psi(-\V{x})$, for almost every $\V{x}\in\RR^3$
and every $\psi\in\HR=L^2(\RR^3_\V{x},\CC^4\otimes\Fock[\HP])$.
Obviously, 
$[\vt\,,\,-i\partial_{x_j}]=[\vt\,,V]=[\vt\,,\,\Hf]=0$, on 
$\dom(|\DO|)\cap\dom(\Hf)$.
Since $\alpha_2$ squares to one and $C\,\alpha_2=-\alpha_2\,C$,
for all entries of $\alpha_2$ are purely imaginary,
we further get $\vt^2=-\id$ and $[\vt\,,\,\alpha_2]=0$. Moreover,
the Dirac matrices $\alpha_0$, $\alpha_1$, and $\alpha_3$ have real 
entries  and $[J\,\alpha_2\,,\,\alpha_j]=J\,\{\alpha_2,\alpha_j\}=0$
by \eqref{Clifford}, whence $[\vt\,,\,\alpha_j]=0$, for $j\in\{0,1,3\}$. 
Finally $[\vt\,,\,e^{\pm i\V{k}\cdot\V{x}}]=0$ implies
$[\vt\,,\,A^{(j)}]=0$ on $\dom(\Hf^{1/2})$, for $j\in\{1,2,3\}$.
It follows that $[\vt\,,\,\DA]=0$ on $\core=\vt\,\core$ and, since
$\DA$ is essentially self-adjoint on $\core$, we obtain
$\vt\,\dom(\DA)=\dom(\DA)$ and $[\vt\,,\,\DA]=0$ on $\dom(\DA)$,
which implies $\vt\,\RA{iy}-\RA{-iy}\,\vt=0$ on $\HR$, for every $y\in\RR$.
Using the representation \eqref{sgn} we conclude that
$[\vt\,,\,\PA]=0$ on $\HR$. 
In particular, $\vt$ can be considered as an operator acting
on $\HRp$.
Furthermore, we obtain
$\NPO{V}\,\vt\,\vp-\vt\,\NPO{V}\,\vp=0$,
for every $\vp\in\core$.
Since $\PA\,\sD$ is a form core for $\NPO{V}$
we can easily extend this commutation relation
and show that $\vt$ maps $\dom(\NPO{V})$ into itself
and
$\NPO{V}\,\vt\,\psi=\vt\,\NPO{V}\,\psi$, for
every $\psi\in\dom(\NPO{V})$.
Hence, by Kramers' degeneracy
theorem every eigenvalue of $\NPO{V}$ is evenly
degenerated.
(In fact, $\NPO{V}\,\phi=E_V\,\phi$ implies
$\NPO{V}\,\vt\,\phi=E_V\,\vt\,\phi$, and $\phi\bot\vt\,\phi$,
since $\SPn{\vt\,\phi}{\phi}=-\SPn{\vt\,\phi}{\vt\,(\vt\,\phi)}
=-\SPn{C\,\phi}{C\,\vt\,\phi}=-\SPn{\vt\,\phi}{\phi}$.)
{\qed}

\smallskip

\noindent
Using similar arguments 
we derive the next lemma which has been referred to
in Sections~\ref{sec-exp} and~\ref{sec-ex-m}.
We use the notation introduced in Subsection~\ref{ssec-mass}
and write
$$
{H}^{\np,-}_{0,m,\ve}\,
:=\,P_{\V{A}_{m,\ve}}^-\,(|D_{\V{A}_{m,\ve}}|+\Hfme)\,P_{\V{A}_{m,\ve}}^-
$$
for the analogs of the no-pair operators on the negative spectral
subspace which appeared in \eqref{nurleen2011} and \eqref{nurleen2012}.
Then $\wh{H}^\np_{0,m,\ve}=\NPO{0,m,\ve}+{H}^{\np,-}_{0,m,\ve}$;
see \eqref{nurleen2011} and \eqref{nurleen2012}.
We unify the notation by setting $\NPO{0,0,0}:=\NPO{0}$, etc.

\begin{lemma}\label{le-H+H-}
Let $e\in\RR$, $\UV\in(0,\infty)$, and $m\grg\ve\grg0$. Then
\begin{equation*}
\Thnp_{m,\ve}\,\stackrel{\mathrm{def.}}{=}\,\inf\spec[\NPO{0,m,\ve}]\,=\,
\inf\spec[{H}^{\np,-}_{0,m,\ve}]\,=\,\inf\spec[\wh{H}^\np_{0,m,\ve}]\,.
\end{equation*}
\end{lemma}

{\proof}
Let $C$ and $R$ be defined as in the preceding proof. 
We introduce the anti-linear
operator $\tau:\HR_m^>\to\HR_m^>$, $\tau:=\alpha_2\,C\,R$.
Similar as in the previous proof we verify
that $\tau\,\DAme=-\DAme\tau$ on $\dom(\DAme)$, thus
$\R{\V{A}_{m,\ve}}{iy}\,\tau=-\tau\,\R{\V{A}_{m,\ve}}{iy}$, thus
$\PAepm\tau=\tau\,\PAemp$ again by \eqref{sgn}. 
Notice that we use the property \eqref{xenia} of the
discretized phase in $\V{A}_{m,\ve}$ to obtain these relations
in the case $\ve>0$.
Consequently, 
$\NPO{0,m,\ve}\,\tau=
\tau\,{H}^{\np,-}_{0,m,\ve}$
on a natural dense domain (again called $\sD$) in $\HR_m^>$ 
with $\tau\sD=\sD$.
By Theorem~\ref{Cor-Friedr-Extens}
we may assume that the no-pair operators in the plus or minus spaces
are essentially self-adjoint on $\PAepm\core$, respectively,
and we readily conclude.
{\qed}


\section{Commutator estimates}
\label{sec-commutators}

\noindent
In this final section we collect some bounds
proved in \cite{MatteStockmeyer2009a}
on the operator norm of various commutators
which have been used repeatedly in the preceding
sections. In the whole section we assume that
$\V{G}_\V{x}$ fulfills Hypothesis~\ref{hyp-G}.


\subsection{Basic estimates}

\noindent
Our estimates on commutators involving the field energy $\Hf$
are based on the next lemma.
The following quantity appears in its statement and in
various estimates below,
\begin{equation}\label{def-deltanu}
\delta_\nu^2\,\equiv\,{\delta_\nu}(E)^2\,:=\,
8\int\,\frac{w_\nu(k,E)^2}{\omega(k)}\,
\|\V{G}(k)\|_\infty^2\,dk\,,\qquad E,\nu>0\,,
\end{equation}
where
$
w_\nu(k,E):=
E^{1/2-\nu}\,((E+\omega(k))^{\nu+1/2}-E^\nu\,(E+\omega(k))^{1/2})
$.
We observe that $w_{1/2}(k,E)\klg\omega(k)$ and, hence, 
$\delta_{1/2}(E)\klg2\,d_1$, $E>0$.
Moreover,
$\delta_{\nu}(E)\klg\delta_\nu(1)$, for $E\grg1.$ We recall the identity
\begin{equation}\label{Hf=dGamma}
\SPb{\Hf^{1/2}\phi}{\Hf^{1/2}\psi}=\int\omega(k)\,
\SPn{a(k)\phi}{a(k)\psi}dk\,,
\quad \phi,\psi\in\dom(\Hf^{1/2})\,,
\end{equation}
which is a consequence of the permutation
symmetry and Fubini's theorem.

\begin{lemma}\label{le-tim}
Let $\nu$, 
$E>0$, and set $\HT:=\Hf+E$. 
Then
\begin{equation}
\big\|\,[\valpha\cdot\V{A}\,,\,\HT^{-\nu}]\,\HT^{\nu}\,\big\|
\,\klg\,
\delta_\nu(E)/E^{1/2}
\,.\label{combound2}
\end{equation}
\end{lemma}

{\proof} (\cite{MatteStockmeyer2009a})
We pick $\phi,\psi\in\core$ and write
\begin{eqnarray}
\lefteqn{
\SPb{\phi}{\nonumber
[\valpha\cdot\V{A}\,,\,\HT^{-\nu}]\,\HT^{\nu}
\,\psi}
}
\\
&=&
\SPb{\phi}{
[\valpha\cdot a(\V{G})\,,\,\HT^{-\nu}]\,\HT^{\nu}
\,\psi}
\,-\,\SPb{[\valpha\cdot a(\V{G})\,,\,\HT^{-\nu}]\,\phi}{
\HT^{\nu}
\,\psi}\,.\label{heidi1}
\end{eqnarray} 
By definition of $a(k)$ and $\Hf$ we have the 
pull-through formula
$a(k)\,\theta(\Hf)\,\psi=\theta(\Hf+\omega(k))\,a(k)\,\psi$,
for almost every $k$ and every Borel function $\theta$ on $\RR$,
which leads to
\begin{align*}
[\,a(k)&\,,\,\HT^{-\nu}\,]\,\HT^{\nu}\,\psi
\\
&=\,
\big\{\big((\HT+\omega(k))^{-\nu}-\HT^{-\nu}\big)(\HT+\omega(k))^{\nu+1/2}
\big\}\,
a(k)\,\HT^{-1/2}\,\psi\,.
\end{align*}
We denote the 
operator $\{\cdots\}$ by $F(k)$. 
Then $F(k)$ is bounded and
\begin{eqnarray*}
\|F(k)\|&\klg&
\int_0^1\sup_{t\grg0}\Big|\,\frac{d}{ds}\,
\frac{(t+E+\omega(k))^{\nu+1/2}}{(t+E+s\,\omega(k))^{\nu}}\Big|\,ds
\\
&=&
-\int_0^1\frac{d}{ds}\,
\frac{(E+\omega(k))^{\nu+1/2}}{(E+s\,\omega(k))^{\nu}}\,ds
\,=\,
w_\nu(k,E)/E^{1/2}
\,.
\end{eqnarray*}
Using these remarks, the Cauchy-Schwarz inequality,
and \eqref{Hf=dGamma},
we obtain
\begin{eqnarray*}
\lefteqn{
\big|\SPb{\phi}{
[\valpha\cdot a(\V{G})\,,\,\HT^{-\nu}]\,\HT^{\nu}\,\psi}\big|
}
\\
&\klg&
\int\|\phi\|\,\|\valpha\cdot\V{G}(k)\|\,
\|F(k)\|\,\big\|\,a(k)\,\HT^{-1/2}\,\psi\,\big\|\,dk
\\
&\klg&\|\phi\|\,
\Big(2\int\frac{\|F(k)\|^2}{\omega(k)}
\,\|\V{G}(k)\|^2_\infty\,dk\Big)^{1/2}
\Big(\int\omega(k)\,\big\|\,a(k)\,\HT^{-1/2}\,\psi\,\big\|^2\,dk\Big)^{1/2}
\\
&\klg&
\frac{\delta_{\nu}(E)}{2E^{1/2}}\:
\|\phi\|\,\big\|\,\Hf^{1/2}\,\HT^{-1/2}\,\psi\,\big\|\,
\,.
\end{eqnarray*}
A similar argument applied to the
second term in \eqref{heidi1} yields
$$
\big|\SPb{[\valpha\cdot a(\V{G})\,,\,\HT^{-\nu}]\,\phi}{
\HT^{\nu}
\,\psi}\big|\,\klg\,\frac{\wt{\delta}_\nu(E)}{2E^{1/2}}\,
\big\|\,\Hf^{1/2}\,\HT^{-1/2}\,\phi\,\big\|\,\|\psi\|\,,
$$
where $\wt{\delta}_\nu(E)$ is defined by \eqref{def-deltanu}
with $w_\nu(k,E)$ replaced by
$$
\wt{w}_\nu(k,E):=
E^{1/2-\nu}(E^\nu\,(E+\omega(k))^{1/2}
-E^{2\nu}\,(E+\omega)^{1/2-\nu})\,.
$$
Evidently, $\wt{w}_\nu\klg w_\nu$, thus $\wt{\delta}_\nu\klg\delta_\nu$,
which concludes the proof.
{\qed}

\smallskip

\noindent
Choosing $E$ large enough, we 
can certainly make to norm in \eqref{combound2} as
small as we please. This observation can be exploited to ensure
that certain Neumann series converge in the next proof which yields
convenient formulas allowing to interchange the field
energy with resolvents of the Dirac operator.

We define $J:[0,1)\to\RR$ by
$J(0):=1$ and
$J(a):=\sqrt{6}/(1-a^2)$, for $a\in (0,1)$,
so that $\|R^F_{\V{A}}(iy)\|\klg J(a)\,(1+y^2)^{-1/2}$,
where $R_\V{A}^0(iy):=\RA{iy}$;
recall \eqref{exp-marah0} and \eqref{exp-marah1}.

\begin{corollary}\label{cor-T-Xi}
Let $y\in\RR$ and $F\in C^\infty(\RR^3_\V{x},\RR)$
such that $|\nabla F|\klg a<1$. 
Assume that $\nu,E>0$ satisfy $\delta_\nu\,J(a)/E^{1/2}<1$, 
and introduce the following operators, 
$T_\nu\,:=\,\ol{[\HT^{-\nu}\,,\,\valpha\cdot\V{A}]\,\HT^{\nu}}$,
\begin{align}
\Xi_{\nu}^F(iy):=\sum_{j=0}^\infty\{-&R_{\V{A}}^F(iy)\,T_\nu\}^j,
\label{def-THT}
\quad\Upsilon_{\nu}^F(iy):=\sum_{j=0}^\infty\{-T_\nu^*\,R_\V{A}^F(iy)\}^j.
\end{align}
Then $\|T_\nu\|\,\klg\,\delta_\nu/E^{1/2}$,
$
\|\Theta_{\nu}^F(iy)\|
\klg(1-\delta_\nu J(a)/E^{1/2})^{-1}
$, $\Theta\in\{\Xi,\Upsilon\}$,
and
\begin{align}
\HT^{-\nu}\,R_\V{A}^F(iy)
\,&=\,\label{eva3}
\Xi_{\nu}^F(iy)
\,R_\V{A}^F(iy)\,\HT^{-\nu}\,,
\\
R_\V{A}^F(iy)\,\HT^{-\nu}
\,&=\,\label{eva3b}
\HT^{-\nu}\,R_\V{A}^F(iy)\,\Upsilon_{\nu}^F(iy)\,,
\end{align}
In particular, $R_\V{A}^F(iy)$ maps $\dom(\Hf^\nu)$
into itself.
\end{corollary}

{\proof}
The following somewhat formal computations show the simple 
idea behind the proof
(see Section~3 of \cite{MatteStockmeyer2009a} for more details),
\begin{align*}
  \HT^{-\nu}\,R_\V{A}^F(iy)
  &=
  R_\V{A}^F(iy)\,\HT^{-\nu}+
R_\V{A}^F(iy)\,\big[\DA+i\valpha\cdot\nabla F-iy\,,\,\HT^{-\nu}\big]\,
  R_\V{A}^F(iy)
\\
  &=R_\V{A}^F(iy)\,\HT^{-\nu}+R_\V{A}^F(iy)\,
\big[\valpha\cdot\V{A},\HT^{-\nu}\big]\,\HT^{\nu}
  \big(\HT^{-\nu}\,R_\V{A}^F(iy)\big),
\end{align*}
which implies that
$R_\V{A}^F(iy)\,\HT^{-\nu}
=\left(1+R_\V{A}^F(iy)\,T_\nu\right)^{-1}\,\HT^{-\nu}\,R_\V{A}^F(iy)$.
The operator inverse appearing here is given by
the Neumann series $\Xi_{\nu}^F(iy)$
when we choose $E$ in $\HT=\Hf+E$ so large that
$\delta_\nu J(a)/E^{1/2}<1$. 
{\qed}

\subsection{Commuting projections with the field energy}\label{exopu}

\begin{lemma}\label{le-sgn}
Let $a,\kappa\in[0,1)$, let $\nu$, $E$, and $F$ be as in
Corollary~\ref{cor-T-Xi}, and assume that $F$ is bounded.
Define
$\cC_{\nu}^F:=e^F\,[\,\SA\,,\,\HT^{-\nu}\,]\,\HT^{\nu}\,e^{-F}$
on $\dom(\Hf^{\nu})$. Then
\begin{equation}\label{com-sgn}
\big\|\,|\DA|^\kappa\,\cC_\nu^F\,\big\|
\,\klg\,(1+a\,J(a))\,
\frac{{\rm const}(\kappa)\,\delta_\nu\,J(a)/E^{1/2}}{
1-\delta_\nu\,J(a)/E^{1/2}}
\,.
\end{equation}
In particular, 
$\SA$ maps $\dom(\Hf^{\nu})$ into itself
and, if $E^{1/2}>\delta_\nu$, then
the following identities hold true on $\dom(\Hf^\nu)$,
where $\cC_\nu:=(\cC_\nu^0)^*\in\LO(\HR)$,
\begin{align}\label{eva99}
\HT^{\nu}\,\SA&=\SA\,\HT^{\nu}+\cC_\nu\,\HT^{\nu}\,,
\qquad
\SA\,\HT^{\nu}=\HT^{\nu}\,\SA+\HT^{\nu}\,\cC^*_\nu\,.
\end{align}
\end{lemma}

{\proof} (\cite{MatteStockmeyer2009a})
Combining \eqref{eva3} with \eqref{sgn} we obtain,
for all $\phi\in\dom(|\DA|^{\kappa})$ and $\psi\in\dom(\Hf^{\nu})$,
\begin{align*}
\big|\SPb{&|\DA|^{\kappa}\,\phi}{\cC_{\nu}^F\,\psi}\big|
\klg
\int_\RR\big|\SPb{|\DA|^{\kappa}\,\phi}{
R_\V{A}^F(iy)\,T_\nu\,\Xi_{\nu}^F(iy)
\,R_\V{A}^F(iy)\,\psi}\big|\,\frac{dy}{\pi}
\\
&\klg\,\|T_\nu\|
\int_\RR\big\|\,|\DA|^{\kappa}\,R_\V{A}^F(iy)\,\phi\,\big\|\,
\|\Xi_{\nu}^F(iy)\|\,\|R_\V{A}^F(iy)\|\,\frac{dy}{\pi}
\;\|\phi\|\,\|\psi\|\,.
\end{align*}
Here we estimate $\|T_\nu\|$ by means of \eqref{combound2} and
we write 
$$
|\DA|^{\kappa}\,R_\V{A}^F(iy)=|\DA|^{\kappa}
\,\R{\V{A}}{iy}\,(\id-i\valpha\cdot\nabla F\,R_\V{A}^F(iy))\,,
$$ 
where $\|\,|\DA|^\kappa\,\RA{iy}\|\klg\const(\kappa)(1+y^2)^{-1/2+\kappa/2}$.
Moreover, 
$\|\Xi_{\nu}^F(iy)\|\klg(1-\delta_\nu J(a)/E^{1/2})^{-1}$, 
$y\in\RR$, by Corollary~\ref{cor-T-Xi}.
Altogether these remarks
yield the asserted estimate. 
Now, the identity 
$\SA\,\HT^{-\nu}
=\HT^{-\nu}\,\SA-\HT^{-\nu}\,(\cC_{\nu}^0)^*$
in $\LO(\HR)$
shows that $\SA$ maps the domain of $\Hf^{\nu}$ into itself
and that the first identity in \eqref{eva99} is valid.
Taking the adjoint of \eqref{eva99} and using 
$[\HT^{\nu}\,\SA]^*=\SA\,\HT^{\nu}$
(which is true since $\HT^{\nu}\,\SA$ is densely defined
and $\SA=\SA^{-1}\in\LO(\HR)$)
we also obtain the second identity in \eqref{eva99}.
{\qed}

\smallskip

\noindent
We can now prove an estimate asserted in the proof
of Theorem~\ref{le-sb-PF4}, namely 
$
\cT=\Re\Theta
\klg\ve\,|\DA|+\ve^{-1}\,\const\,d_1/E^{1/2}
$,
for $\ve\in(0,1]$ and $E\grg(4d_1)^2$,
where $\Theta:=[|\DA|,\HT^{-1/2}]\HT^{1/2}$.
In fact, this follows easily from Corollary~\ref{cor-T-Xi} and
Lemma~\ref{le-sgn} since
$\Theta=|\DA|^{1/2}\SA\{|\DA|^{1/2}\cC_{1/2}^0\}+T_{1/2}\cC_{1/2}^0-T_{1/2}\SA$.



\subsection{Double commutators}

\begin{lemma}\label{exp-le-dc}
Let  $a\in[0,1)$ and let $F$ satisfy \eqref{hyp-F<>}. Moreover,
let $\nu,E>0$ such that $\delta_\nu\,J(a)/E^{1/2}\klg1/2$.
Then
\begin{align}
\big\|\,|\DA|\,\big[\chi_1\,e^F,\,[\PA\,,\,\chi_2\,e^{-F}]\big]\,\big\|
&\klg
J(a)\prod_{i=1,2}
(a+\|\nabla\chi_i\|_\infty)\,,
\label{exp-dc-DA}
\\
\big\|\HT^{\nu}\,\big[\chi_1\,e^F,\,[\PA\,,\,\chi_2\,
e^{-F}]\big]\,\HT^{-\nu}\big\|
&\klg
8\,J(a)\prod_{i=1,2}
(a+\|\nabla\chi_i\|_\infty)\,.
\label{exp-dc-HT}
\end{align}
Moreover, for every self-adjoint multiplication operator $V$
in $L^2(\RR^3)$ satisfying $H^1(\RR^3)\subset\dom(V)$,
we find some constant $C_V\in(0,\infty)$, depending only on $V$,
such that
\begin{align}
\big\|\,|V|\,\big[\chi_1\,e^F,\,[\PA\,,\,\chi_2\,e^{-F}]\big]
\,\HT^{-1/2}\big\|
&\klg
C_V J(a)\prod_{i=1,2}
(a+\|\nabla\chi_i\|_\infty)
\,.
\label{exp-dc-VC}
\end{align}
In \eqref{exp-dc-VC} we also assume that $E\grg(4d_1\,J(a))^2$ and $E\grg1$.
\end{lemma}

{\proof} (\cite{MatteStockmeyer2009a})
Let $\phi,\psi\in\core$, $\|\phi\|=\|\psi\|=1$.
First, we derive a bound on
$$
I_{\phi,\psi}\,:=\,
\int_{\RR}\Big|\SPB{|\DA|\,\phi}{\HT^\nu\,
\big[\,\chi_1\,e^F\,,\,[\RA{iy}\,,\,\chi_2\,e^{-F}]\,\big]\,
\HT^{-\nu}\,\psi}\Big|\,\frac{dy}{2\pi}\,.
$$
Expanding the double commutator we get
$$
\big[\,\chi_1\,e^F\,,\,[\RA{iy}\,,\,\chi_2\,e^{-F}]\,\big]
\,=\,\eta(\chi_1,\chi_2,F\,;y)\,+\,
\eta(\chi_2,\chi_1,-F\,;y)\,,
$$
where
\begin{eqnarray*}
\lefteqn{
\eta(\chi_1,\chi_2,F\,;y)
}
\\
&:=&
\RA{iy}\,\valpha\cdot(\nabla\chi_1+\chi_1\,\nabla F)\,e^F\,\RA{iy}\,e^{-F}
\,\valpha\cdot(\nabla\chi_2-\chi_2\,\nabla F)\,\RA{iy}
\,.
\end{eqnarray*}
We obtain
\begin{align}
\nonumber
&\int_{\RR}\Big|\SPB{|\DA|\,\phi}{\HT^\nu\,
\eta(\chi_1,\chi_2,F\,;y)\,
\HT^{-\nu}\,\psi}\Big|\,\frac{dy}{2\pi}
\\
&\klg\nonumber
\int_{\RR}\Big|\SPB{\phi}{
|\DA|\,\RA{iy}\,\Upsilon_{\nu}^0(iy)\,
\valpha\cdot(\nabla\chi_1+\chi_1\,\nabla F)
\,\times
\\
& \qquad\;\times\nonumber
\,R_\V{A}^F(iy)\,\Upsilon_{\nu}^F(iy)\,
\valpha\cdot(\nabla\chi_2-\chi_2\,\nabla F)\,\RA{iy}\,
\Upsilon_{\nu}^0(iy)\,\psi
}\Big|\,\frac{dy}{2\pi}
\\
&\klg\label{exp-yvette1}
\frac{(a+\|\nabla\chi_1\|)(a+\|\nabla\chi_2\|)}{(1-\delta_\nu/E^{1/2})^2}
\cdot
\frac{J(a)}{1-\delta_\nu\,J(a)/E^{1/2}}
\int_{\RR}\frac{dy}{2\pi(1+y^2)}\,.
\end{align}
A bound analogous to \eqref{exp-yvette1}
holds true when the roles of
$\chi_1$ and $\chi_2$ are interchanged and $F$ is
replaced by $-F$. Consequently, $I_{\phi,\psi}$ is bounded
by two times the right hand side of \eqref{exp-yvette1}.
Altogether
this shows that \eqref{exp-dc-DA} and \eqref{exp-dc-HT} hold true.
(Just ignore $|\DA|$ or $\HT$, respectively, in the above argument.)
By the closed graph theorem we further have
$\|\,|V|\,\psi\|^2\klg\const(V)\,(\|\nabla\psi\|^2+\|\psi\|^2)$,
$\psi\in H^1(\RR^3)$. Hence,
\eqref{exp-dc-VC} follows from \eqref{exp-dc-DA}, \eqref{exp-dc-HT},
and the inequality
\begin{equation*}
\big\|\,|V|\,\vp\,\big\|^2\,\klg\,
\const(V)\Big(\big\|\,|\DA|\,\vp\,\big\|^2+\big\|\,\HT^{1/2}\,\vp\,\big\|^2\Big)\,,
\qquad \vp\in\core\,,
\end{equation*}
which holds true, for $E\grg d_1^2$, by
\eqref{dia3}, \eqref{bea1}, and 
$$
\|\,|V|\,\vp\|^2\klg\const(V)\,(\|\nabla\llb\vp\rrb\,\|^2+\|\vp\|^2)\,,
\quad
\vp\in\sD\,.
$$
\qed


\section*{Acknowledgments}
This work has been partially supported
by the DFG (SFB/TR12).
O.M. and E.S. thank the Institute for Mathematical Sciences
and the Center for Quantum Technologies of the National University
of Singapore for their generous hospitality.


\def\cprime{$'$} \def\cprime{$'$} \def\cprime{$'$} \def\cprime{$'$}
  \def\cprime{$'$}

\end{document}